\documentclass[runningheads]{llncs}

\usepackage{eccv}

\usepackage{eccvabbrv}

\usepackage{graphicx}
\usepackage{booktabs}

\usepackage[accsupp]{axessibility}  

\usepackage{hyperref}

\usepackage{orcidlink}

\usepackage{bm}
\usepackage{caption}
\usepackage{tabularx}
\usepackage{alphalph}

\begin{document}

\title{DragVideo: Interactive Drag-style Video Editing}

\titlerunning{DragVideo}

\author{
    Yufan Deng\thanks{Equal contribution. The order of authorship was determined alphabetically.}\inst{1}\orcidlink{0009-0008-2899-3055} \and
    Ruida Wang$^\star$\inst{1}\orcidlink{0009-0005-1497-6914} \and
    Yuhao Zhang$^\star$\inst{1}\orcidlink{0009-0008-5137-1211} \and \\
    Yu-Wing Tai\inst{2}\orcidlink{0000-0002-3148-0380} \and
    Chi-Keung Tang\inst{1}\orcidlink{0000-0001-7155-2919}
}

\authorrunning{Deng et al.}

\institute{Hong Kong University of Science and Technology, \\ Clear Water Bay, Kowloon, Hong Kong \and
Dartmouth College Hanover, NH 03755, USA \\
\email{\{ydengbd, rwangbr, yzhanglp\}@connect.ust.hk} \\
\email{yu-wing.tai@darthmouth.edu} \qquad
\email{cktang@cs.ust.hk}
}

\maketitle

\newcommand\tsWidth{0.00cm}
\newcommand\tsmWidth{0.15cm}
\newcommand\tsHeight{-0.05cm}

\begin{center}
    \vspace{-0.1in}
    \captionsetup{type=figure}
    \begin{subfigure}{\linewidth}
        \centering
        \subfloat{\includegraphics[width = 0.114\linewidth]{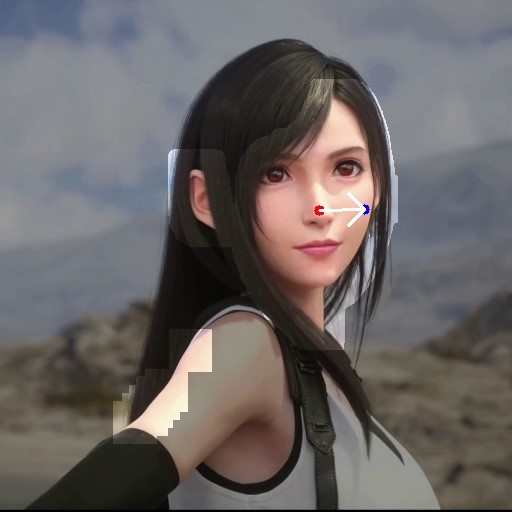}}\hspace{\tsWidth}
        \subfloat{\includegraphics[width = 0.114\linewidth]{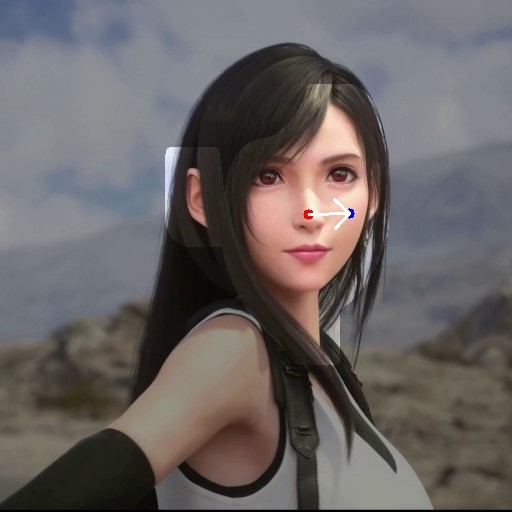}}\hspace{\tsWidth}
        \subfloat{\includegraphics[width = 0.114\linewidth]{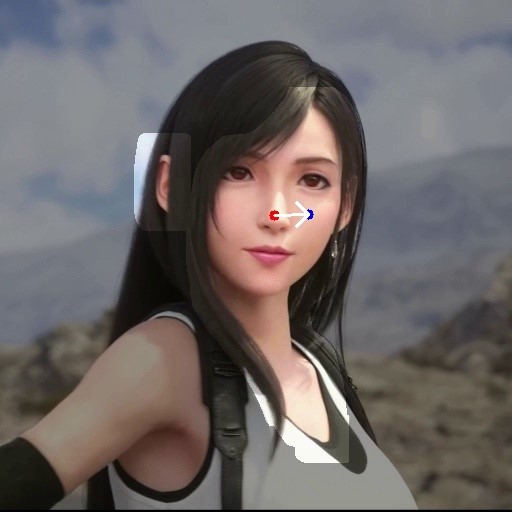}}\hspace{\tsWidth}
        \subfloat{\includegraphics[width = 0.114\linewidth]{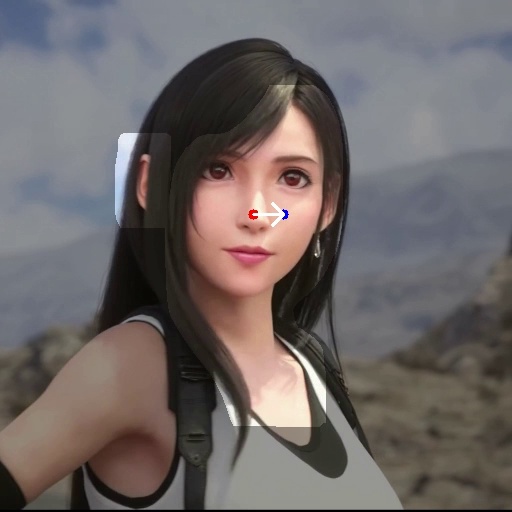}}\hspace{\tsmWidth}
        \subfloat{\includegraphics[width = 0.114\linewidth]{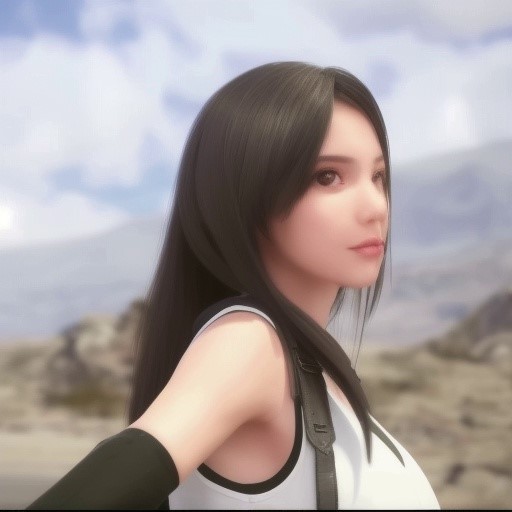}}\hspace{\tsWidth}
        \subfloat{\includegraphics[width = 0.114\linewidth]{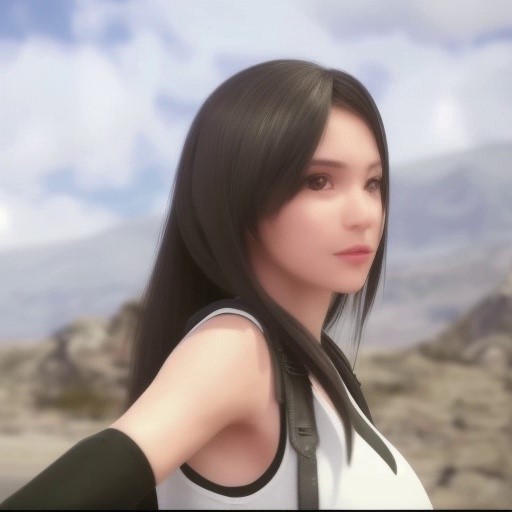}}\hspace{\tsWidth}
        \subfloat{\includegraphics[width = 0.114\linewidth]{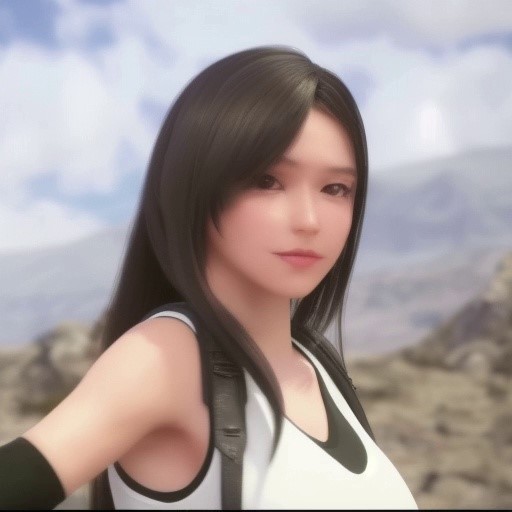}}\hspace{\tsWidth}
        \subfloat{\includegraphics[width = 0.114\linewidth]{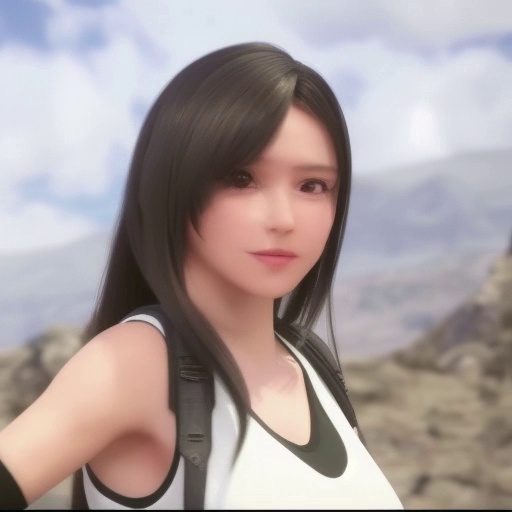}}
        \vspace{-0.1cm}
        \addtocounter{subfigure}{-8}
        \caption{Turn face}
        \label{subfig:tifaturn}
    \end{subfigure}

    \vspace{\tsHeight}
    \begin{subfigure}{\linewidth}
        \centering
        \subfloat{\includegraphics[width = 0.114\linewidth]{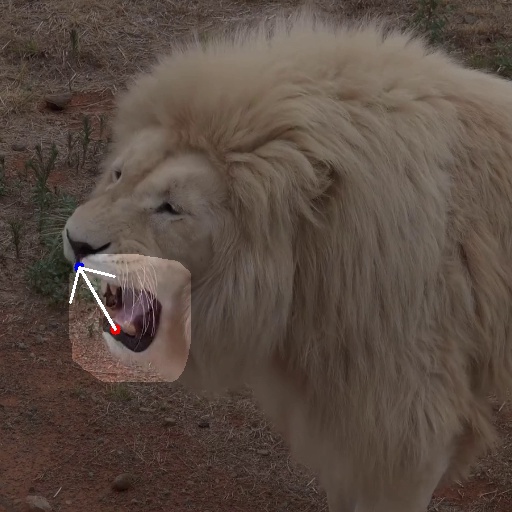}}\hspace{\tsWidth}
        \subfloat{\includegraphics[width = 0.114\linewidth]{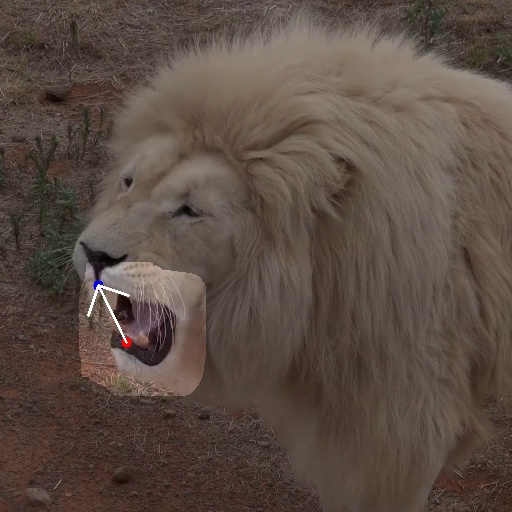}}\hspace{\tsWidth}
        \subfloat{\includegraphics[width = 0.114\linewidth]{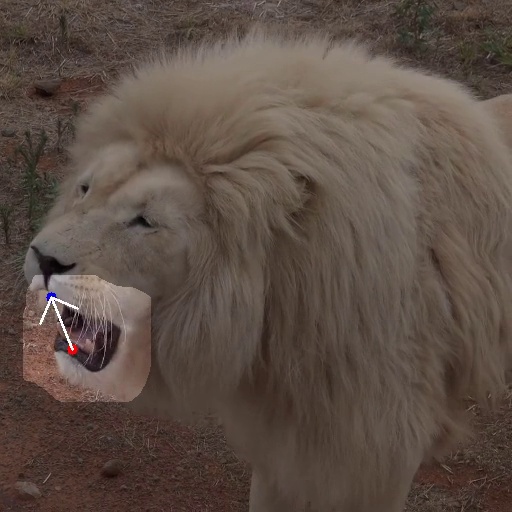}}\hspace{\tsWidth}
        \subfloat{\includegraphics[width = 0.114\linewidth]{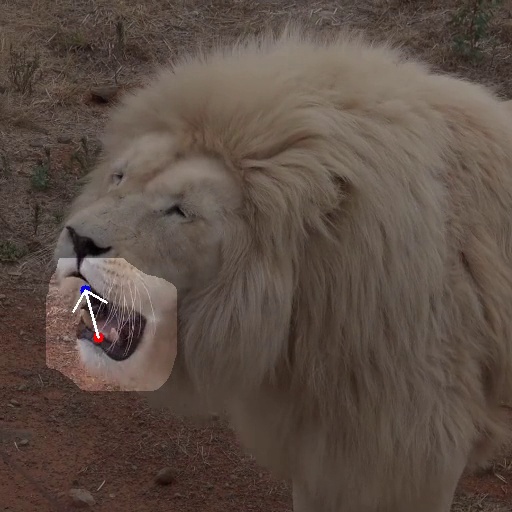}}\hspace{\tsmWidth}
        \subfloat{\includegraphics[width = 0.114\linewidth]{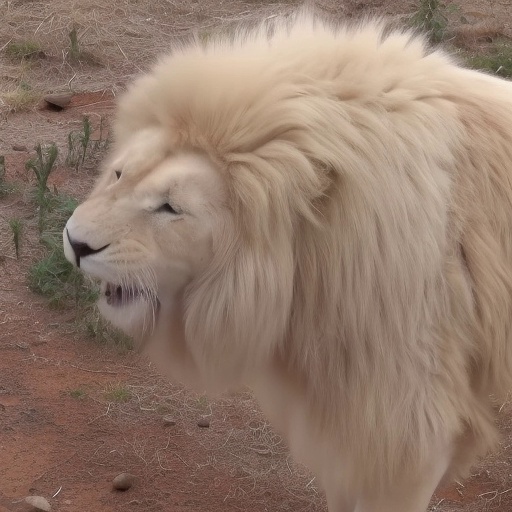}}\hspace{\tsWidth}
        \subfloat{\includegraphics[width = 0.114\linewidth]{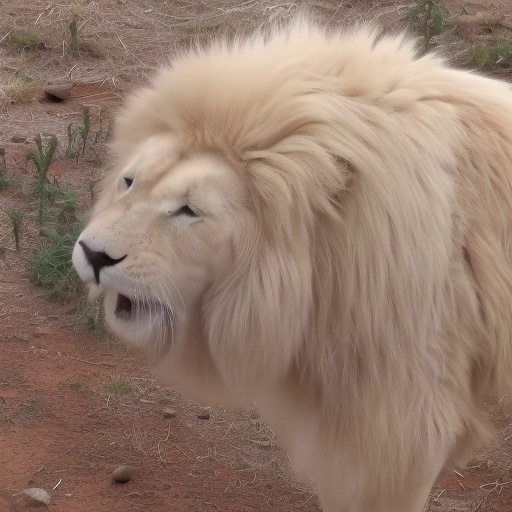}}\hspace{\tsWidth}
        \subfloat{\includegraphics[width = 0.114\linewidth]{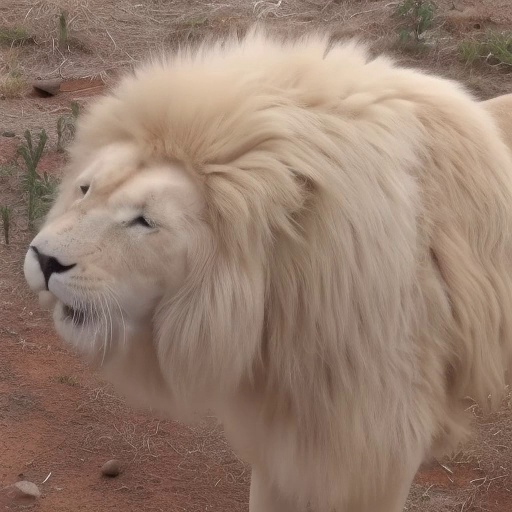}}\hspace{\tsWidth}
        \subfloat{\includegraphics[width = 0.114\linewidth]{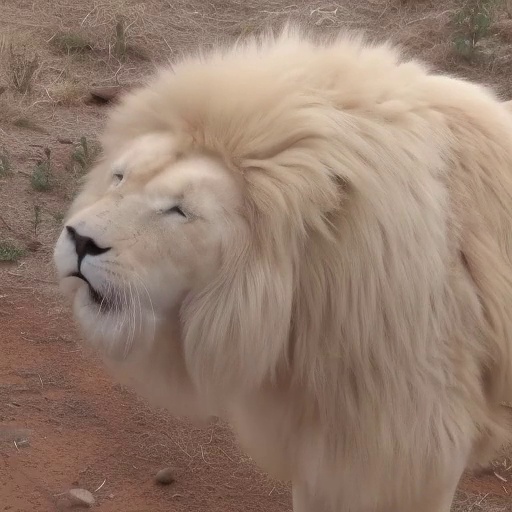}}
        \vspace{-0.1cm}
        \addtocounter{subfigure}{-8}
        \caption{Close mouth}
        \label{subfig:lion}
    \end{subfigure}

    \vspace{\tsHeight}
    \begin{subfigure}{\linewidth}
        \centering
        \subfloat{\includegraphics[width = 0.114\linewidth]{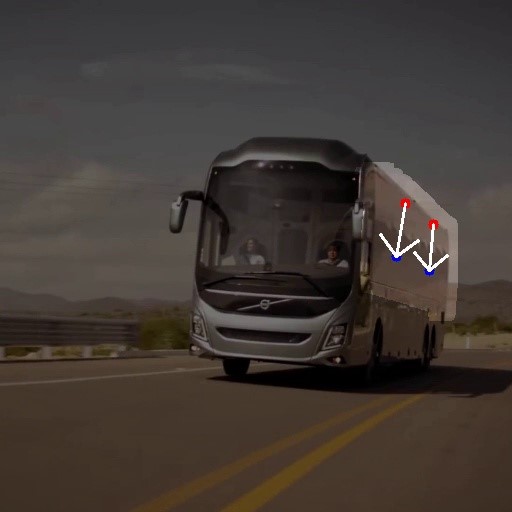}}\hspace{\tsWidth}
        \subfloat{\includegraphics[width = 0.114\linewidth]{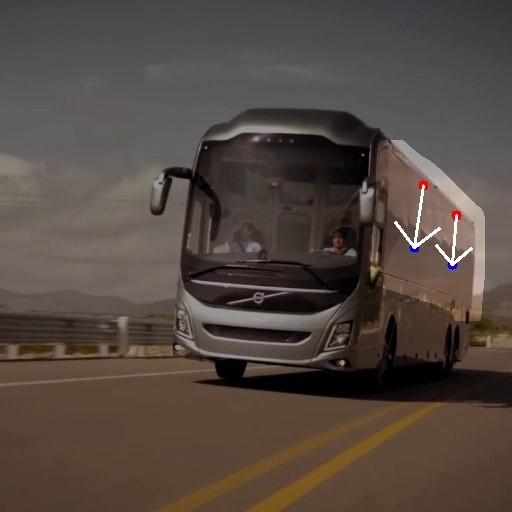}}\hspace{\tsWidth}
        \subfloat{\includegraphics[width = 0.114\linewidth]{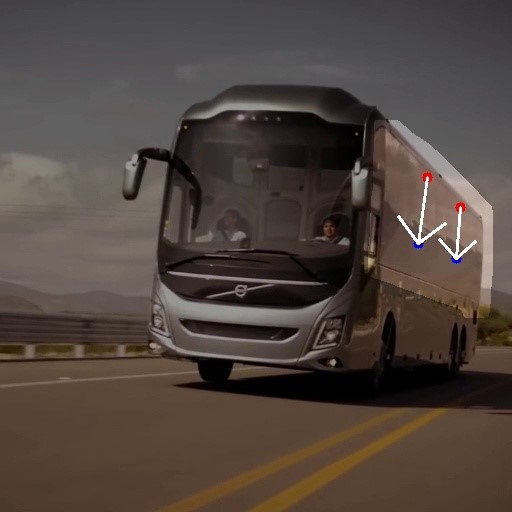}}\hspace{\tsWidth}
        \subfloat{\includegraphics[width = 0.114\linewidth]{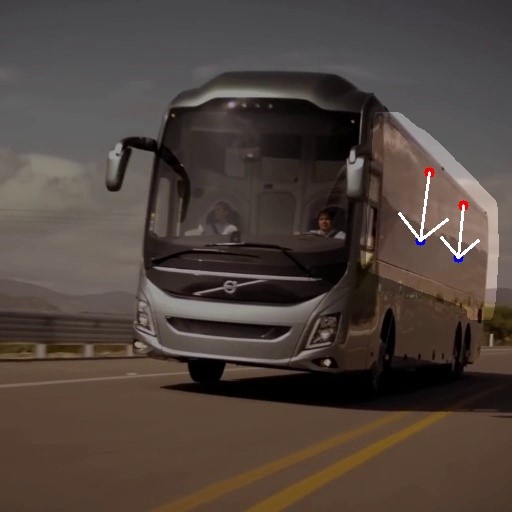}}\hspace{\tsmWidth}
        \subfloat{\includegraphics[width = 0.114\linewidth]{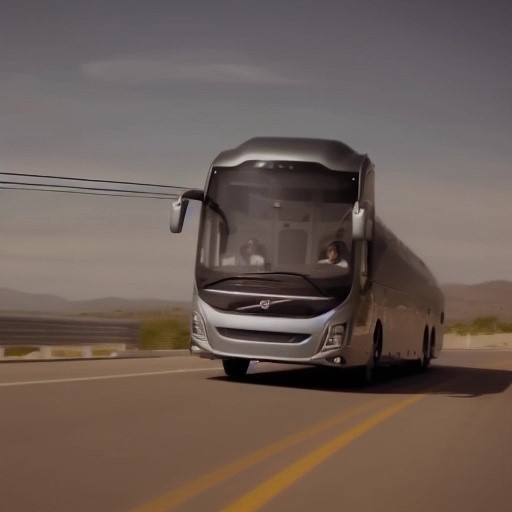}}\hspace{\tsWidth}
        \subfloat{\includegraphics[width = 0.114\linewidth]{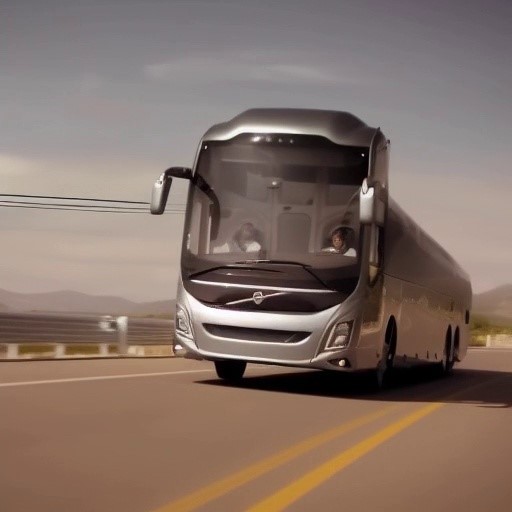}}\hspace{\tsWidth}
        \subfloat{\includegraphics[width = 0.114\linewidth]{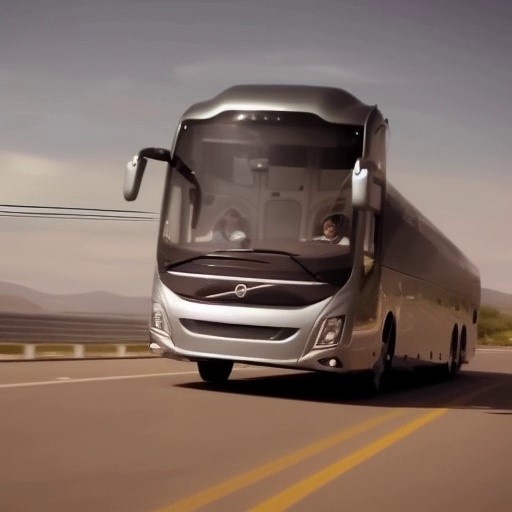}}\hspace{\tsWidth}
        \subfloat{\includegraphics[width = 0.114\linewidth]{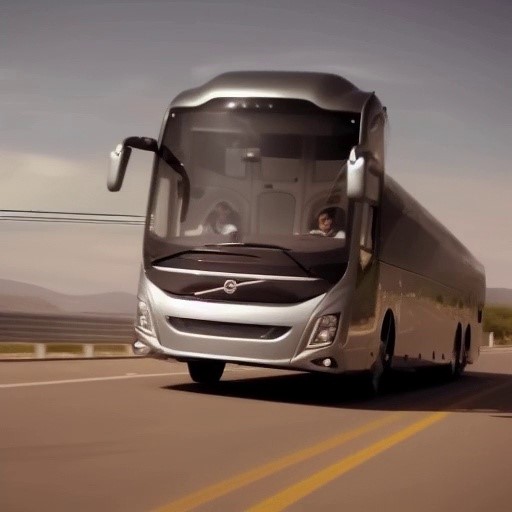}}
        \vspace{-0.1cm}
        \addtocounter{subfigure}{-8}
        \caption{Squeeze bus}
        \label{subfig:bus}
    \end{subfigure}
    
    \vspace{\tsHeight}
    \begin{subfigure}{\linewidth}
        \centering
        \subfloat{\includegraphics[width = 0.114\linewidth]{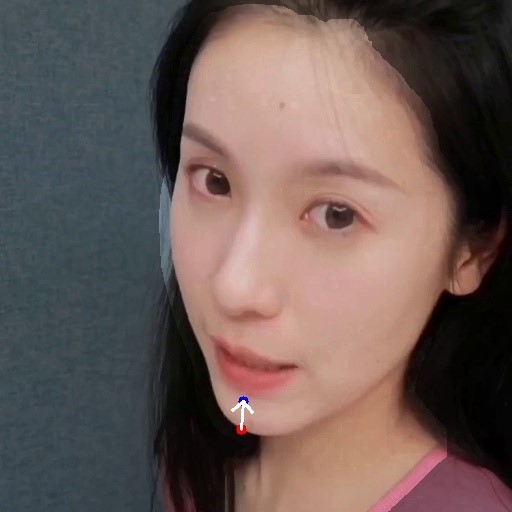}}\hspace{\tsWidth}
        \subfloat{\includegraphics[width = 0.114\linewidth]{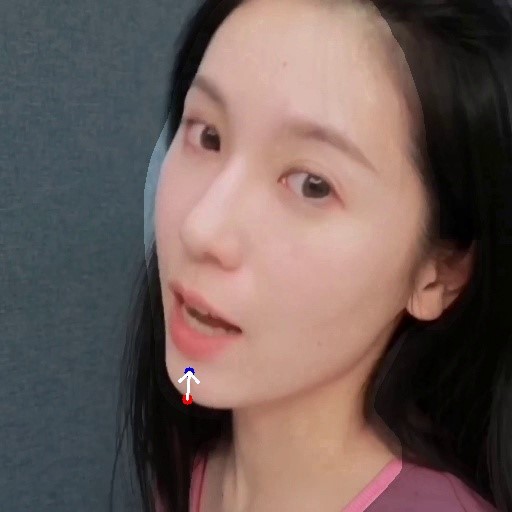}}\hspace{\tsWidth}
        \subfloat{\includegraphics[width = 0.114\linewidth]{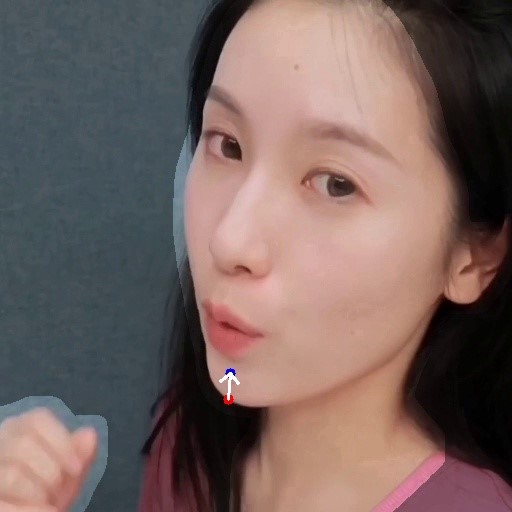}}\hspace{\tsWidth}
        \subfloat{\includegraphics[width = 0.114\linewidth]{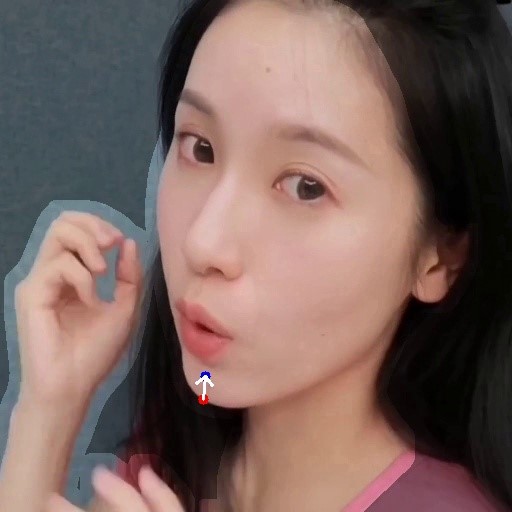}}\hspace{\tsmWidth}
        \subfloat{\includegraphics[width = 0.114\linewidth]{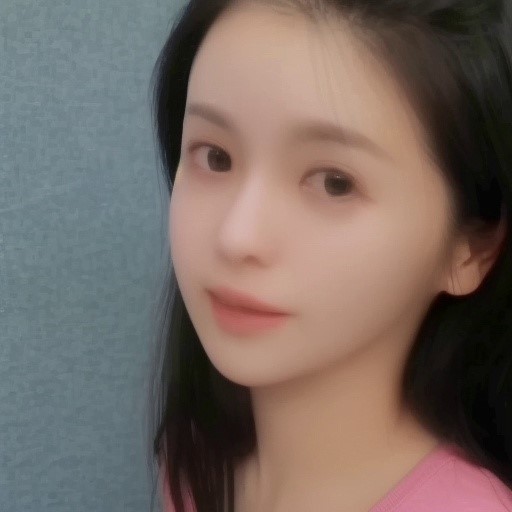}}\hspace{\tsWidth}
        \subfloat{\includegraphics[width = 0.114\linewidth]{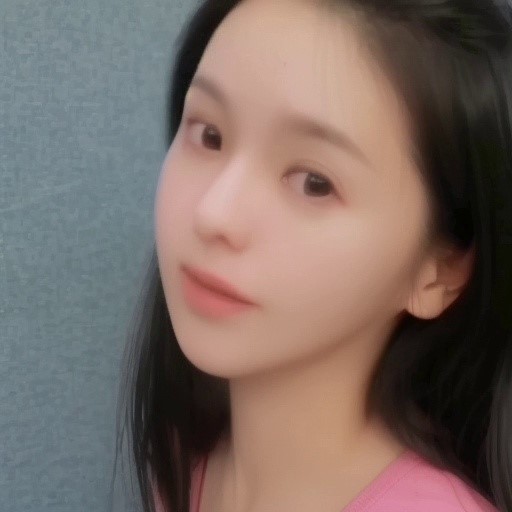}}\hspace{\tsWidth}
        \subfloat{\includegraphics[width = 0.114\linewidth]{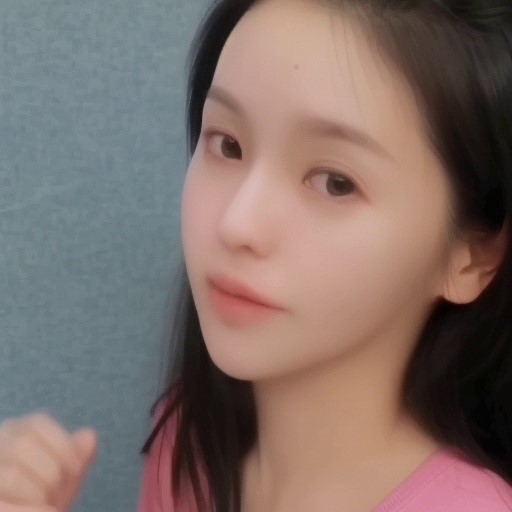}}\hspace{\tsWidth}
        \subfloat{\includegraphics[width = 0.114\linewidth]{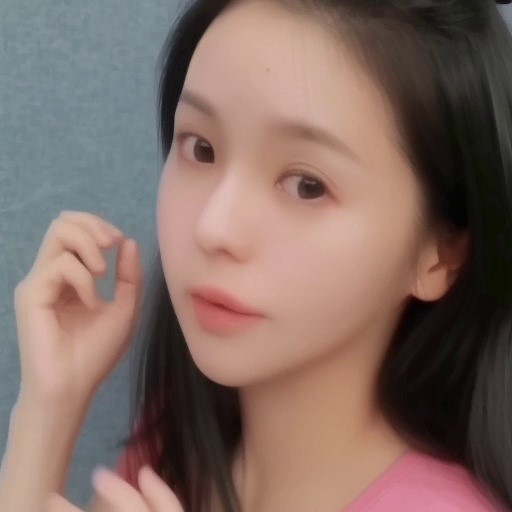}}
        \vspace{-0.1cm}
        \addtocounter{subfigure}{-8}
        \caption{Shorten face}
        \label{subfig:faceshort}
    \end{subfigure}
 \vspace{-0.25in}
    \caption{Results of DragVideo. Left four frames are propagated editing instructions (points and masks). Right four frames are edited output. Our results achieves natural, accurate, spatio-temporal consistent edit without noticeable distortion/artifacts. }
    \vspace{-0.1in}
    \label{fig:teaser}
\end{center}
\begin{abstract}

Video generation models have shown their superior ability to generate photo-realistic video. However, how to accurately control (or edit) the video remains a formidable challenge. The main issues are: 1) how to perform direct and accurate user control in editing; 2) how to execute editings like changing shape, expression, and layout without unsightly distortion and artifacts to the edited content; and 3) how to maintain spatio-temporal consistency of video after editing. To address the above issues, we propose \textbf{DragVideo}, a general drag-style video editing framework. Inspired by DragGAN~\cite{draggan}, DragVideo addresses issues 1) and 2) by proposing the drag-style video latent optimization method which gives desired control by updating noisy video latent according to drag instructions through video-level drag objective function. We amend issue 3) by integrating the video diffusion model with sample-specific LoRA and Mutual Self-Attention in DragVideo to ensure the edited result is spatio-temporally consistent. We also present a series of testing examples for drag-style video editing and conduct extensive experiments across a wide array of challenging editing cases, showing DragVideo can edit video in an intuitive, faithful-to-user-intention manner, with nearly unnoticeable distortion and artifacts, while maintaining spatio-temporal consistency. While traditional prompt-based video editing fails to do the former two and directly applying image drag editing fails in the last, DragVideo's versatility and generality are emphasized. Project page: \url{https://dragvideo.github.io/}
\keywords{Video Editing \and Diffusion Model}

\end{abstract}

\section{Introduction}
\label{sec:intro}

Nowadays, powerful video generation networks have attracted great attention in both the industry and academia. However, in many applications, visual content, no matter whether generated or actual videos, needs to be accurately edited to satisfy real-world needs. That is why drag-style editing has gained significant attention since the debut of DragGAN~\cite{draggan}, a powerful technique for pixel-level interactive editing that is accurate and largely free of noticeable distortions/artifacts using intuitive drag instructions.

\begin{figure*}[h]
    \captionsetup[subfigure]{labelformat=empty}
    \vspace{-0.2in}
    \centering
    \begin{subfigure}[b]{0.99\linewidth}
        \centering
        \includegraphics[width=0.99\linewidth]{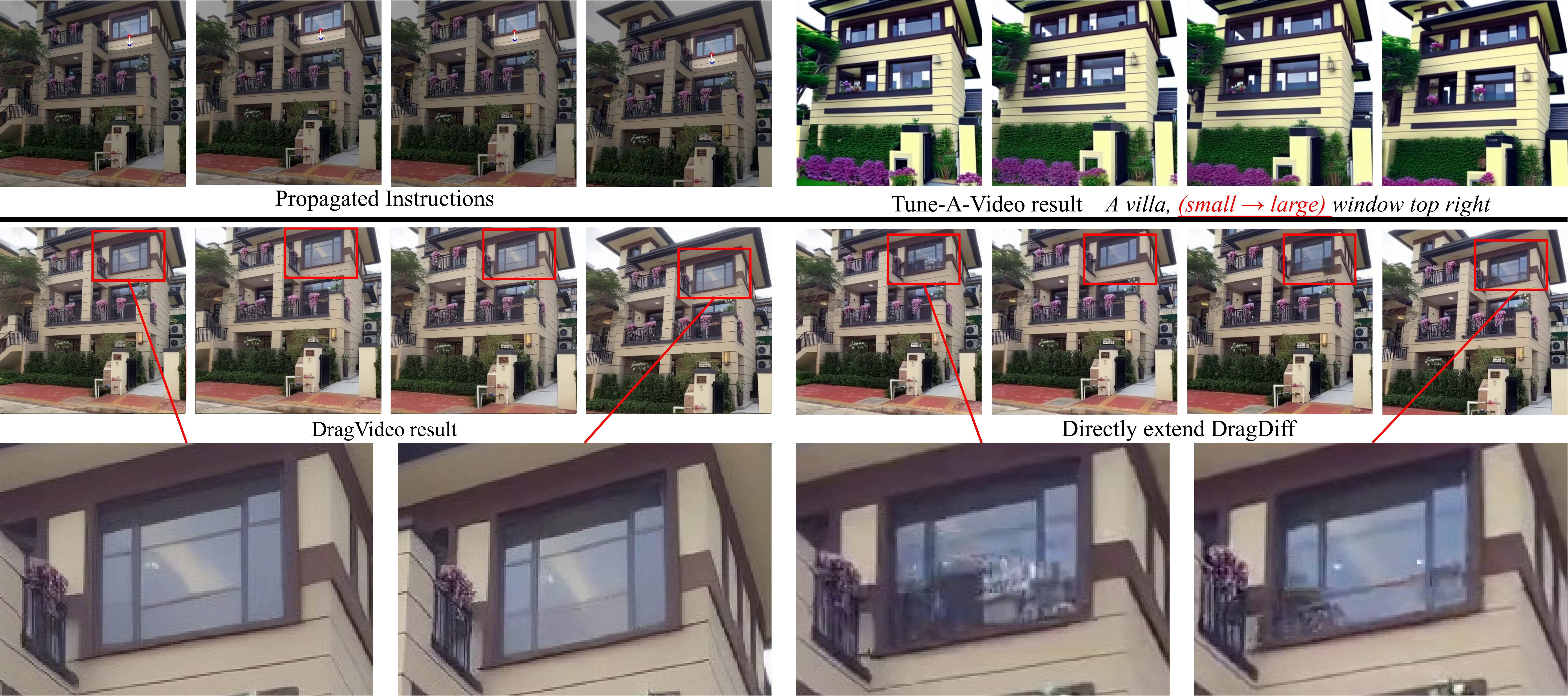}
    \end{subfigure}
    \vspace{-0.1in}
    \caption{
        We intend to enlarge the window of the building. Directly extending DragDiff (bottom right) to video faces the temporal inconsistent problem; Using prompt-based editing (Tune-A-Video, top right) can introduce unintended style alterations without achieving the goal. \textit{DragVideo} editing (bottom left) addresses the mentioned problems.
    }
    \vspace{-0.2in}
    \label{fig:probElab}
\end{figure*}

Despite the impressive precise control and faithful-to-original editing result of DragGAN and the following models (DragDiff and DragonDiff~\cite{dragdiff, dragondiff}), their video extension has yet to be explored. 
It is widely known that directly extending static image methods to videos may face serious spatio-temporal inconsistency. 
The state-of-the-art video editing tools, including Tune-a-video~\cite{wu2023tune}, CoDeF~\cite{ouyang2023codef}, Rerender A Video~\cite{yang2023rerender}, VideoComposer~\cite{wang2023videocomposer}, and Edit-a-video~\cite{shin2023edit} tries to address the spatio-temporal consistency in video editing, but they mainly focused on style changes by changing prompts, which is inaccurate, often with unintended style alterations and artifacts, as illustrated in Fig.\ref{fig:probElab}.
.

To address the above challenges,  
we propose \textit{DragVideo}, a novel framework that performs accurate drag-style video editing while maintaining comparably good spatio-temporal consistency. From the users' point of view, they only need to ``drag'' on objects to be edited, together with indicating editable region for the first and last frames, \textit{DragVideo} will then finish the editing automatically.

\begin{figure*}[t]
    \captionsetup[subfigure]{labelformat=empty}
    \centering
    \vspace{-0.1in}
    \begin{subfigure}[b]{0.9\linewidth}
        \centering
        \includegraphics[width=0.99\linewidth]{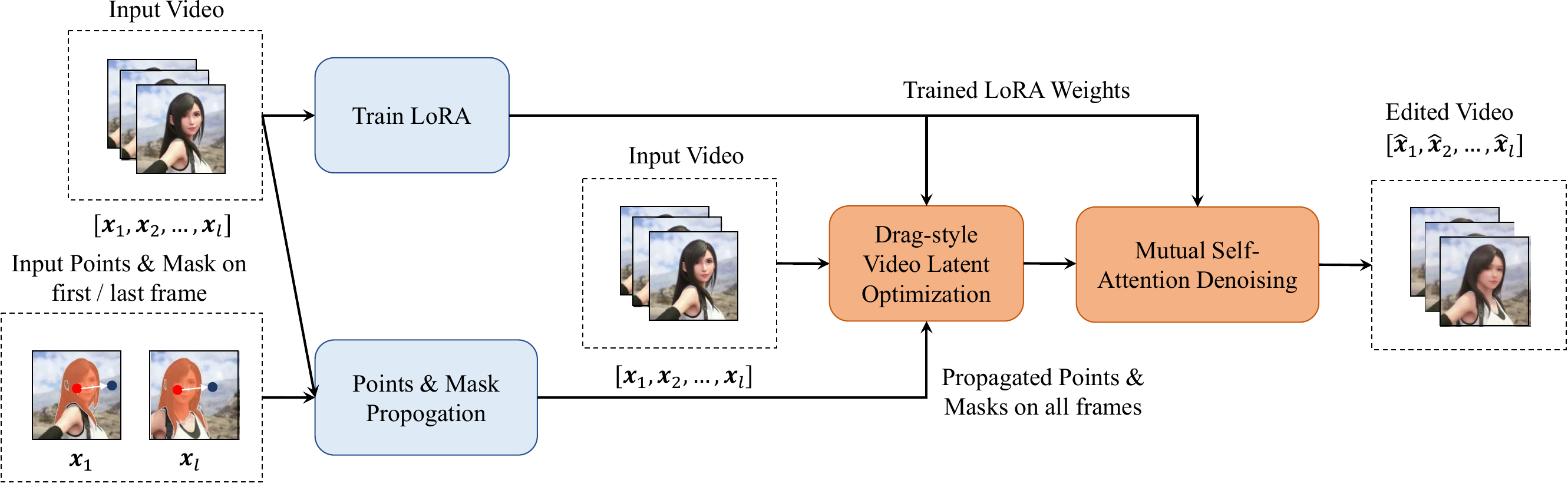}
       
    \end{subfigure}
    \caption{
    \textbf{Overview of DragVideo:} Given an input video of length $l$, \textit{DragVideo} firstly train a Sample-specific LoRA for the video, then propagate user-given points and masks. After that, \textit{DragVideo} process drag by Drag-style Video Latent Optimization. Finally, denoise the noisy video latent through Mutual-Self Attention.
    }
    \label{fig:main}
    \vspace{-0.1in}
\end{figure*}
\begin{figure*}[t]
    \captionsetup[subfigure]{labelformat=empty}
    \centering
    \begin{subfigure}[b]{0.99\linewidth}
        \centering
        \includegraphics[width=0.99\linewidth]{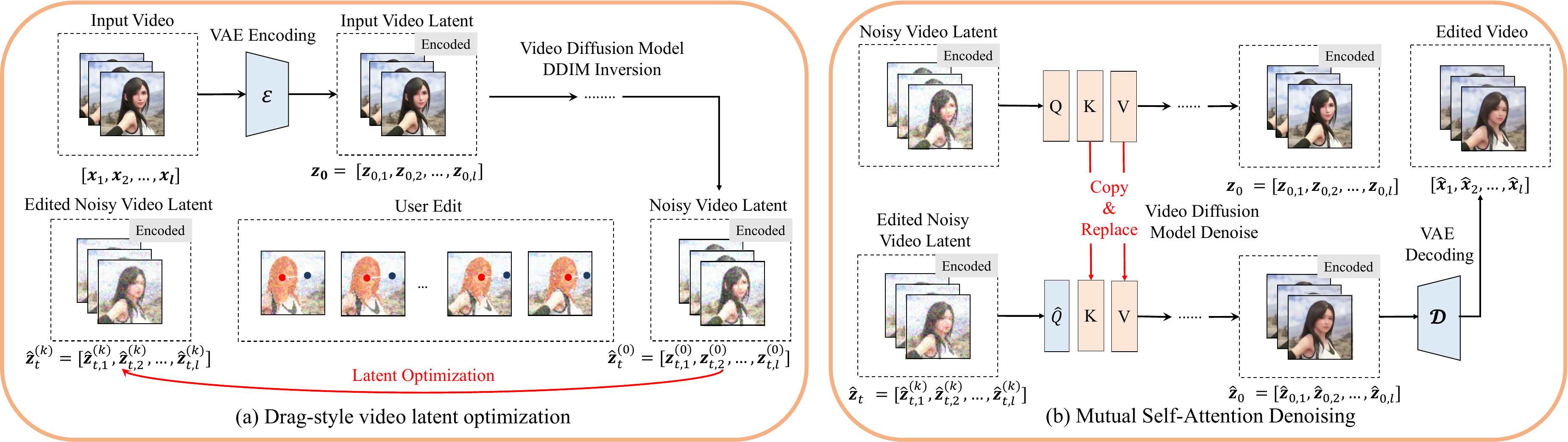}
       
    \end{subfigure}
    \caption{
    \textbf{Core Components of DragVideo:} Drag-style Video Latent Optimization firstly performs DDIM Inversion to video latent, then optimizes the latent by video-level drag objective function to perform drag editing. Finally, \textit{DragVideo} removes noise from drag-edited video latent by Mutual Self-Attention Denoising.
    }
    \label{fig:mainmodule}
    \vspace{-0.2in}
\end{figure*}

\textit{DragVideo} consists of the following stages, as shown in Fig.\ref{fig:main}. First, we train a sample-specific LoRA~\cite{lora} for the video to ease the spatial inconsistency problem and ensure faithful reconstruction. Then, we apply PIPs \cite{harley2022particle} and TAMs \cite{yang2023track} to propagate the drag instruction (points and masks) to all frames. After that, we optimize the latent to perform drag-style editing while maintaining temporal consistency. 
This is achieved by using the video-level drag objective function and video diffusion model to optimize the noisy video latent based on drag instructions (Fig.\ref{fig:mainmodule} (a)). Finally, we derive the edited video by denoising the edited video latent using the video latent diffusion model with Mutual Self-Attention Mechanism \cite{mutualattn} and previously trained LoRA to preserve identity between edited and original video (Fig.\ref{fig:mainmodule} (b)).

We conduct exhaustive quantitative and qualitative experiments together with user studies on \textit{DragVideo} with multiple baselines. The findings indicate that directly extending DragDiff to video encounters significant spatio-temporal inconsistency, and prompt-based editing fails to achieve accurate and artifact-free results. In contrast, DragVideo seamlessly circumvents these challenges, demonstrating superior performance without the aforementioned drawbacks.

In summary, our major contributions are:
\vspace{-0.1in}
\begin{enumerate}
    \item We propose \textit{DragVideo}, the first end-to-end drag-style video editing framework, that achieves faithful and intuitive editing, with almost unnoticeable distortion/artifacts while preserving spatio-temporal consistency, on a single RTX-4090 or RTX-A6000 GPU.
    \item A new drag-style video latent optimization module that integrates video diffusion models, with video-level drag-objective functions for performing effective drag-style video editing. 
    \item A wide array of quantitative and qualitative experiments, as well as user studies to validate the effectiveness and temporal consistency of DragVideo. DragVideo is a clear winner over the baselines.

\end{enumerate}
\vspace{-0.1in}

\section{Related work}
\label{sec:relate_work}

\subsection{Video Diffusion Models}

The rapid development of Diffusion Models has significantly impacted image generation, beginning with DDPM~\cite{ho2020denoising}, followed by the introduction of a text-image joint feature space by CLIP~\cite{radford2021learning}.
Empowered by DDIM~\cite{ddim}, an efficient parallel training technique, stable-diffusion~\cite{rombach2022high} achieves a general and high-quality image synthesis with text guidance, namely text-to-image (T2I) model.

Given these advancements, researchers began to explore the potential of stable diffusion models for video generation.
For example, Tune-a-Video \cite{wu2023tune} proposed a one-shot video generation methodology with minor architectural modifications and sub-network tuning. Text2Video-Zero \cite{khachatryan2023text2video} enhanced a pre-trained T2I model via latent warping under a predefined affine matrix, offering a training-free method for video generation. Pix2Video~\cite{ceylan2023pix2video} leverages a pre-trained image diffusion model to do training-free text-guided video editing. AnimateDiff \cite{guo2023animatediff} introduced a pre-trained plug-in motion module to T2I to capture video motion, resulting in promising results for generating motion-consistent, text-guided videos. Meanwhile, many other methods use diverse condition embedding to control the video generation process \cite{tanveer2024anamodiff,tanveer2024anamodiff, chen2023control, zhang2023controlvideo}. The recent emergence of Sora \cite{videoworldsimulators2024} shows that the limit of the transformer-based video diffusion model is yet to be reached.
Using prompt-to-prompt editing and DDIM Inversion \cite{mokady2022null}, video diffusion models \cite{bai2024uniedit, khachatryan2023text2video, guo2023animatediff, ouyang2023codef} have excelled in text-guided style-level editing. However, current video diffusion model-based editing cannot achieve accurate and intuitive control of edited content, which makes our DragVideo important.

\subsection{Drag-Style Editing}

The recent introduction of DragGAN ~\cite{draggan} brought forward a novel drag-style editing method on static images, wherein the user provides one or multiple paired (\emph{handle}, \emph{target}) points. 
DragGAN uses the motion supervision loss computed on the intermediate feature map of pre-trained StyleGAN2's ~\cite{stylegan} decoder to iteratively optimize the latent code,
achieving impressive image editing results. 
This drag-style editing for static images can also be extended to the diffusion model by performing the optimization process in the U-Net's decoder before the denoising of the randomly sampled noisy latent code or the one obtained after DDIM inversion~\cite{dragdiff, dragondiff}. DragDiffusion~\cite{dragdiff} employs LoRA~\cite{lora} and MSA~\cite{mutualattn} to improve the spatial consistency between the edited image and the original one. Meanwhile, DragonDiffusion~\cite{dragondiff} incorporates cosine-similarity-based drag objective function based on the area masks to conduct more general drag editing. Also, CNS-Edit~\cite{hu2024cns} proposes coupled neural shape representation to perform drag-style editing on 3D objects. Despite the success of accurate image editing, when directly extended to video, DragDiffusion suffers from a serious temporal consistency problem; \textit{DragVideo} proposes technical contributions to address this problem by developing video-level drag-style editing

\subsection{Points and Mask Tracking}
 
In recent years, significant improvements in points tracking across frames in video have been made, such as RAFT~\cite{teed2020raft}, TAPNET~\cite{doersch2022tap}, and state-of-the-art OmniMotion~\cite{wang2023tracking}. Among them, Persistent Independent Particles (PIPs)~\cite{harley2022particle} studies pixel tracking as a long-range motion estimation problem. 
This method demonstrated moderate robustness toward long-term tracking challenges. 
Meanwhile, the Segment Anything Model (SAM)~\cite{kirillov2023segment} 
has delivered commendable results in rapid image segmentation. Trained on over 1 billion segmentation masks, SAM has paved the way for the Track-Anything Model (TAM)~\cite{yang2023track}. 
TAM integrates SAM and Xmen~\cite{cheng2022xmem} 
in a cyclical process to track the target object masks as provided by SAM and utilize SAM to refine the mask details predicted by Xmen. The points and mask tracking are closely related to \textit{DragVideo} since it ensures a smooth editing experience by allowing users to only put drag instructions and masks on first and last frames without repeated work on intermediate frames.

\section{Methodology}
\label{sec:meth}

This section presents an in-depth technical exposition of \textit{DragVideo}. 
The workflow is depicted in Fig.\ref{fig:main}, which processes the input video and point pairs to execute the ``drag'' operation on the video. Section~\ref{meth:unet} describes the preliminaries of the video diffusion models we use. Section~\ref{meth:lora} elucidates the Sample-specific LoRA fine-tuning, integral for enhancing the preservation of personal identity in the edited video. Section~\ref{meth:prop} provides descriptions of the propagation of the user's point pairs throughout the entire video. Section~\ref{meth:drag} describes Drag-style video latent optimization, one of the core components that drag-style editing of video works. Lastly, Section~\ref{meth:MSA} details the employment of the Mutual Self-Attention technique, which is another core part of \textit{DragVideo} that help ensuring consistency between the input and the output videos.

\subsection{Video Diffusion model}\label{meth:unet}

Video Diffusion models aim at generating high-quality and spatio-temporal consistent videos following the standard pattern of diffusion models, which are trained on a large number of videos. 
Their impressive generative ability shows their potential to perform high-quality drag-style editing. A video diffusion model is suitable for \textit{DragVideo} since its video latent space has pixel-level correspondence, which will be used for the video-level drag objective function. In this paper, we use AnimateDiff~\cite{guo2023animatediff} as the video diffusion model. In particular, the Motion Module proposed in AnimateDiff can enhance any T2I model to a video diffusion model, which gives us a large potential to perform editing videos in almost all domains and styles. Notably, the \textit{DragVideo} framework is not limited to AnimateDiff, it is a general video editing framework that can be applied to any video diffusion model as long as the latent bares pixel-level information, which gives the model good space to enhance the usability of newly emerged video diffusion models such as \cite{videoworldsimulators2024}.

\subsection{Sample-specific LoRA}\label{meth:lora}

The first step for \textit{DragVideo} is training a sample-specific LoRA within the video diffusion model, as shown in Fig.\ref{fig:main}. The LoRA module can capture crucial features from the original video, which ensures the preservation of necessary fidelity to the original video during the denoising process and avoids the spatial inconsistency problem. 

Sample-specific LoRA's training process adheres to the standard training procedures of the stable-diffusion model \cite{ddim}. Formally, the objective function for the training task is defined as: 
\begin{equation}
\mathcal{L}_\text{LoRA} (z, \Delta\theta) = \mathbb{E} [| \epsilon - \epsilon_{\theta + \Delta \theta} (\alpha_t \bm{z} + \sigma_t \epsilon)|^2],
\end{equation}
where $\theta$ and $\Delta \theta$ represent the parameters of the video U-Net and LoRA, respectively, $\bm{z}$ denotes the video latent, $\epsilon \sim \mathcal{N} (\mathbf{0}, \mathbf{I})$ is the randomly sampled noise added to the video latent, $\epsilon_{\theta + \Delta \theta} (\cdot)$ signifies the noise predicted by the LoRA-enhanced Video U-Net, and $\alpha_t$ and $\sigma_t$ are hyperparameters of the DDIM noise scheduler at step $t$, with $t$ being randomly sampled from the total steps of the scheduler. We update the LoRA parameters by executing gradient descent on $\Delta \theta$ based on the objective function $\mathcal{L}_\text{LoRA}$.

\subsection{Point and Mask Propagation}\label{meth:prop}

The point and mask propagation provides users a smooth experience in editing video. Given an input video, users only need to put the handle and target points on the first and last frames. Handle points reside on an object for dragging, while target points indicate the future locations of the handle points after dragging. Handle points are automatically used to generate default mask for editable areas, which ensures other parts are not affected by the editing. Given an input video with the user-supplied paired ({\em handle}, {\em target}) points and masks on the start frame ($x_1$) and end frame ($x_l$) of the video (named drag instruction), \textit{DragVideo} adopts Persistent Independent Particles (PIPs)~\cite{harley2022particle} for point tracking and Track-Anything Model (TAM)~\cite{yang2023track} for mask propagation. Both are existing tools that have been tested to have stable long-term tracking consistency. After tracking, each frame has its corresponding handle-target point pairs and mask.

\subsection{Drag-style video latent optimization}\label{meth:drag}

This section provides a comprehensive description of the core part of \textit{DragVideo} optimize video latent based on video-level drag objective function and propagated drag instructions provided by Section~\ref{meth:prop}, see Fig.\ref{fig:main} and Fig.\ref{fig:mainmodule} (a). First, to facilitate high-level editing with drag instructions, we perform DDIM Inversion to add noise by a video diffusion model. Then we iteratively perform motion supervision on noisy video latent and embedded point-tracking to perform drag-style editing

\subsubsection{DDIM Inversion}\label{drag:DDIMinv}

Past research~\cite{mokady2022null} has shown that editing on a noisy latent (i.e., $z_t$ with large $t$) allows higher-level editing. This not only applies to prompt-based editing; drag-style editing follows the same pattern. Therefore, 
we use DDIM Inversion~\cite{ddim, mokady2022null} to add back the noise predicted by the video diffusion model to the video latent. That is, given an input video latent $\bm{z}_0$, we obtain the $t$-th step noisy latent, denoted as $\bm{z}_t$, which will be used in drag-style editing.

\subsubsection{Motion Supervision}\label{drag:motion}

Inspired by DragGAN~\cite{draggan} and DragDiffusion~\cite{dragdiff}, we propose the video-level motion supervision loss function as the drag objective function for \textit{DragVideo}. 
Our objective function can perform pixel-level accurate editing without additional training of neural networks nor geometry information. As suggested by DIFT~\cite{tang2023emergent}, most diffusion models', including AnimateDiff's, intermediate features exhibit a significant feature and location correspondence that can be utilized for motion supervision. We denote $\mathbf{F} (\bm{z}_t)$ as the feature output by the video diffusion model from noisy video latent of $\bm{z}_t$. 

In our implementation, we opt for the second and third layers of AnimateDiff U-Net's output as a feature map. In order to enhance motion supervision, $\mathbf{F}$ is resized by linear interpolation, i.e., 
\begin{equation}
    \mathbf{F}: \mathbb{R}^{l \times c \times h_\text{latent} \times w_\text{latent}} \rightarrow \mathbb{R}^{l \times c \times \frac{h}{2} \times \frac{w}{2}}
\end{equation} 
where $h_\text{latent}, w_\text{latent}$ are the heights and widths of VAE encoded video latent, and $h, w$ are the heights and widths of the original video and the $\frac{h}{2}, \frac{w}{2}$ dimension directly follows from DragDiffusion~\cite{dragdiff}. We denote in the $k$-th iteration, the $j$-th handle point at frame $i$ as $p_{i, j}^{(k)}$, where $p_{i, j}^{(0)}$ is the initial handle point. 
The latent $\hat{\bm{z}}_t^{(k)} $ is incrementally optimized in the $k+1$-th iteration to move the patch around $p_{i, j}^{(k)}$ toward $t_{i, j}$. 
Denote $\mathcal{B}_r (p_{i, j}^{(k)})$ as a small circle of area with radius $r$ (a hyperparameter) around $p_{i, j}^{(k)}$, and $\hat{\bm{z}}_t^{(k)}$ as the edited latent code for $k$-th iteration. Then, our motion supervision loss $\mathcal{L} (\hat{\bm{z}}_{t}^{(k)})$ for optimizing the video latent is given by 

\begin{equation}
    \begin{aligned}
        \sum_{i = 1}^l \sum_{j = 1}^n \sum_{q \in \mathcal{B}_r (p_{i, j}^{(k)})} 
        & \|\mathbf{F}_{q + d_{i, j}^{(k)}} (\hat{\bm{z}}_t^{(k)}) - \text{sg}(\mathbf{F}_q (\hat{\bm{z}}_t^{(k)})) \|_1 +   \\
        & \lambda \|(\hat{\bm{z}}_{t - 1}^{(k)} - \text{sg}(\bm{z}_{t - 1})) * (\mathbb{I} - M)\|_1
    \end{aligned}
    \label{eqn:loss}
\end{equation}
where sg$(\cdot)$ is the stop gradient operator, i.e. the argument will not be backward propagated. This ensures $\mathcal{B}_r (p_{i, j}^{(k)})$ is moved toward a location centered at $(p_{i, j}^{(k)} + d_{i, j}^{(k)})$ but not the other way, where $d_{i, j}^{(k)} = \frac{t_{i, j} - p_{i, j}^{(k)}}{\|t_{i, j} - p_{i, j}^{(k)}\|_2}$ is the normalized vector  pointing from $p_{i, j}^{(k)}$ to $t_{i, j}$. The $i$ sums up all frames in the video and $j$ sums up all points in one frame. As the components of $q$ are not integer, we obtain $\mathbf{F}_{q + d_{i, j}^{(k)}} (\hat{\bm{z}}_t^{(k)})$ via bilinear interpolation. In the second term, we apply regularization to the video latent from the binary mask $M$ obtained by mask propagation (Section~\ref{meth:prop}) to ensure the update lies within the masked region.

For each motion supervision step, this loss is used to optimize the edited latent code $\hat{\bm{z}}_t^{(k)}$ for one time: 
\begin{equation}
    \hat{\bm{z}}_{t}^{(k + 1)} = \hat{\bm{z}}_{t}^{(k)} - \eta \cdot \frac{\partial}{\partial \hat{\bm{z}}_t^{(k)}}\mathcal{L} (\hat{\bm{z}}_{t}^{(k)})
\end{equation}
where $\eta$ is the learning rate for latent optimization. By performing the above optimization via motion supervision, we can incrementally ``drag'' the handle point to the target point one step at a time.

\subsubsection{Embedded Point Tracking}\label{drag:point}
 
After each step of motion supervision, an updated noisy video latent $\hat{\bm{z}}_{t + 1}^{(k + 1)}$ and new feature map $\mathbf{F}(\hat{\bm{z}}_{t + 1}^{(k + 1)})$ are produced. Since motion supervision only updates video latent but does not give precise new location of handle points $p_{i, j}^{(k + 1)} $. Thus, to perform the next step of the motion supervision update, we need to embed a point tracking method to track the new handle points $p_{i, j}^{(k + 1)}$. To distinguish from handle/target point tracking in Section~\ref{meth:prop}, we name this point tracking as \textit{Embedded Point Tracking}. Since video U-Net's feature map $\mathbf{F}$ contains rich positional information~\cite{dragdiff, tang2023emergent}, inspired by DragGAN~\cite{draggan}, we utilize $\mathbf{F} (\hat{\bm{z}}_t^{(k + 1)}) $ and $\mathbf{F}(\bm{z}_t) $ to track new handle points $q_{i, j}^{(k + 1)}$ by nearest neighbor method within the square patch $\Omega (q_{i, j}^{(k)}, r') = \{ (x, y) : |x - x_{i, j}^{(k)}| \leqslant r', |y - y_{i, j}^{(k)}| \leqslant r'\} $, where $x_{i, j}^{(k)}$ and $y_{i, j}^{(k)}$ is the $x$ and $y$ coordinate of $q_{i, j}^{(k)}$ respectively and $r'$ is a hyperparameter. The tracking method is as follows:
\begin{equation}
    p_{i, j}^{(k + 1)} = \text{argmin}_{q \in \Omega(p_{i, j}^{(k)}, r')} \{ \| F_q (\hat{\bm{z}}_t^{(k + 1)}) - F_{p_{i, j}^{(0)}} (\bm{z}_t)\|_1\}
\end{equation}

\subsection{Mutual Self-Attention Video Denoising}\label{meth:MSA}

Finally, we perform denoising of the dragged video via Mutual Self-Attention (MSA) controlled denoising. This step reconstructs the drag-edited video from the edited noisy video latent, representing another essential core component of our approach as illustrated in Fig.\ref{fig:main} and Fig.\ref{fig:mainmodule} (b).
Simply applying DDIM denoising~\cite{ddim}, even with Sample-specific LoRA, on the dragged noisy video latent may easily lead to an undesirable identity shift, degradation in quality, and spatial inconsistency from the original video. This problem can be attributed to a lack of guidance from the original video during the denoising process. In~\cite{mutualattn}, a prompt-based image editing method can preserve the identity of the original image by changing the self-attention in image diffusion into cross-attention using $K, V$ from original latent, namely Mutual Self-Attention. Inspired by them, we propose upgrading the MSA to the video diffusion model. 
For the U-Net module in AnimateDiff, given input $x$, the attention used in~\cite{mutualattn} can be represented as $y = \text{softmax} ( \frac{Q(x)K(x)^T}{\sqrt{d}} ) V(x)$ for some $d$. In video-level MSA, with the input of attention for original video latent is $x$ and the input of attention for edited video latent is $\hat{x}$, we replace the keys and values for the edited output with $K(x), V(x)$:
\begin{equation}
    \hat{y} = \text{softmax} (\frac{Q(\hat{x}) K(x)^T }{\sqrt{d}}) V(x). 
\end{equation}
Thus, we need to perform the denoising process for both the original video latent $\bm{z}_t$ and edited video latent $\hat{\bm{z}}_t$, while utilizing $\bm{z}_t$ as MSA guidance for $\hat{z}_t$. In doing so, a more coherent denoise can be achieved.

\section{Experiments}
\label{sec:exp}
\newcommand\rsWidth{0.00cm}
\newcommand\rsmWidth{0.15cm}
\newcommand\rsHeight{-0.05cm}

\begin{figure*}[th]
    \centering
    \vspace{\rsHeight}
    \begin{subfigure}{\linewidth}
        \centering
        \subfloat{\includegraphics[width = 0.114\linewidth]{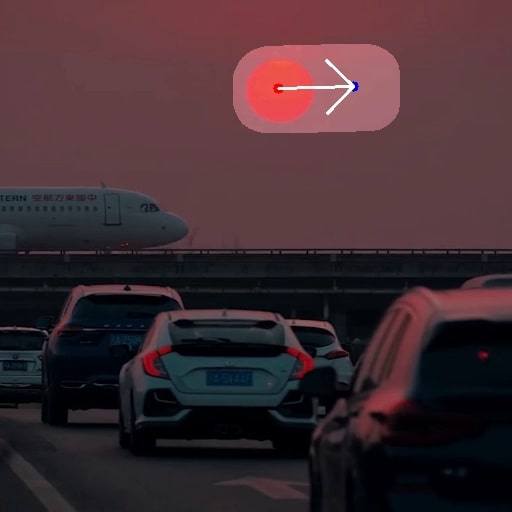}}\hspace{\rsWidth}
        \subfloat{\includegraphics[width = 0.114\linewidth]{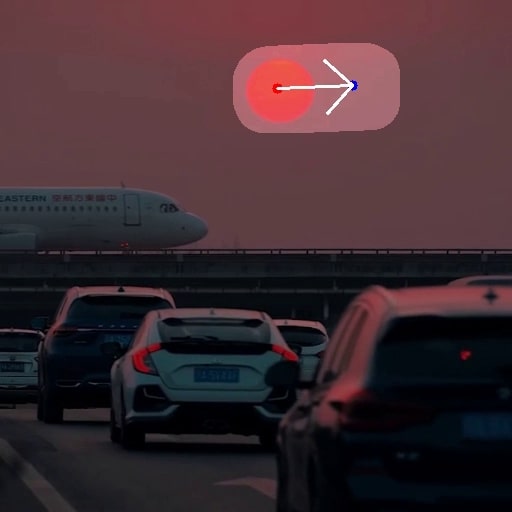}}\hspace{\rsWidth}
        \subfloat{\includegraphics[width = 0.114\linewidth]{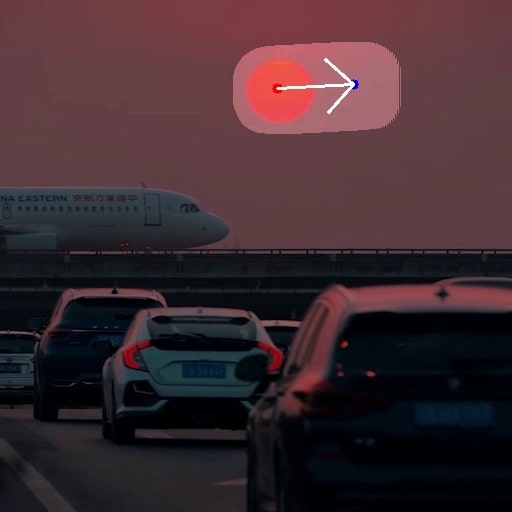}}\hspace{\rsWidth}
        \subfloat{\includegraphics[width = 0.114\linewidth]{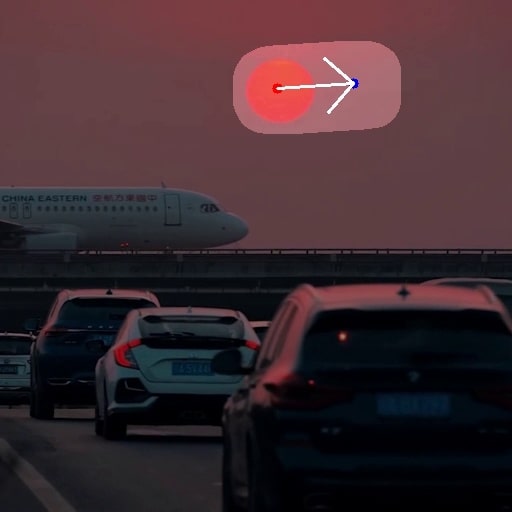}}\hspace{\rsmWidth}  
        \subfloat{\includegraphics[width = 0.114\linewidth]{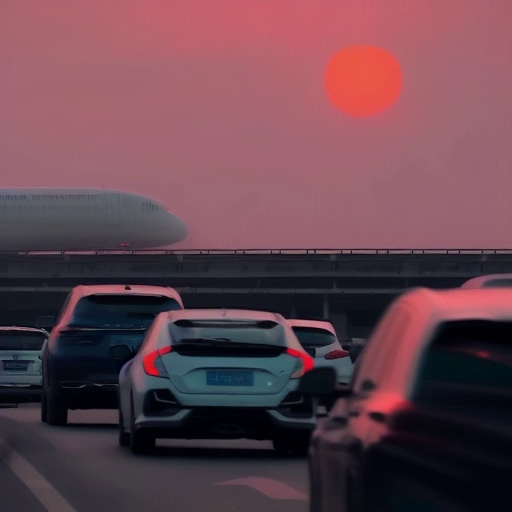}}\hspace{\rsWidth}
        \subfloat{\includegraphics[width = 0.114\linewidth]{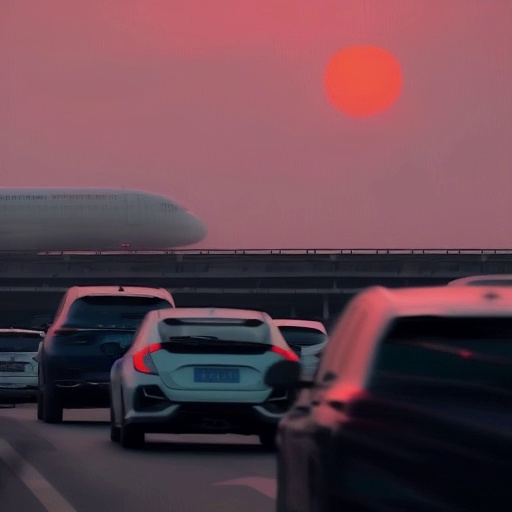}}\hspace{\rsWidth}
        \subfloat{\includegraphics[width = 0.114\linewidth]{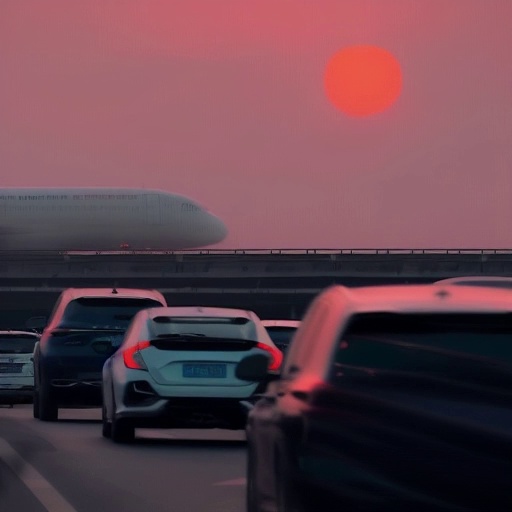}}\hspace{\rsWidth}  
        \subfloat{\includegraphics[width = 0.114\linewidth]{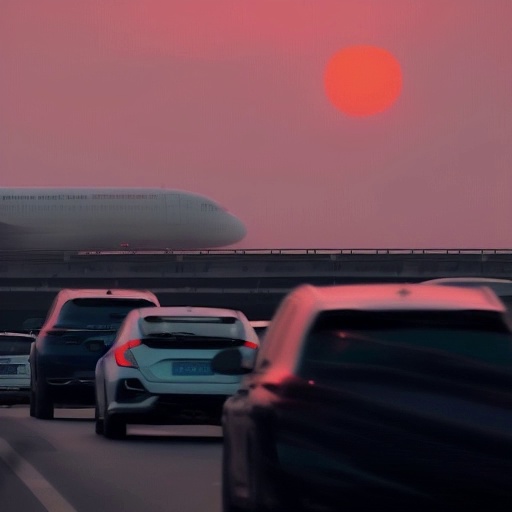}}\hspace{\rsWidth}
        \vspace{-0.1cm}
        \addtocounter{subfigure}{-8}
        \caption{Move the sun}
        \label{subfig:vila}
    \end{subfigure}

    \vspace{\rsHeight}
    \begin{subfigure}{\linewidth}
        \centering
        \subfloat{\includegraphics[width = 0.114\linewidth]{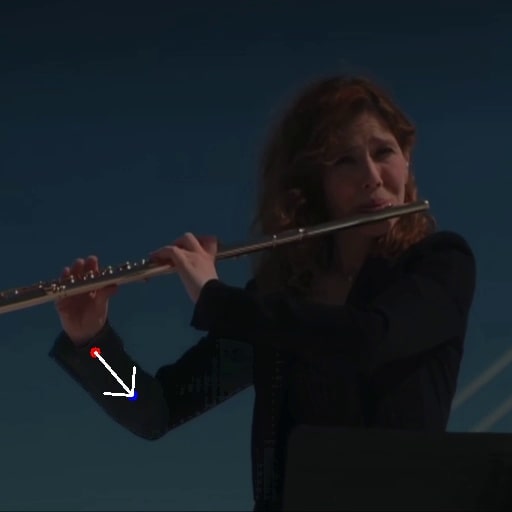}}\hspace{\rsWidth}
        \subfloat{\includegraphics[width = 0.114\linewidth]{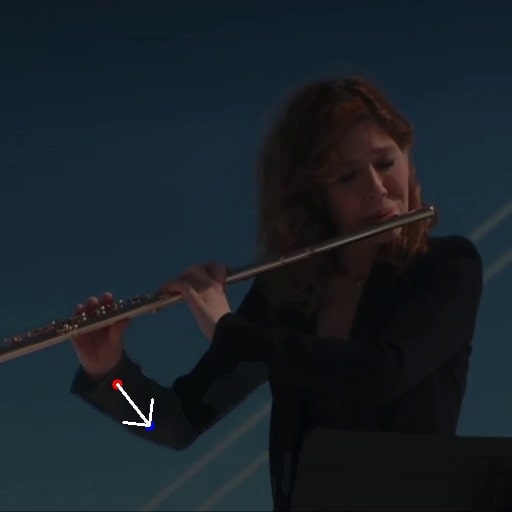}}\hspace{\rsWidth}
        \subfloat{\includegraphics[width = 0.114\linewidth]{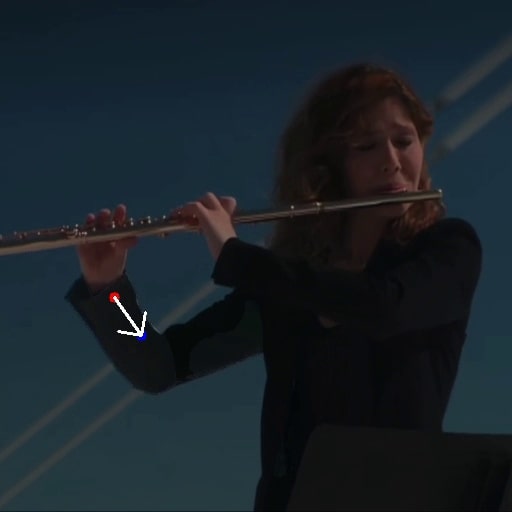}}\hspace{\rsWidth}
        \subfloat{\includegraphics[width = 0.114\linewidth]{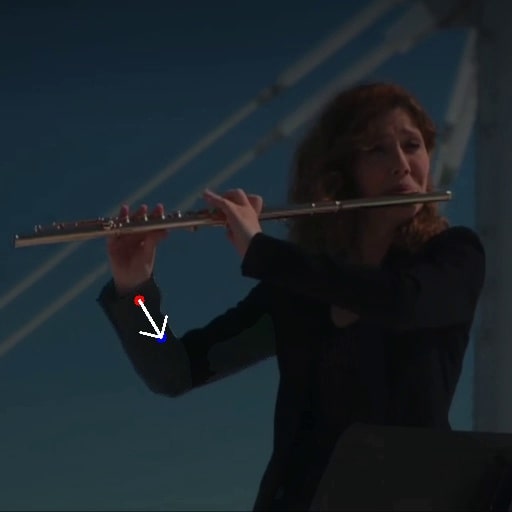}}\hspace{\rsmWidth}  
        \subfloat{\includegraphics[width = 0.114\linewidth]{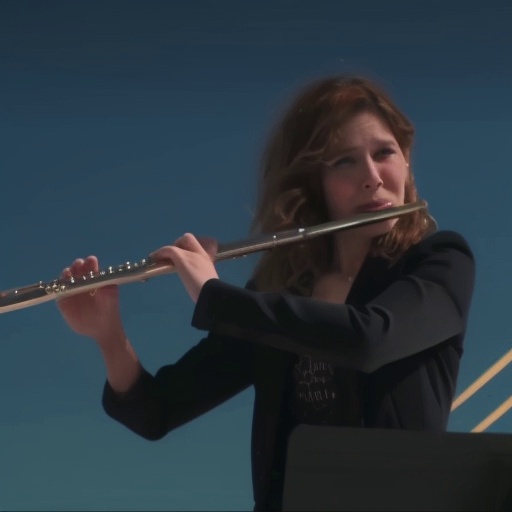}}\hspace{\rsWidth}
        \subfloat{\includegraphics[width = 0.114\linewidth]{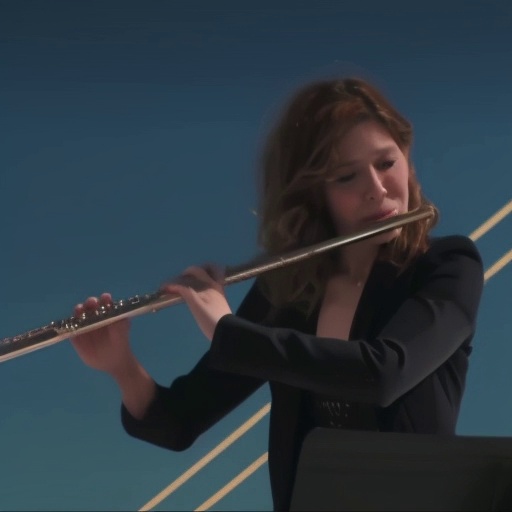}}\hspace{\rsWidth}
        \subfloat{\includegraphics[width = 0.114\linewidth]{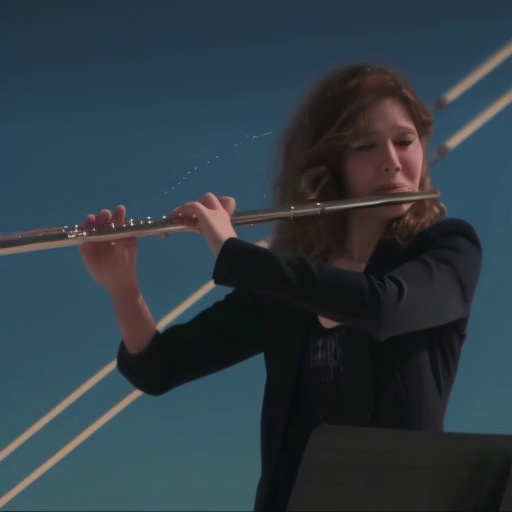}}\hspace{\rsWidth}  
        \subfloat{\includegraphics[width = 0.114\linewidth]{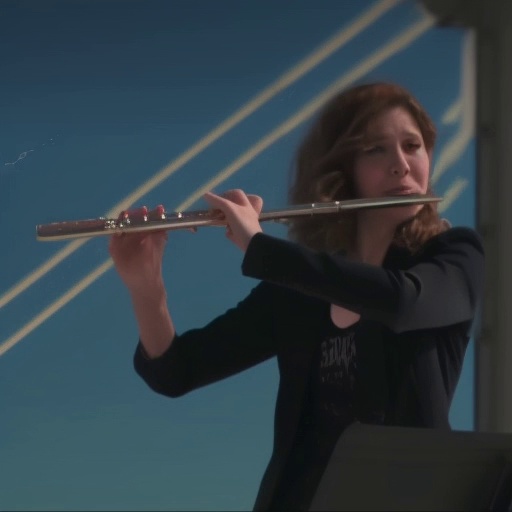}}\hspace{\rsWidth}
        \vspace{-0.1cm}
        \addtocounter{subfigure}{-8}
        \caption{Shorten sleeves}
        \label{subfig:sleeves}
    \end{subfigure}

    \vspace{\rsHeight}
    \begin{subfigure}{\linewidth}
        \centering
        \subfloat{\includegraphics[width = 0.114\linewidth]{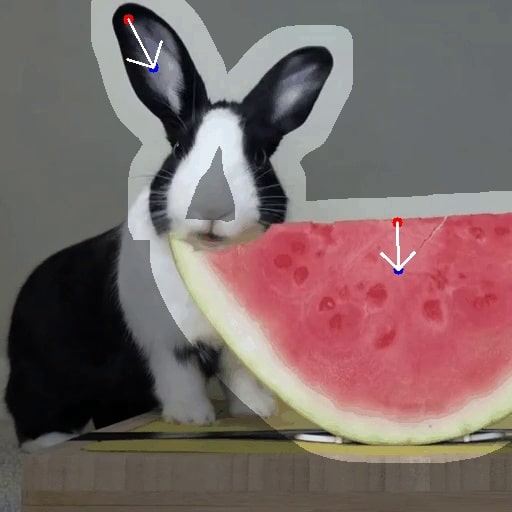}}\hspace{\rsWidth}
        \subfloat{\includegraphics[width = 0.114\linewidth]{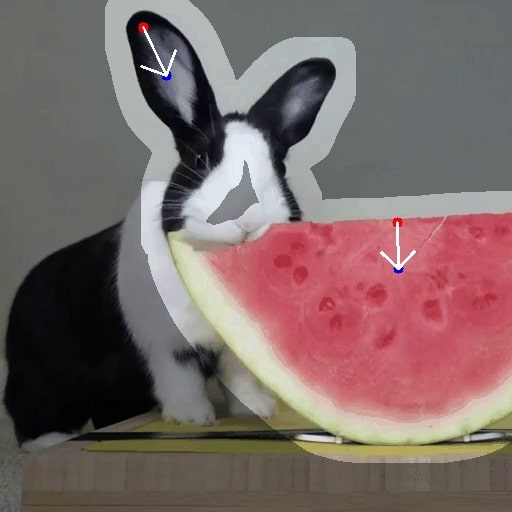}}\hspace{\rsWidth}
        \subfloat{\includegraphics[width = 0.114\linewidth]{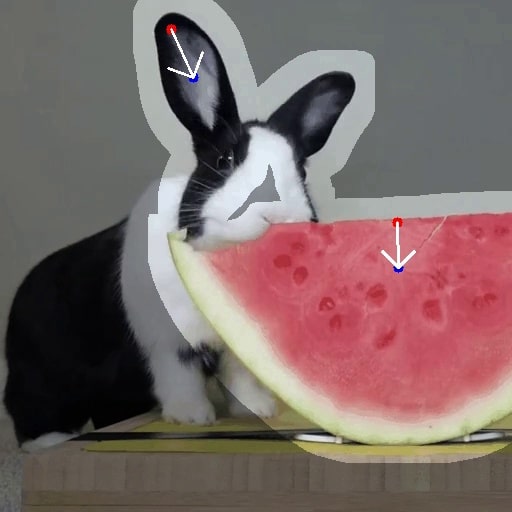}}\hspace{\rsWidth}
        \subfloat{\includegraphics[width = 0.114\linewidth]{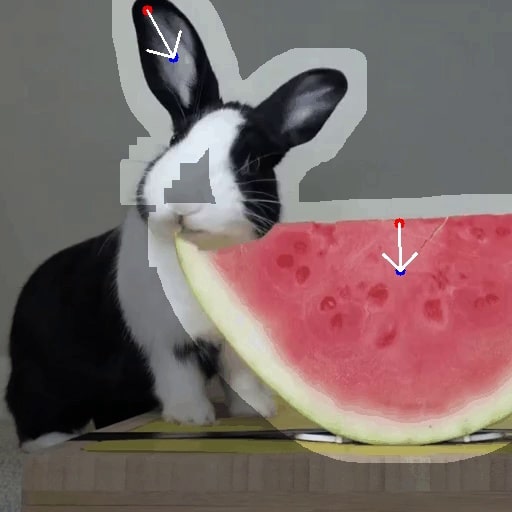}}\hspace{\rsmWidth}  
        \subfloat{\includegraphics[width = 0.114\linewidth]{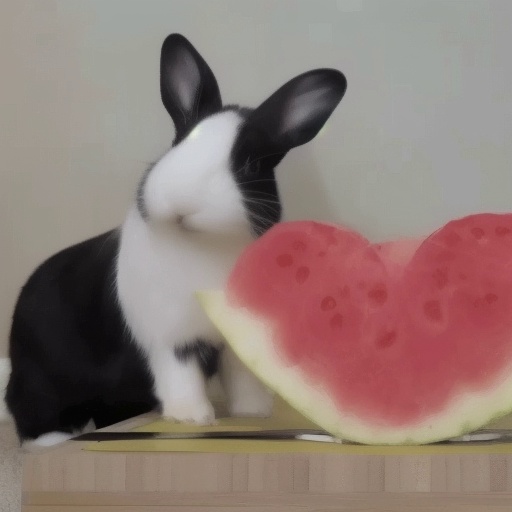}}\hspace{\rsWidth}
        \subfloat{\includegraphics[width = 0.114\linewidth]{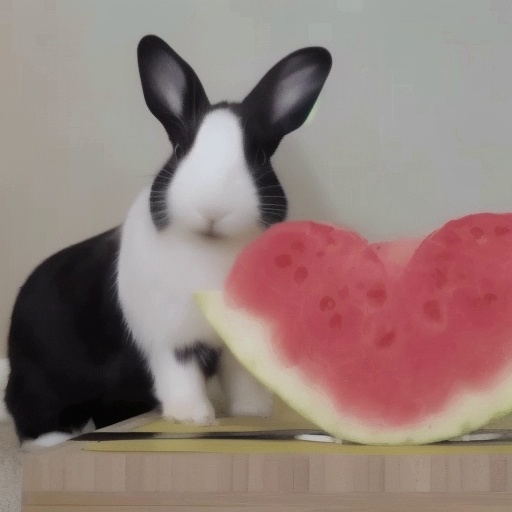}}\hspace{\rsWidth}
        \subfloat{\includegraphics[width = 0.114\linewidth]{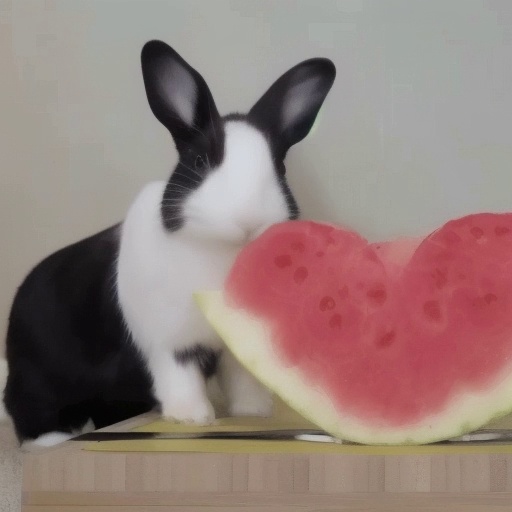}}\hspace{\rsWidth}  
        \subfloat{\includegraphics[width = 0.114\linewidth]{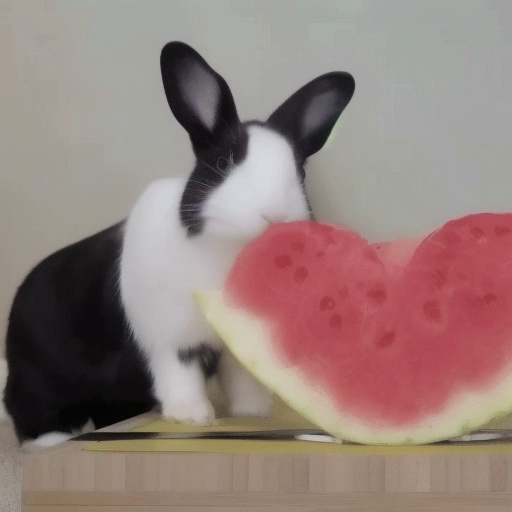}}\hspace{\rsWidth}
        \vspace{-0.1cm}
        \addtocounter{subfigure}{-8}
        \caption{ Shorten the ears, change the shape of watermelon}
        \label{subfig:rabbit}
    \end{subfigure}

    \vspace{\rsHeight}
    \begin{subfigure}{\linewidth}
        \centering
        \subfloat{\includegraphics[width = 0.114\linewidth]{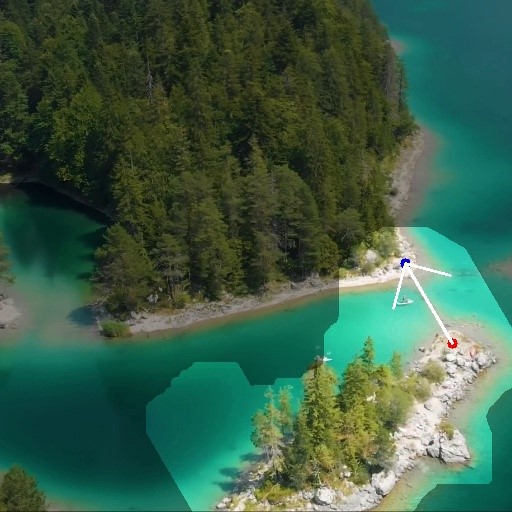}}\hspace{\rsWidth}
        \subfloat{\includegraphics[width = 0.114\linewidth]{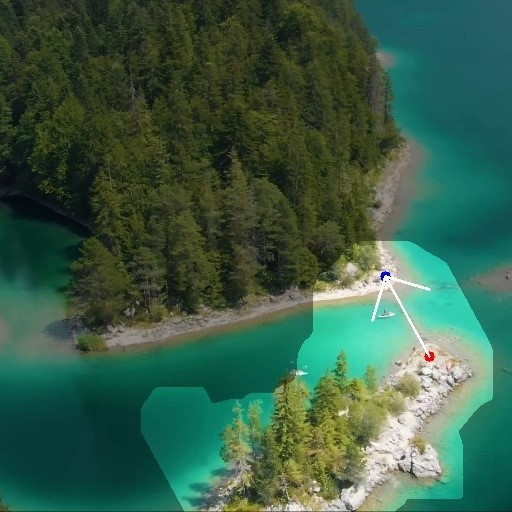}}\hspace{\rsWidth}
        \subfloat{\includegraphics[width = 0.114\linewidth]{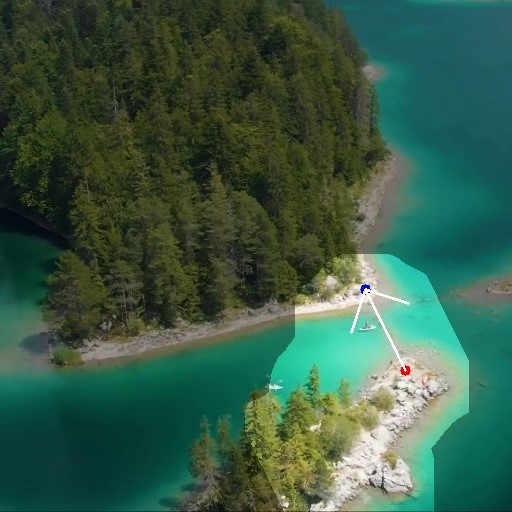}}\hspace{\rsWidth}
        \subfloat{\includegraphics[width = 0.114\linewidth]{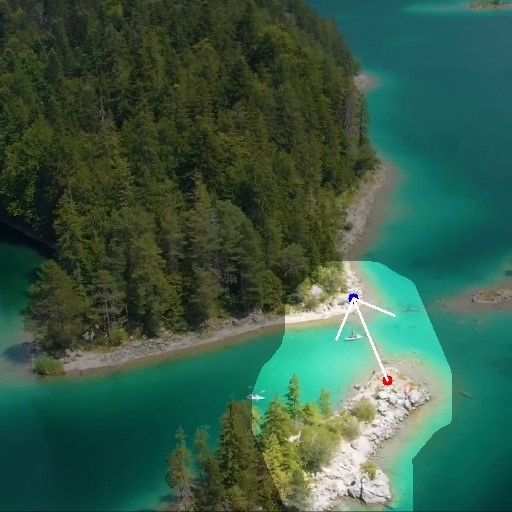}}\hspace{\rsmWidth}  
        \subfloat{\includegraphics[width = 0.114\linewidth]{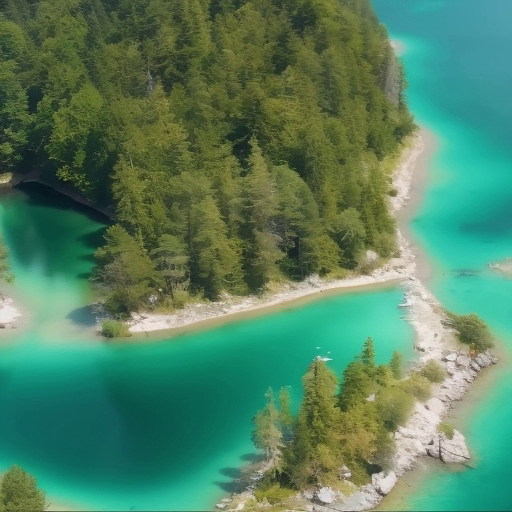}}\hspace{\rsWidth}
        \subfloat{\includegraphics[width = 0.114\linewidth]{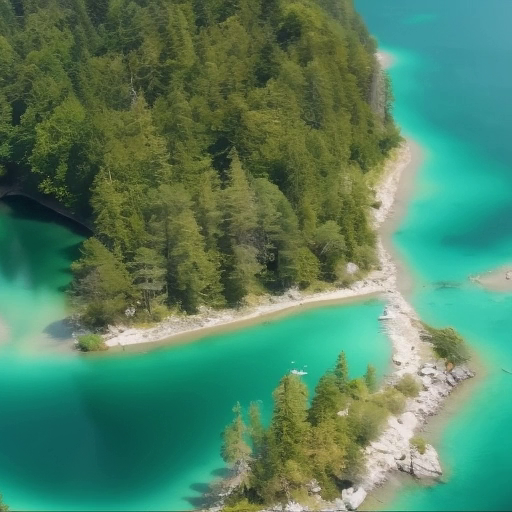}}\hspace{\rsWidth}
        \subfloat{\includegraphics[width = 0.114\linewidth]{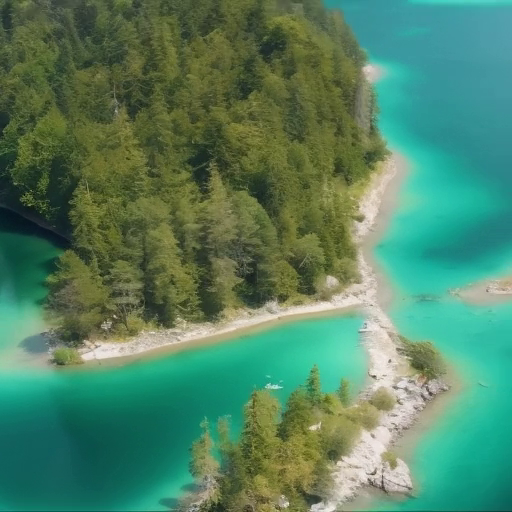}}\hspace{\rsWidth}  
        \subfloat{\includegraphics[width = 0.114\linewidth]{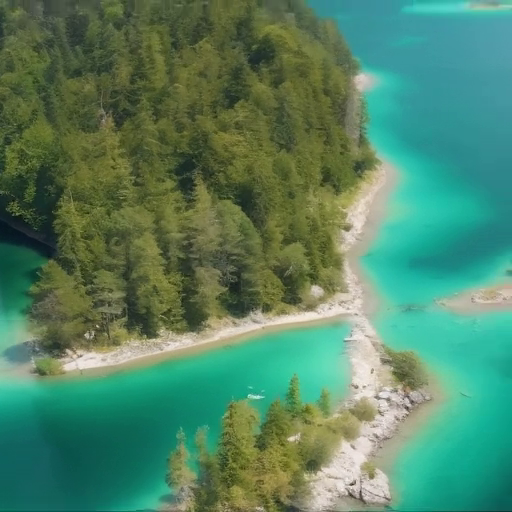}}\hspace{\rsWidth}
        \vspace{-0.1cm}
        \addtocounter{subfigure}{-8}
        \caption{Connect island}
        \label{subfig:island}
    \end{subfigure}

    \begin{subfigure}{\linewidth}
        \centering
        \subfloat{\includegraphics[width = 0.114\linewidth]{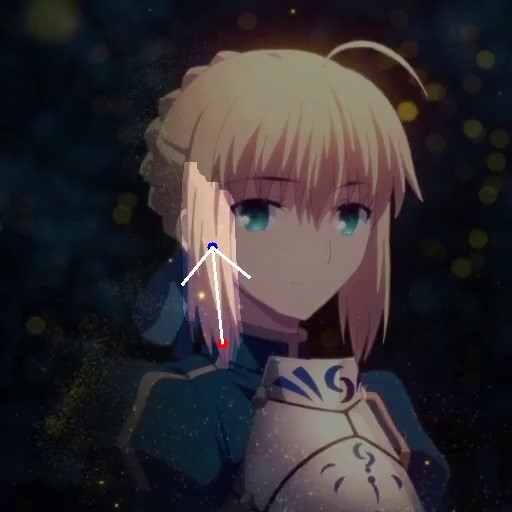}}\hspace{\rsWidth}
        \subfloat{\includegraphics[width = 0.114\linewidth]{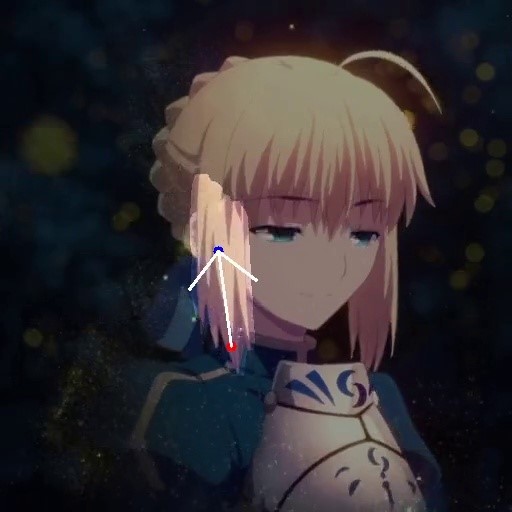}}\hspace{\rsWidth}
        \subfloat{\includegraphics[width = 0.114\linewidth]{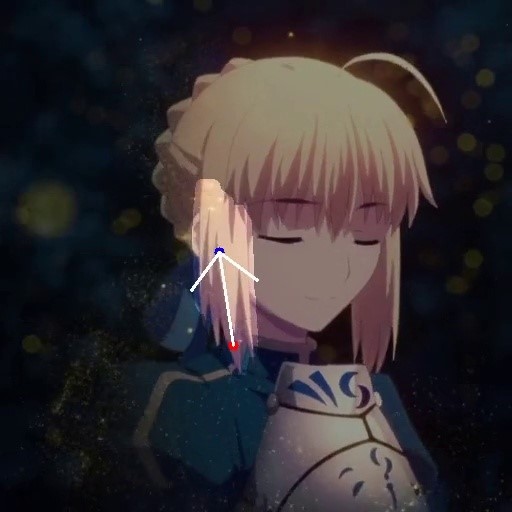}}\hspace{\rsWidth}
        \subfloat{\includegraphics[width = 0.114\linewidth]{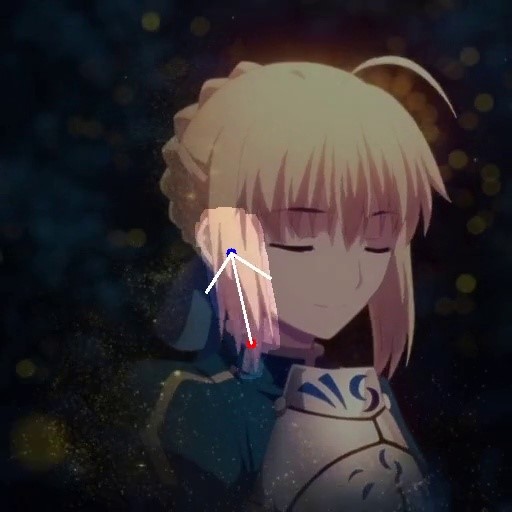}}\hspace{\rsmWidth}
        \subfloat{\includegraphics[width = 0.114\linewidth]{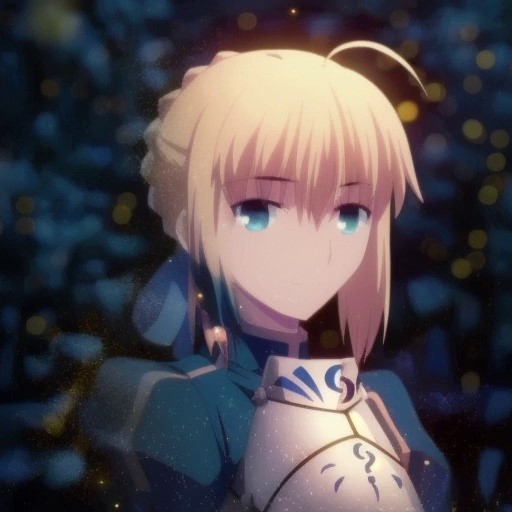}}\hspace{\rsWidth}
        \subfloat{\includegraphics[width = 0.114\linewidth]{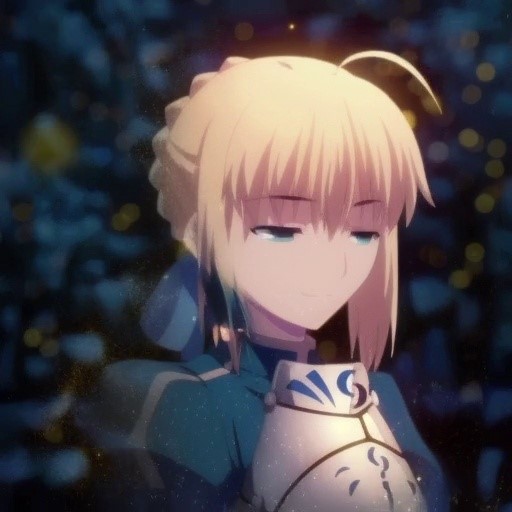}}\hspace{\rsWidth}
        \subfloat{\includegraphics[width = 0.114\linewidth]{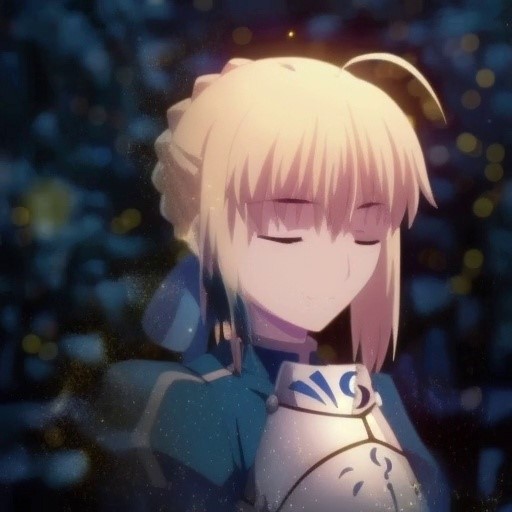}}\hspace{\rsWidth}
        \subfloat{\includegraphics[width = 0.114\linewidth]{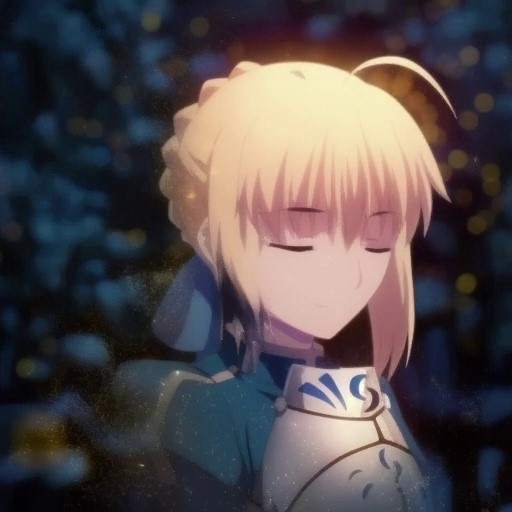}}\hspace{\rsWidth}
        \vspace{-0.1cm}
        \addtocounter{subfigure}{-8}
        \caption{Shorten hair}
        \label{subfig:saber}
    \end{subfigure}

    \vspace{\rsHeight}
    \begin{subfigure}{\linewidth}
        \centering
        \subfloat{\includegraphics[width = 0.114\linewidth]{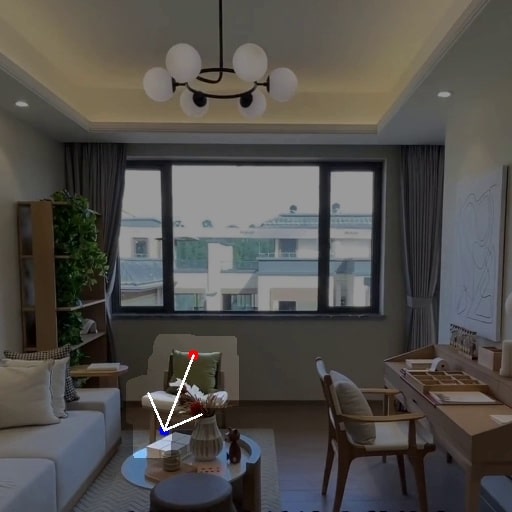}}\hspace{\rsWidth}
        \subfloat{\includegraphics[width = 0.114\linewidth]{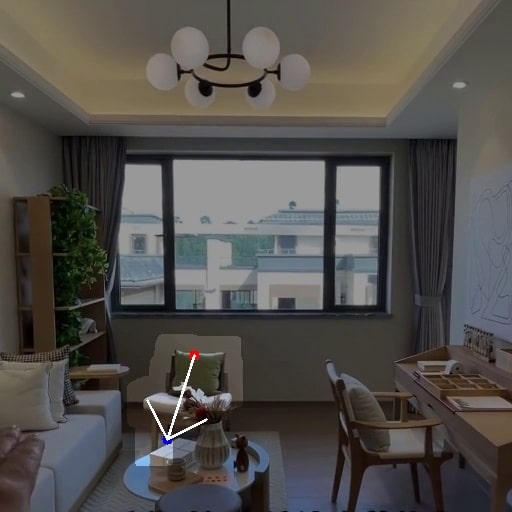}}\hspace{\rsWidth}
        \subfloat{\includegraphics[width = 0.114\linewidth]{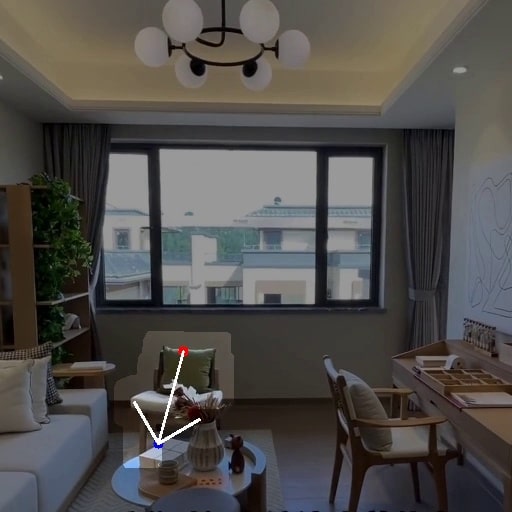}}\hspace{\rsWidth}
        \subfloat{\includegraphics[width = 0.114\linewidth]{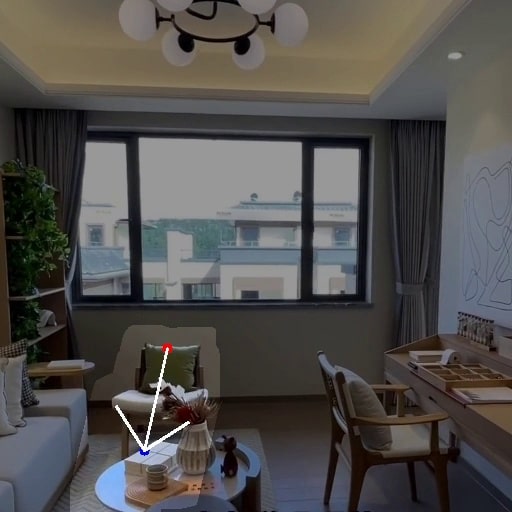}}\hspace{\rsmWidth}  
        \subfloat{\includegraphics[width = 0.114\linewidth]{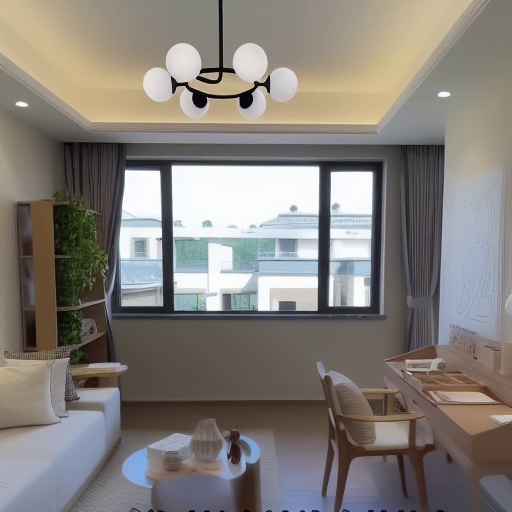}}\hspace{\rsWidth}
        \subfloat{\includegraphics[width = 0.114\linewidth]{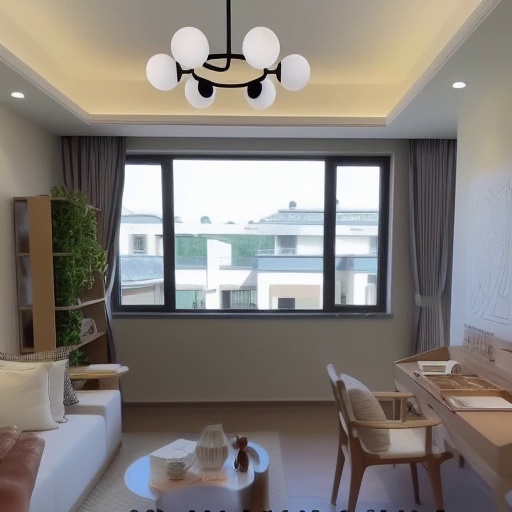}}\hspace{\rsWidth}
        \subfloat{\includegraphics[width = 0.114\linewidth]{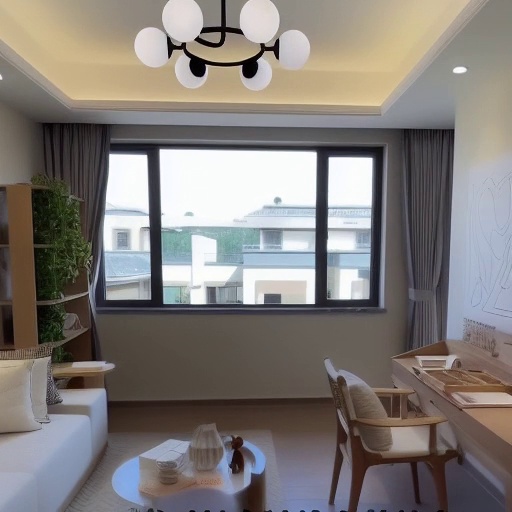}}\hspace{\rsWidth}  
        \subfloat{\includegraphics[width = 0.114\linewidth]{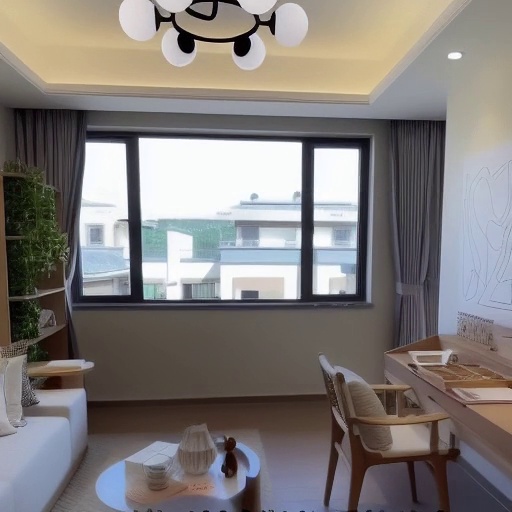}}\hspace{\rsWidth}
        \vspace{-0.1cm}
        \addtocounter{subfigure}{-8}
        \caption{Remove chair}
        \label{subfig:room}
    \end{subfigure}

    \vspace{\rsHeight}
    \begin{subfigure}{\linewidth}
        \centering
        \subfloat{\includegraphics[width = 0.114\linewidth]{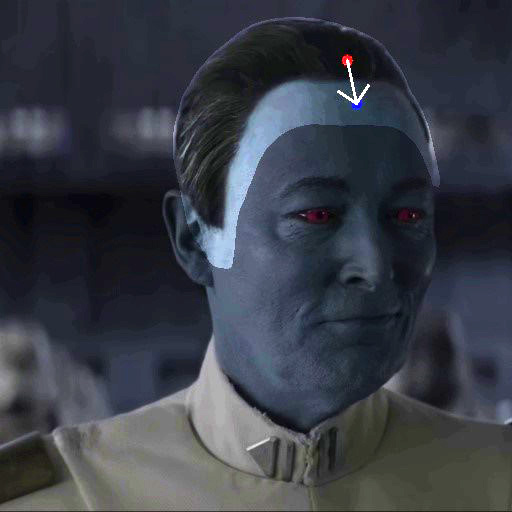}}\hspace{\rsWidth}
        \subfloat{\includegraphics[width = 0.114\linewidth]{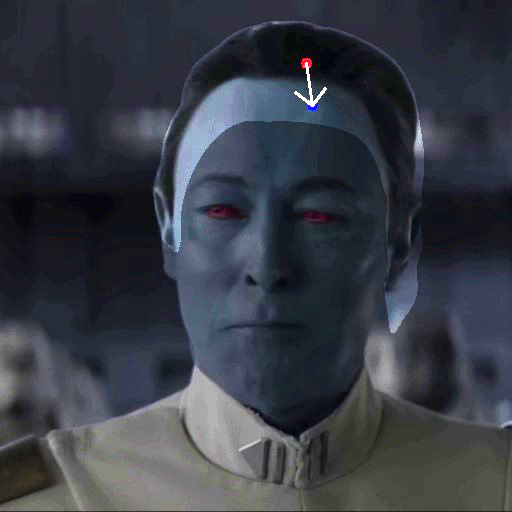}}\hspace{\rsWidth}
        \subfloat{\includegraphics[width = 0.114\linewidth]{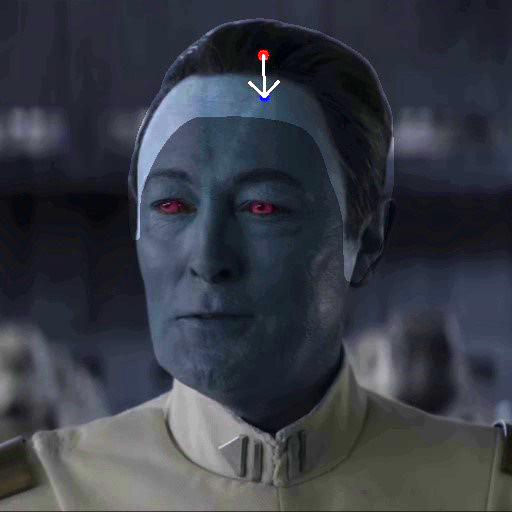}}\hspace{\rsWidth}
        \subfloat{\includegraphics[width = 0.114\linewidth]{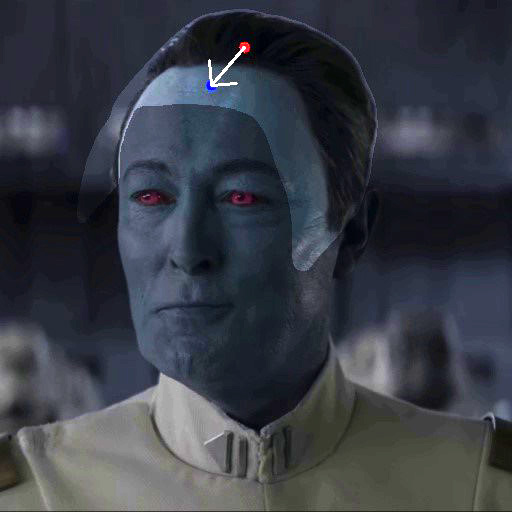}}\hspace{\rsmWidth}  
        \subfloat{\includegraphics[width = 0.114\linewidth]{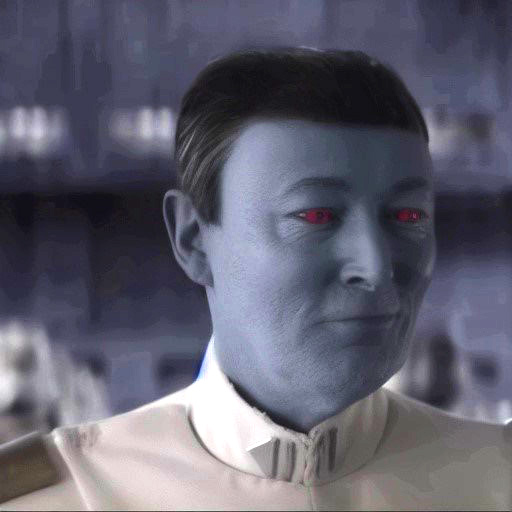}}\hspace{\rsWidth}
        \subfloat{\includegraphics[width = 0.114\linewidth]{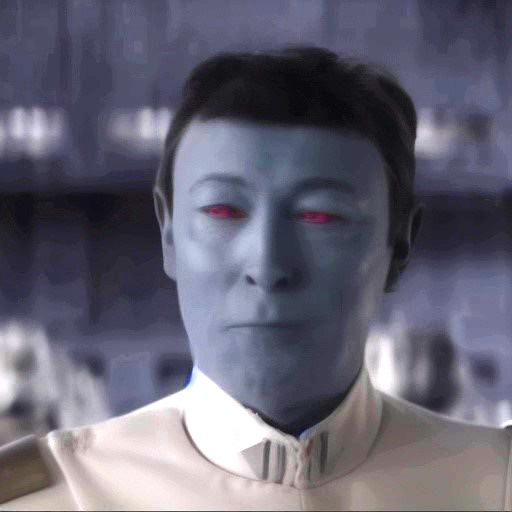}}\hspace{\rsWidth}
        \subfloat{\includegraphics[width = 0.114\linewidth]{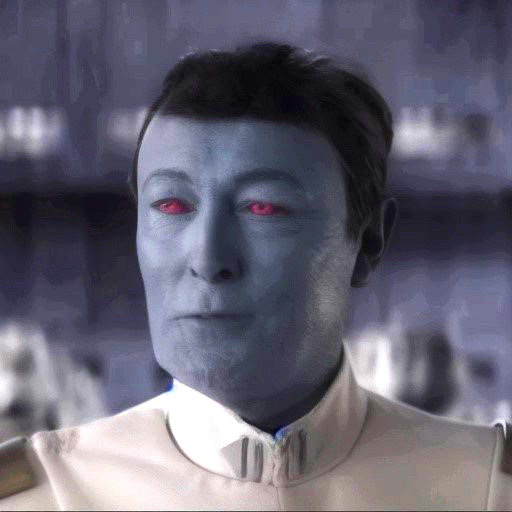}}\hspace{\rsWidth}  
        \subfloat{\includegraphics[width = 0.114\linewidth]{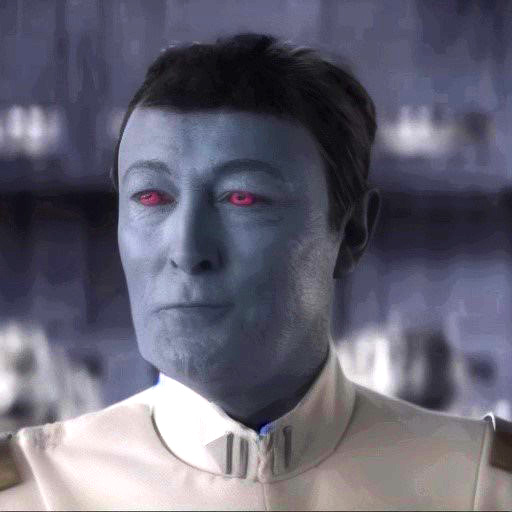}}\hspace{\rsWidth}
        \vspace{-0.1cm}
        \addtocounter{subfigure}{-8}
        \caption{Flatten hairline}
        \label{subfig:starwar}
    \end{subfigure}

    \caption{More results of DragVideo. Left four frames are original frame with propagated editing instructions (red points are handle points, blue points are target points, lighted area is mask). Right four frames are edited output.}
    \label{fig:result}
    \vspace{-0.2in}
\end{figure*}
\newcommand\bsWidth{0.00cm}
\newcommand\bsmWidth{0.15cm}
\newcommand\bsHeight{0.00cm}

\begin{figure*}[t]
    \centering
    \begin{subfigure}{0.47\linewidth}
        \centering
        \includegraphics[width=\linewidth]{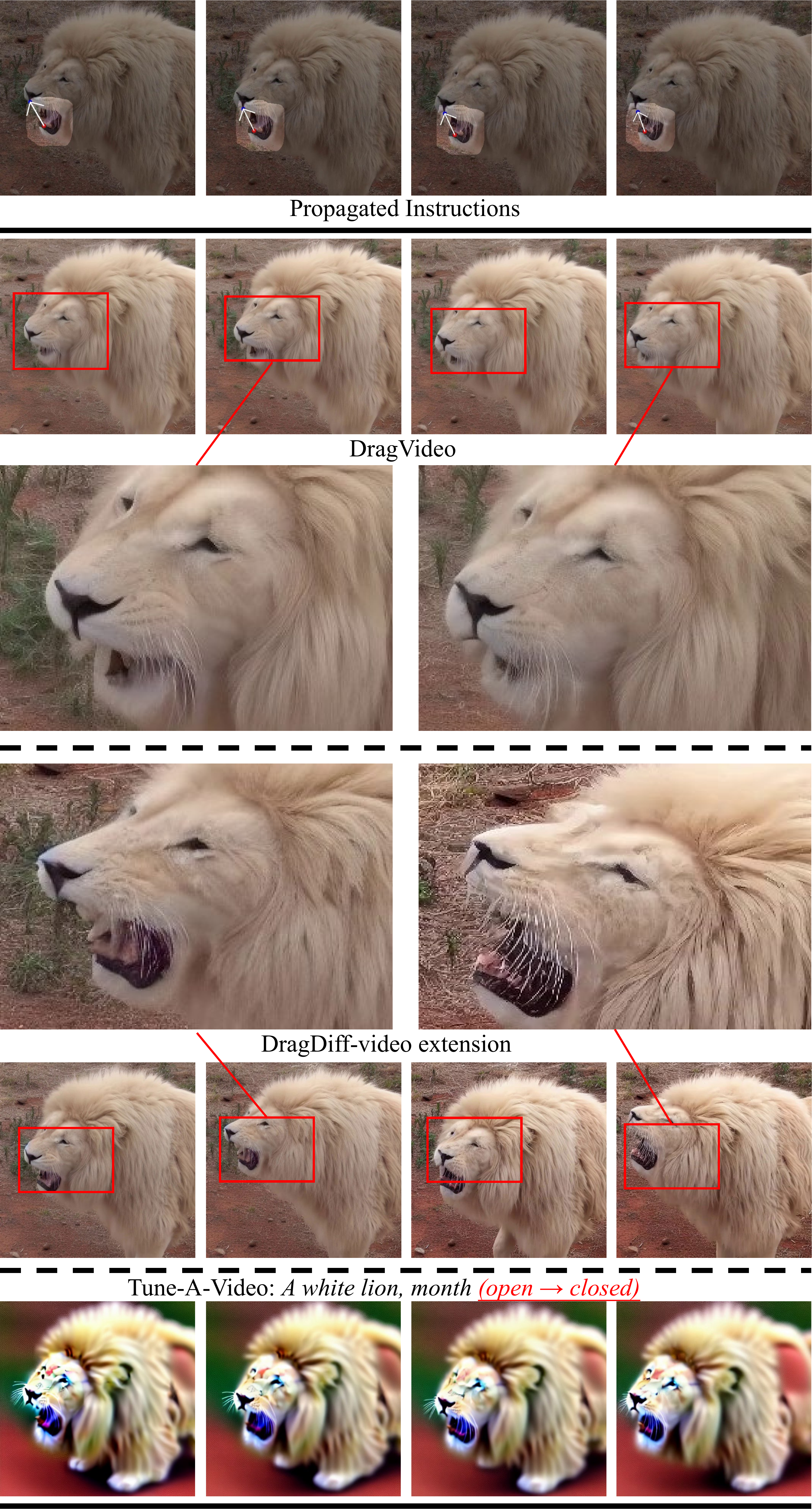}
        \caption{Close mouth}
        \label{subfig:cliff_baseline}
    \end{subfigure}\hspace{\bsmWidth}
    \begin{subfigure}{0.47\linewidth}
        \centering
        \includegraphics[width=\linewidth]{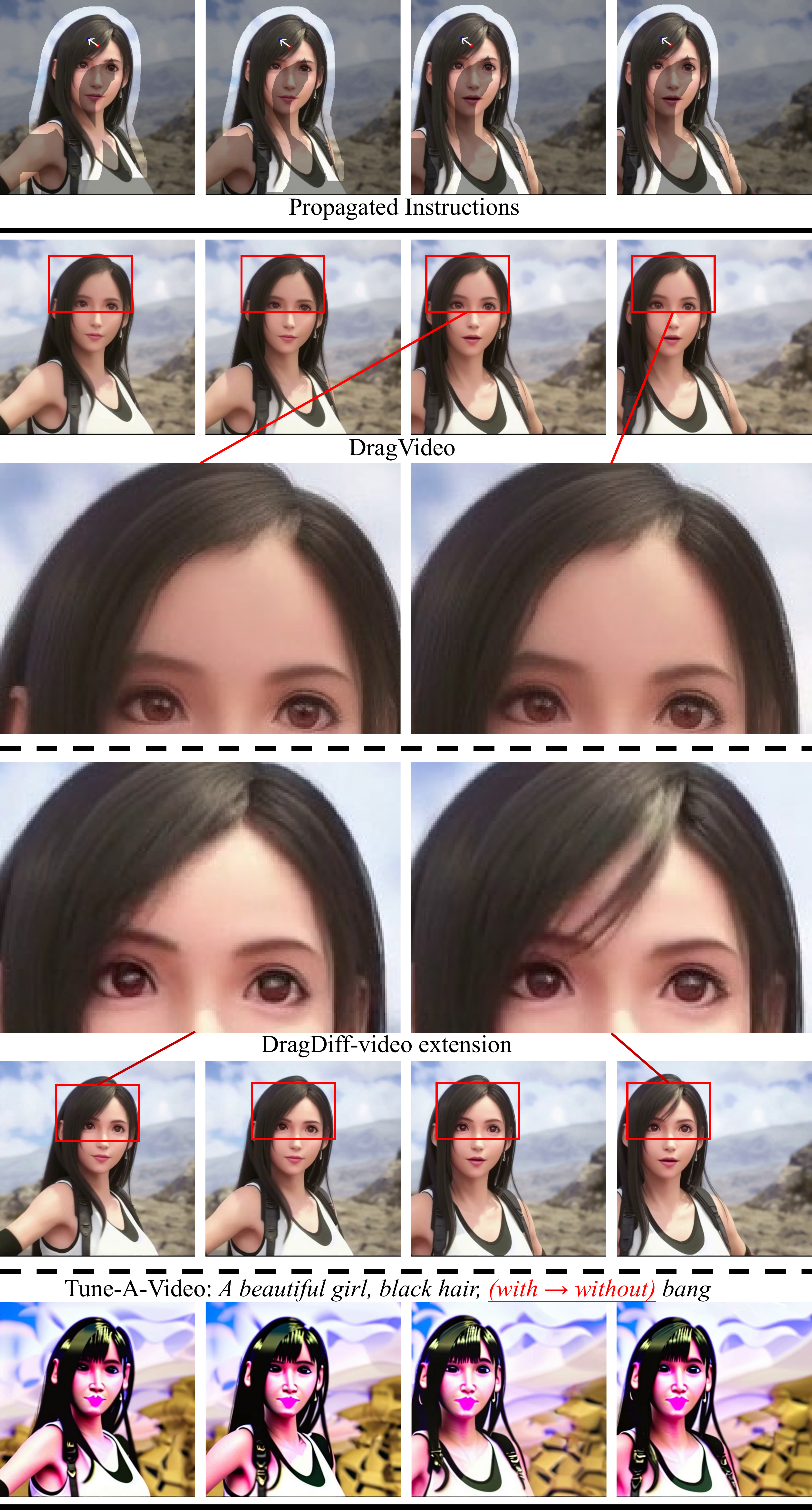}
        \caption{hair parting}
        \label{subfig:tifahair}
    \end{subfigure}
    \caption{Comparison between two baselines and DragVideo. First row is original frames with propagated editing instructions. DragVideo editing (second row with zoom) achieves decent results. Directly extending DragDiff to video (thrid row with zoom) faces the temporal inconsistent problem. Using prompt-based editing Tune-A-Video (last row) can introduce unintended style alterations without achieving the goal.}
    \label{fig:baseline}
    \vspace{-0.3in}
\end{figure*}

\subsection{Implementation Details}\label{exp:imp}
In our experiments, we adopt Stable Diffusion v1.5~\cite{rombach2022high} and its inherited models as base models for motion modules~\cite{guo2023animatediff} in AnimateDiff to construct our video diffusion model. Since there is no widely recognized purely transformer-based open-source video diffusion model, we leave the application of \textit{DragVideo} to them for future work. As aforementioned in Section~\ref{meth:lora}, we introduce LoRA into the projection matrices of query, key, and value in all of the attention modules within our video U-Net. The LoRA rank is set to 16 and the batch size to 12 as default. We employed AdamW~\cite{loshchilov2017decoupled} optimizer with a learning rate of $5\times{10^{-4}}$, and trained LoRA for 100 epochs before commencing drag editing. 

During the latent optimization phase, we set the inverse step to 50 for DDIM scheduler and start latent optimization at the 40th step. In our experiments, we do not apply classifier-free guidance (CFG)~\cite{ho2022classifier} in either DDIM inversion or DDIM denoising processes, as CFG tends to amplify inversion errors. The default batch size for drag optimization is set to 12. We set the drag-style latent optimization step to 40 while the latent learning rate is set to 0.01, $\lambda=0.1$ in Eq.~(\ref{eqn:loss}). The user can modify these parameters on our web user interface (UI) to obtain different results. The computation cost for \textit{DragVideo} does not exceed a single RTX-4090 or RTX-A6000, and the end-to-end processing time for 16 frames (2-4 seconds of video, max capacity for AnimateDiff) is around 5-10 minutes including LoRA training.

\subsection{Experiment Setup}\label{exp:set}

\subsubsection{Datasets}
With the recent debut of drag-style editing on images, there is no benchmark video datasets available with masks and point pairs for drag-style video editing. Thus we annotate an evaluation dataset on around 30 publicly available videos. To thoroughly test the efficacy of our method, we collect a wide range of examples, including pets, faces, furniture, scenery, and so on. The majority of the samples in our evaluation dataset are presented in the figures in paper and supplementary materials.

\subsubsection{Baselines}

As \textit{DragVideo} is the first approach to drag-style video editing, there are no universally approved baselines. Thus, based on the problems that \textit{DragVideo} intends to address, we implement two baselines. The first baseline is directly applying Tune-A-Video, a prompt-based video editing baseline. To ensure fairness, we carefully design prompts to make it as close to our target as possible. Another baseline is directly applying DragDiffusion~\cite{dragdiff} on every frame of the video, named DragDiff-video extension. We make a fair comparison by using the same set of masks and points for this baseline and \textit{DragVideo}. We have also tried to directly extend DragGAN~\cite{draggan} to video but almost all the videos crushed due to the limited ability of GAN. For more comparison between our result and the baseline, kindly refer to the supplementary materials.

\subsection{Qualitative Evaluation}

In this section, we conduct a thorough assessment of our \textit{DragVideo} framework's efficacy through an array of comprehensive experiments encompassing a broad spectrum of editing tasks.

\subsubsection{Results} 
Fig.\ref{fig:teaser} and Fig.\ref{fig:result} exhibit examples of our editing results. Evidently, our \textit{DragVideo} framework facilitates high-quality, drag-based editing on real-world videos. We can see \textit{DragVideo} achieve accurate editing by effectively moving handle points to target points with reasonable generated content in editing. The result is also free from noticeable artifacts as well as preserving relatively good spatio-temporal consistency. A wide array of examples illustrates that \textit{DragVideo} addresses three key problems in video editing in a unified. For additional qualitative results or video file results, kindly refer to supplementary materials.

\subsubsection{Baselines Comparison}

We offer a comparative analysis with baselines in Fig.\ref{fig:probElab} and Fig.\ref{fig:baseline}. We can see from the zoomed results that directly extending the image level DragDiff to video causes temporal inconsistency and identity shift. And if we compare with prompt-based editing, we can see that Tune-A-Video has a limited ability to preserve realistic details in the video; also the edit is not accurate and there are strong noticeable artifacts in the edited videos.

\begin{table}[t]
    \vspace{-0.1in}
    \centering
    \begin{tabular}{l|c|c|c|c}
        \toprule
        Evaluation Metrics        & {\ }DragDiff Ext{\ } & {\ }TuneAVideo{\ } & {\ }DragVideo{\ } & {\ }(Original Video)\\
        \midrule
        CLIP Consistency $\uparrow$ & 0.9833  & 0.9829   & \bf 0.9893 & (0.9903) \\
        RAFT Optical Flow $\downarrow${\ } & 3.3807  & 2.7075    & \bf 2.3780 & (2.3463) \\
        FVD Diff to Original $\downarrow$ & 400.89            & 2682.37          & \bf 397.43 & -\\
        \bottomrule
    \end{tabular}
    \caption{Quantitative evaluation of temporal consistency between neighbor frames in terms of optical flow and CLIP score similarity. FVD measures the overall difference from original videos. Scores of (Original Video) before editing is listed for reference}
    \label{tab:eval}
    \vspace{-0.3in}
\end{table}

\subsection{Quantitative Evaluation}
In this section, we evaluate the temporal consistency of the outputs from \textit{DragVideo}, DragDiff extension, and Tune-A-Video \cite{wu2023tune}. We adopt the CLIP consistency evaluation metric from \cite{wu2023tune}, which is computing the cosine similarity of CLIP \cite{radford2021learning} image embeddings between each consecutive two frames. Detailed calculation is:
\begin{equation}
    \mathit{score_{CLIP}} = \frac{\sum_{i=1}^{l-1} {\mathrm{sim}}({\mathrm{clip}}(f_i),{\mathrm{clip}}(f_{i+1}))}{(l-1)} \quad
\end{equation}
The second evaluation metric is the maximum optical flow between each consecutive two frames obtained by RAFT \cite{teed2020raft}, which can represent the level of jittering. Detailed calculation is:
\begin{equation}
    \mathit{score_{RAFT}} = \frac{\sum_{i=1}^{l-1} \max_{u\in w, v\in h} \|(\Delta x_{u,v}^{(i, i+1)}, \Delta y_{u,v}^{(i, i+1)})\|_2}{(l-1)}
\end{equation}
For both metrics, the averaged scores of 20 sample outputs are reported in Table \ref{tab:eval}. \textit{DragVideo} achieves less jittering, and thus better temporal consistency than the baseline and Tune-A-Video. Furthermore, we conduct the paired t-test to test the significance of \textit{DragVideo}. In CLIP consistency, the p-value for \textit{DragVideo} (DV) / DragDiff-Video (DDV) is 1.27\%, and for DV / Tune-A-Video (TV) is 1.27\%. In RAFT consistency, the p-value for DV / DDV is 0.06\%, and for DV / TV is 2.30\%. They are smaller than the normal 5\% of the significance level and we can draw the conclusion that \textit{DragVideo} is significantly more consistent.

In addition, we apply FVD~\cite{unterthiner2018towards} to evaluate the temporal consistency but it cannot be separated from video quality. The point and mask propagation is very robust; we omit it as they are not
the major contribution.

\newcommand\usWidth{0.0cm}

\begin{figure*}[t]
\vspace{-0.1in}
    \centering
    
    \begin{subfigure}[b]{0.48\linewidth}
        \centering
        \includegraphics[width=\linewidth]{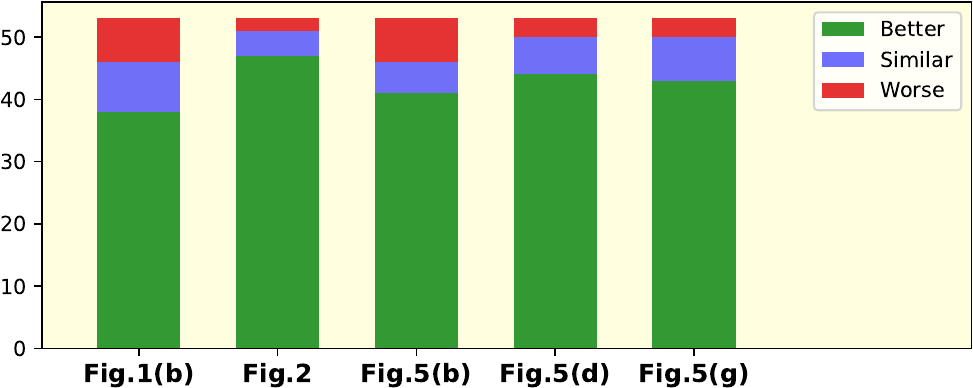}
        \caption{Consistency Study}
        \label{subfig:con_table}
    \end{subfigure}
    \hspace{\usWidth}
    \begin{subfigure}[b]{0.48\linewidth}
        \centering
        \includegraphics[width=\linewidth]{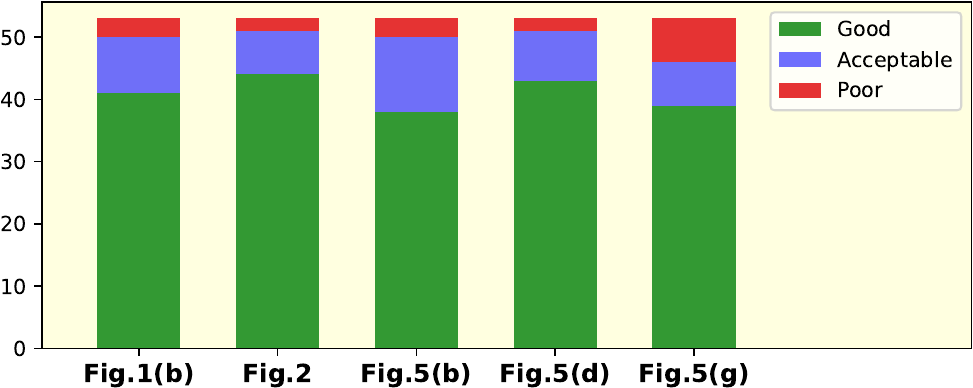}
        \caption{Effectiveness Study}
        \label{subfig:eff_table}
    \end{subfigure}

    \vspace{-0.10in}
    \caption{The User Study result of 53 subjects. (a) measures whether DragVideo performs better/similar/worse as baseline. (b) measures whether DragVideo's edits are good/acceptable/poor. Number of subjects for each option regarding five sample outputs are represented by different color portions.}\label{fig:UserStudy}
    \vspace{-0.3in}
    
\end{figure*}

\subsection{User studies}\label{exp:user}

To further validate the \textit{DragVideo} framework, we perform user studies on both effectiveness and temporal consistencies. The survey on the effectiveness was designed to let participants evaluate the generated results of \textit{DragVideo} on a scale of 1-10 in terms of satisfaction where 10 means a very satisfactory result and 1 means very unsatisfactory. The survey on temporal consistency asks participants to compare the consistency of results from \textit{DragVideo} and DragDiff-video extension using the same 1-10 scale, where 10 means \textit{DragVideo} is better and 1 means DragDiff-video extension is better. The sequence of baseline and \textit{DragVideo} is organized randomly.

We collected 53 surveys from users of diverse backgrounds. The results for user studies are shown in Fig.\ref{fig:UserStudy}. In the plot, we utilize green to represent scores 7-10, blue for 5-6, and red for 1-4. The findings indicate that our method is highly favored by users, which aligns with the quantitative evaluation presented in the paper and underscores the effectiveness of our model in achieving accurate and temporally consistent editing. After performing statistical analysis, the average 98\% confidence interval for effective study is score (7.47, 8.51); for consistency, the score interval is (7.74, 8.96), which demonstrates our result is statistically significant outperform baselines. For a comprehensive statistical analysis of the user study like p-value and 98\% confidence interval, please refer to the supplementary materials.

\newcommand\abWidth{-0.0cm}
\newcommand\abmWidth{0.15cm}
\newcommand\abHeight{0.00cm}

\begin{figure*}[t]
    \vspace{-0.1in}
    \centering
    \captionsetup[subfloat]{labelformat=empty}

    \subfloat[Propagated Instructions]{
        \subfloat{\includegraphics[width = 0.113\linewidth]{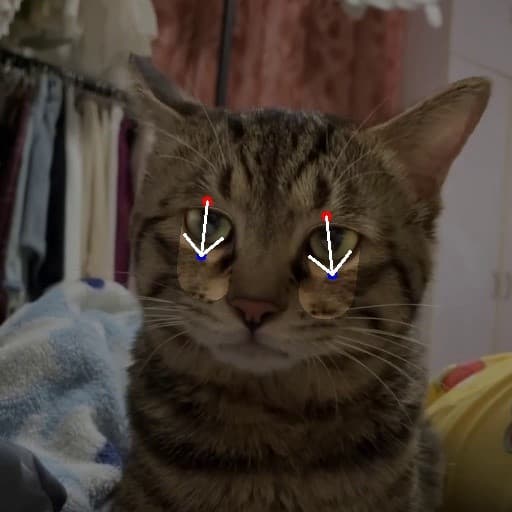}}\hspace{\abWidth}
        \subfloat{\includegraphics[width = 0.113\linewidth]{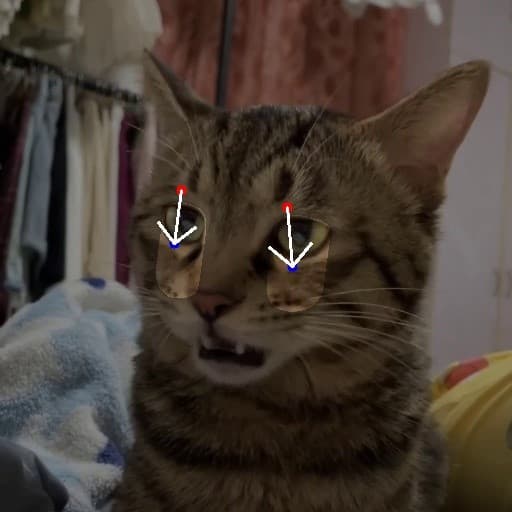}}\hspace{\abWidth}
        \subfloat{\includegraphics[width = 0.113\linewidth]{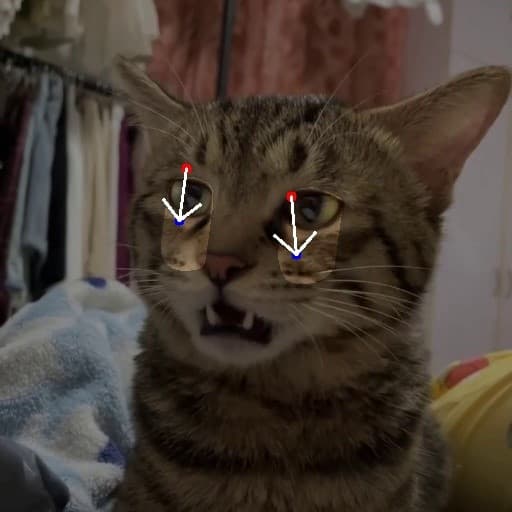}}\hspace{\abWidth}
        \subfloat{\includegraphics[width = 0.113\linewidth]{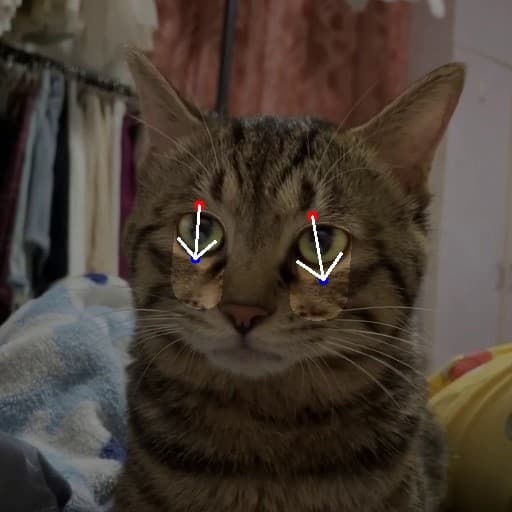}}\hspace{\abmWidth}
    }
    \subfloat[Edited Output]{
        \subfloat{\includegraphics[width = 0.113\linewidth]{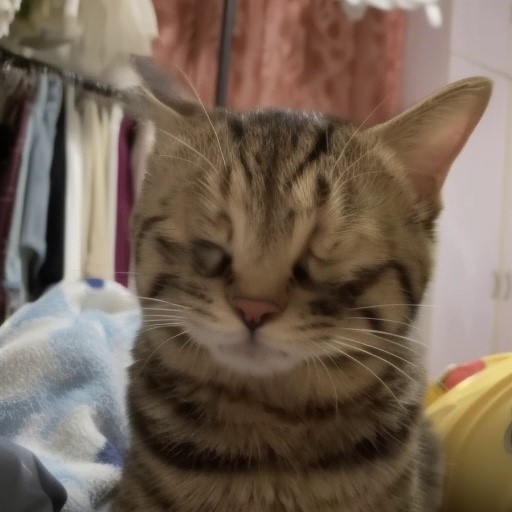}}\hspace{\abWidth}
        \subfloat{\includegraphics[width = 0.113\linewidth]{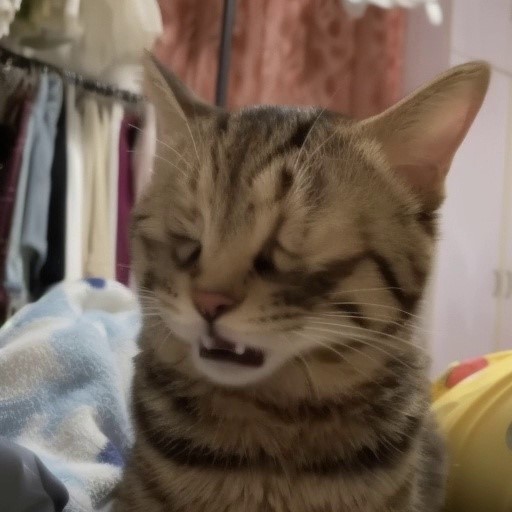}}\hspace{\abWidth}
        \subfloat{\includegraphics[width = 0.113\linewidth]{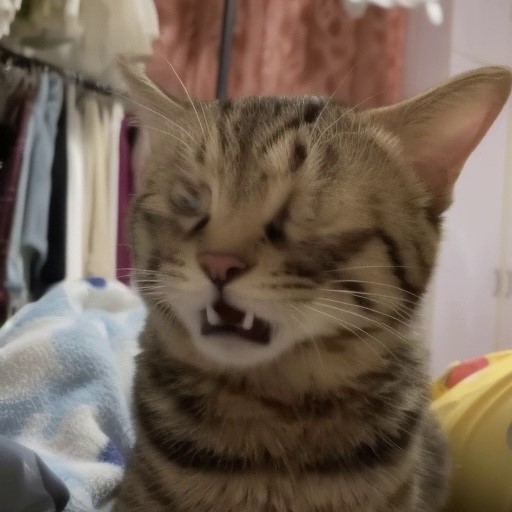}}\hspace{\abWidth}
        \subfloat{\includegraphics[width = 0.113\linewidth]{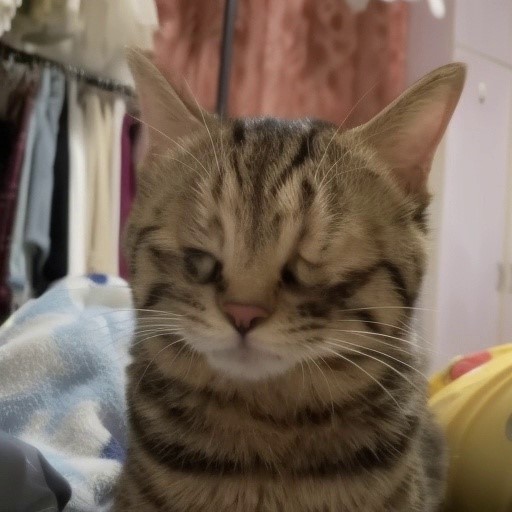}}
    }
    
    \vspace{\abHeight}
    \subfloat[without LoRA]{
        \subfloat{\includegraphics[width = 0.113\linewidth]{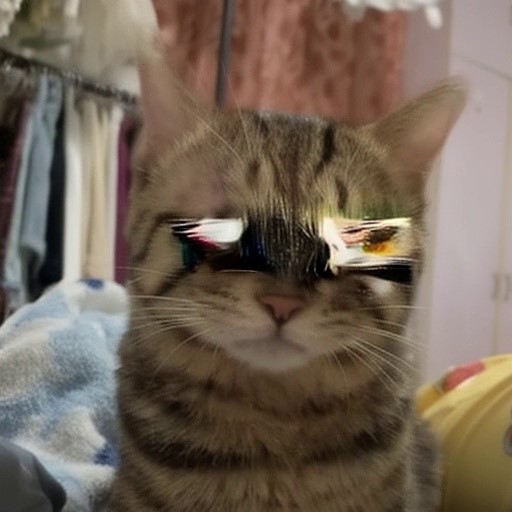}}\hspace{\abWidth}
        \subfloat{\includegraphics[width = 0.113\linewidth]{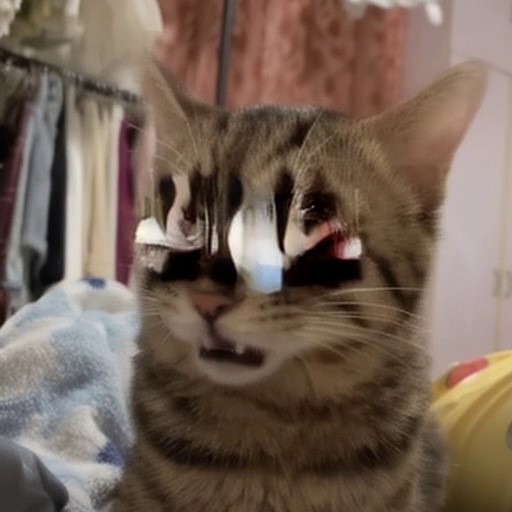}}\hspace{\abWidth}
        \subfloat{\includegraphics[width = 0.113\linewidth]{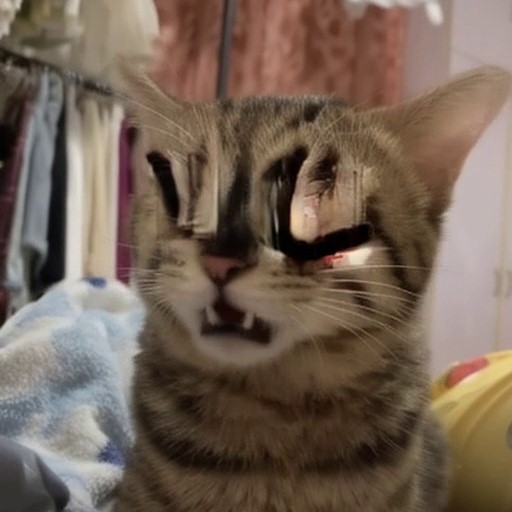}}\hspace{\abWidth}
        \subfloat{\includegraphics[width = 0.113\linewidth]{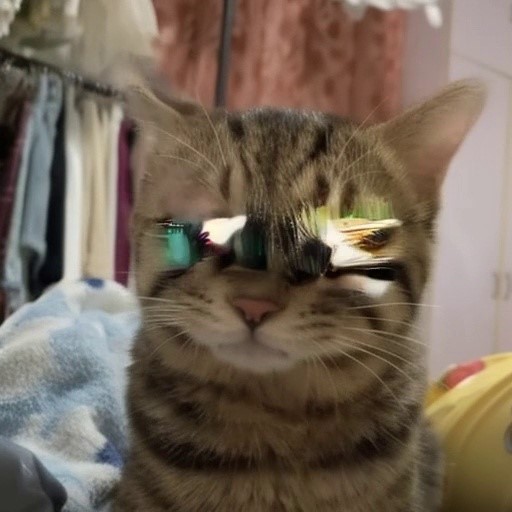}}\hspace{\abmWidth}
    }
    \subfloat[without MSA]{
        \subfloat{\includegraphics[width = 0.113\linewidth]{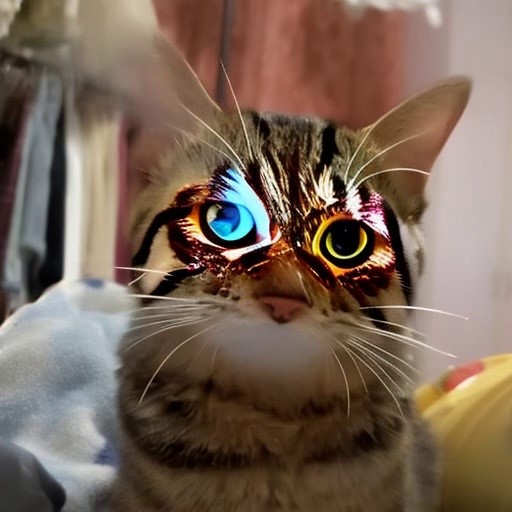}}\hspace{\abWidth}
        \subfloat{\includegraphics[width = 0.113\linewidth]{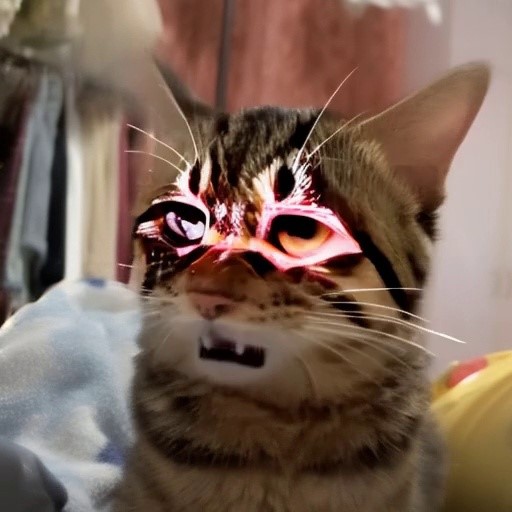}}\hspace{\abWidth}
        \subfloat{\includegraphics[width = 0.113\linewidth]{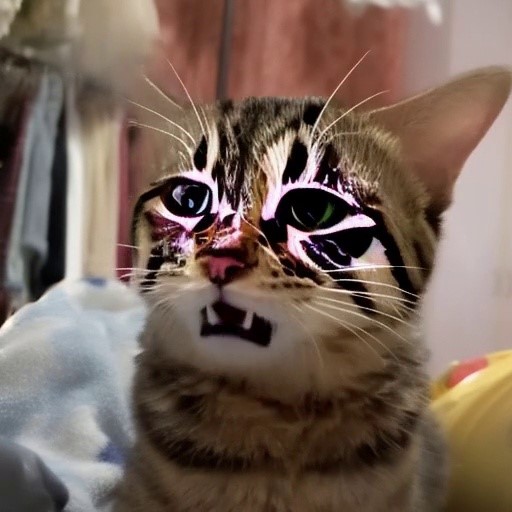}}\hspace{\abWidth}  
        \subfloat{\includegraphics[width = 0.113\linewidth]{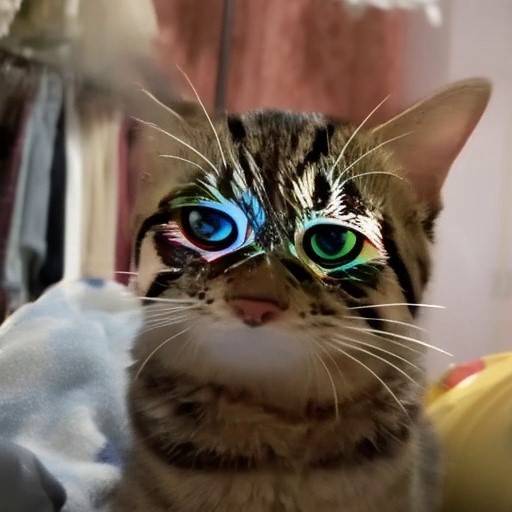}}
    }
\vspace{-0.15in}
    \caption{Ablation study of DragVideo. Missing LoRA (bottom left) and missing MSA (bottom right) reduce the quality of DragVideo's outputs. DragVideo with all components (top right) achieve better results.}
\vspace{-0.15in}
    \label{fig:ablation}
\end{figure*}

\subsection{Ablation Study}

In order to test the effectiveness of Sample specific LoRA and Mutual Self-Attention's effectiveness in preserving spatial consistency in \textit{DragVideo}. Here we present the ablation of them in Fig.\ref{fig:ablation}, where we can see the edited area deviates substantially from the original frame without LoRA while becoming unreasonable without MSA. Missing LoRA makes \textit{DragVideo} unable to preserve the original style while missing MSA allows the optimized latent to destroy intermediate features during denoising.

\section{Discussion}
\label{sec:conc}

\textbf{Limitations} \, Despite the promising results of \textit{DragVideo}, there remain two limitations. First, even when leveraging Sample-specific LoRA and MSA techniques, a minority of the edited outputs still exhibit some blurriness, and spatial inconsistency, suggesting room for further optimization in visual quality. Second, our framework, inheriting the computational challenges of the diffusion model and the drag objective function, still incurs high computational costs. 
This points to a need for improvements in computational efficiency. As a pioneering effort in drag-style video editing, \textit{DragVideo} sets the stage for future work. \\
\\
\textbf{Conclusion} \, 
This paper introduces \textit{DragVideo}, a user-centric and intuitive solution that faithfully captures the user's intentions for drag-style video editing. By harnessing the rich information embedded in video diffusion models, our method allows users to directly and effortlessly manipulate video content beyond text-guided style change. As the first technical endeavor into drag-style video editing with 
careful design, validated by a wide range of qualitative, quantitative experiments, and user studies, \textit{DragVideo} overcomes the limitations inherent in direct frame-by-frame DragImage extension or current text-guided video editing, where distortion and spatio-temporal inconsistency are still quite apparent. 
Our experiments demonstrate the versatility and broad applicability of our approach, validating \textit{DragVideo} as a powerful tool for video editing tasks. With the rapid development of video diffusion models, \textit{DragVideo} contributes the next strong baseline for drag-style video editing, where long-range editing with more powerful generative capability is the future research direction.

\section*{Acknowledgements}
This work was supported in part by Dartmouth College A$\&$S Startup fund.

\bibliographystyle{splncs04}
\bibliography{main}

\begin{thebibliography}{10}
\providecommand{\url}[1]{\texttt{#1}}
\providecommand{\urlprefix}{URL }
\providecommand{\doi}[1]{https://doi.org/#1}

\bibitem{bai2024uniedit}
Bai, J., He, T., Wang, Y., Guo, J., Hu, H., Liu, Z., Bian, J.: Uniedit: A unified tuning-free framework for video motion and appearance editing. arXiv preprint arXiv:2402.13185  (2024)

\bibitem{videoworldsimulators2024}
Brooks, T., Peebles, B., Holmes, C., DePue, W., Guo, Y., Jing, L., Schnurr, D., Taylor, J., Luhman, T., Luhman, E., Ng, C., Wang, R., Ramesh, A.: Video generation models as world simulators  (2024), \url{https://openai.com/research/video-generation-models-as-world-simulators}

\bibitem{mutualattn}
Cao, M., Wang, X., Qi, Z., Shan, Y., Qie, X., Zheng, Y.: Masactrl: Tuning-free mutual self-attention control for consistent image synthesis and editing. arXiv preprint arXiv:2304.08465  (2023)

\bibitem{ceylan2023pix2video}
Ceylan, D., Huang, C.H.P., Mitra, N.J.: Pix2video: Video editing using image diffusion. In: Proceedings of the IEEE/CVF International Conference on Computer Vision. pp. 23206--23217 (2023)

\bibitem{chen2023control}
Chen, W., Wu, J., Xie, P., Wu, H., Li, J., Xia, X., Xiao, X., Lin, L.: Control-a-video: Controllable text-to-video generation with diffusion models. arXiv preprint arXiv:2305.13840  (2023)

\bibitem{cheng2022xmem}
Cheng, H.K., Schwing, A.G.: Xmem: Long-term video object segmentation with an atkinson-shiffrin memory model. In: European Conference on Computer Vision. pp. 640--658. Springer (2022)

\bibitem{doersch2022tap}
Doersch, C., Gupta, A., Markeeva, L., Recasens, A., Smaira, L., Aytar, Y., Carreira, J., Zisserman, A., Yang, Y.: Tap-vid: A benchmark for tracking any point in a video. Advances in Neural Information Processing Systems  \textbf{35},  13610--13626 (2022)

\bibitem{guo2023animatediff}
Guo, Y., Yang, C., Rao, A., Wang, Y., Qiao, Y., Lin, D., Dai, B.: Animatediff: Animate your personalized text-to-image diffusion models without specific tuning. arXiv preprint arXiv:2307.04725  (2023)

\bibitem{harley2022particle}
Harley, A.W., Fang, Z., Fragkiadaki, K.: Particle video revisited: Tracking through occlusions using point trajectories. In: European Conference on Computer Vision. pp. 59--75. Springer (2022)

\bibitem{ho2020denoising}
Ho, J., Jain, A., Abbeel, P.: Denoising diffusion probabilistic models. Advances in neural information processing systems  \textbf{33},  6840--6851 (2020)

\bibitem{ho2022classifier}
Ho, J., Salimans, T.: Classifier-free diffusion guidance. arXiv preprint arXiv:2207.12598  (2022)

\bibitem{lora}
Hu, E.J., Shen, Y., Wallis, P., Allen-Zhu, Z., Li, Y., Wang, S., Wang, L., Chen, W.: Lora: Low-rank adaptation of large language models. arXiv preprint arXiv:2106.09685  (2021)

\bibitem{hu2024cns}
Hu, J., Hui, K.H., Liu, Z., Zhang, H., Fu, C.W.: Cns-edit: 3d shape editing via coupled neural shape optimization. arXiv preprint arXiv:2402.02313  (2024)

\bibitem{stylegan}
Karras, T., Laine, S., Aittala, M., Hellsten, J., Lehtinen, J., Aila, T.: Analyzing and improving the image quality of stylegan. In: Proceedings of the IEEE/CVF conference on computer vision and pattern recognition. pp. 8110--8119 (2020)

\bibitem{khachatryan2023text2video}
Khachatryan, L., Movsisyan, A., Tadevosyan, V., Henschel, R., Wang, Z., Navasardyan, S., Shi, H.: Text2video-zero: Text-to-image diffusion models are zero-shot video generators. arXiv preprint arXiv:2303.13439  (2023)

\bibitem{kirillov2023segment}
Kirillov, A., Mintun, E., Ravi, N., Mao, H., Rolland, C., Gustafson, L., Xiao, T., Whitehead, S., Berg, A.C., Lo, W.Y., et~al.: Segment anything. arXiv preprint arXiv:2304.02643  (2023)

\bibitem{loshchilov2017decoupled}
Loshchilov, I., Hutter, F.: Decoupled weight decay regularization. arXiv preprint arXiv:1711.05101  (2017)

\bibitem{mokady2022null}
Mokady, R., Hertz, A., Aberman, K., Pritch, Y., Cohen-Or, D.: Null-text inversion for editing real images using guided diffusion models. arXiv preprint arXiv:2211.09794  (2022)

\bibitem{moore1989introduction}
Moore, D.S., McCabe, G.P.: Introduction to the practice of statistics. WH Freeman/Times Books/Henry Holt \& Co (1989)

\bibitem{dragondiff}
Mou, C., Wang, X., Song, J., Shan, Y., Zhang, J.: Dragondiffusion: Enabling drag-style manipulation on diffusion models. arXiv preprint arXiv:2307.02421  (2023)

\bibitem{ouyang2023codef}
Ouyang, H., Wang, Q., Xiao, Y., Bai, Q., Zhang, J., Zheng, K., Zhou, X., Chen, Q., Shen, Y.: Codef: Content deformation fields for temporally consistent video processing. arXiv preprint arXiv:2308.07926  (2023)

\bibitem{draggan}
Pan, X., Tewari, A., Leimk{\"u}hler, T., Liu, L., Meka, A., Theobalt, C.: Drag your gan: Interactive point-based manipulation on the generative image manifold. In: ACM SIGGRAPH 2023 Conference Proceedings. pp. 1--11 (2023)

\bibitem{radford2021learning}
Radford, A., Kim, J.W., Hallacy, C., Ramesh, A., Goh, G., Agarwal, S., Sastry, G., Askell, A., Mishkin, P., Clark, J., et~al.: Learning transferable visual models from natural language supervision. In: International conference on machine learning. pp. 8748--8763. PMLR (2021)

\bibitem{rombach2022high}
Rombach, R., Blattmann, A., Lorenz, D., Esser, P., Ommer, B.: High-resolution image synthesis with latent diffusion models. In: Proceedings of the IEEE/CVF conference on computer vision and pattern recognition. pp. 10684--10695 (2022)

\bibitem{dragdiff}
Shi, Y., Xue, C., Pan, J., Zhang, W., Tan, V.Y., Bai, S.: Dragdiffusion: Harnessing diffusion models for interactive point-based image editing. arXiv preprint arXiv:2306.14435  (2023)

\bibitem{shin2023edit}
Shin, C., Kim, H., Lee, C.H., Lee, S.g., Yoon, S.: Edit-a-video: Single video editing with object-aware consistency. arXiv preprint arXiv:2303.07945  (2023)

\bibitem{ddim}
Song, J., Meng, C., Ermon, S.: Denoising diffusion implicit models. arXiv preprint arXiv:2010.02502  (2020)

\bibitem{tang2023emergent}
Tang, L., Jia, M., Wang, Q., Phoo, C.P., Hariharan, B.: Emergent correspondence from image diffusion. arXiv preprint arXiv:2306.03881  (2023)

\bibitem{tanveer2024anamodiff}
Tanveer, M., Wang, Y., Wang, R., Zhao, N., Mahdavi-Amiri, A., Zhang, H.: Anamodiff: 2d analogical motion diffusion via disentangled denoising. arXiv preprint arXiv:2402.03549  (2024)

\bibitem{teed2020raft}
Teed, Z., Deng, J.: Raft: Recurrent all-pairs field transforms for optical flow. In: Computer Vision--ECCV 2020: 16th European Conference, Glasgow, UK, August 23--28, 2020, Proceedings, Part II 16. pp. 402--419. Springer (2020)

\bibitem{unterthiner2018towards}
Unterthiner, T., Van~Steenkiste, S., Kurach, K., Marinier, R., Michalski, M., Gelly, S.: Towards accurate generative models of video: A new metric \& challenges. arXiv preprint arXiv:1812.01717  (2018)

\bibitem{wang2023tracking}
Wang, Q., Chang, Y.Y., Cai, R., Li, Z., Hariharan, B., Holynski, A., Snavely, N.: Tracking everything everywhere all at once. arXiv preprint arXiv:2306.05422  (2023)

\bibitem{wang2023videocomposer}
Wang, X., Yuan, H., Zhang, S., Chen, D., Wang, J., Zhang, Y., Shen, Y., Zhao, D., Zhou, J.: Videocomposer: Compositional video synthesis with motion controllability. arXiv preprint arXiv:2306.02018  (2023)

\bibitem{wu2023tune}
Wu, J.Z., Ge, Y., Wang, X., Lei, S.W., Gu, Y., Shi, Y., Hsu, W., Shan, Y., Qie, X., Shou, M.Z.: Tune-a-video: One-shot tuning of image diffusion models for text-to-video generation. In: Proceedings of the IEEE/CVF International Conference on Computer Vision. pp. 7623--7633 (2023)

\bibitem{yang2023track}
Yang, J., Gao, M., Li, Z., Gao, S., Wang, F., Zheng, F.: Track anything: Segment anything meets videos. arXiv preprint arXiv:2304.11968  (2023)

\bibitem{yang2023rerender}
Yang, S., Zhou, Y., Liu, Z., Loy, C.C.: Rerender a video: Zero-shot text-guided video-to-video translation. arXiv preprint arXiv:2306.07954  (2023)

\bibitem{zhang2023controlvideo}
Zhang, Y., Wei, Y., Jiang, D., Zhang, X., Zuo, W., Tian, Q.: Controlvideo: Training-free controllable text-to-video generation. arXiv preprint arXiv:2305.13077  (2023)

\end{thebibliography}

\clearpage
\setcounter{page}{1}
\begin{appendix}

\section{User study analysis}

This section provides further statistical analysis of the user study. Following \cite{moore1989introduction}, when the sample size exceeds 30 (we have 53 surveys received), the central limit theorem fits well. This allows the assumption that the distribution of sample means converges to a normal distribution. The results of our user study, including mean and 98\% confidence intervals (c.i.) are presented in Tables \ref{tab:statAnalCons} and \ref{tab:statAnalEffec}. Additionally, since the mean converges to the normal distribution, we conducted a t-test to statistically evaluate whether \textit{DragVideo} significantly outperforms the baseline. As delineated in Section\ref{exp:user}, our hypothesis are defined as follows:

\[
\begin{aligned}
    \text{H}_0 \text{:} & \text{ The mean score of the sample is equal to 6, } \\
                        & \text{ indicating DragVideo is equivalence or inferiority to the baseline} \\
    \text{H}_1 \text{:} & \text{ The mean score of the sample is greater than 6, } \\
                        & \text{ indicating DragVideo is superior to the baseline} \\
\end{aligned}
\]

The critical value for decision-making in the t-test is the t-value. With 53 survey responses (degrees of freedom = 52), a t-value exceeding 2.40 allows us to reject H$_0$ at a 99\% confidence level, indicating \textit{DragVideo}'s superiority with a maximum error rate of 1\%.

The analyses confirm that \textit{DragVideo} surpasses baseline methods in terms of temporal consistency and effectiveness with statistical significance.

\begin{table}
    \centering
    \begin{tabular}{cccc}
        \toprule
        Sample for study                             & Mean      & 98\% c.i.        & t-value \\
        \hline
        Close mouth Fig.\ref{subfig:lion}          & 7.84      & (7.10, 8.57)     & 4.86 \\
        Enanrging window Fig.\ref{fig:probElab}      & 8.90      & (8.39, 9.41)     & 11.12 \\
        Shorten sleeves Fig.\ref{subfig:sleeves}      & 8.38      & (7.76, 8.99)     & 7.51 \\
        Connect island Fig.\ref{subfig:island}       & 8.62      & (8.13, 9.11)     & 10.32 \\
        Flatten hairline Fig.\ref{subfig:starwar}    & 8.82      & (8.28, 9.37)     & 10.20 \\
        \bottomrule
    \end{tabular}
    \caption{User study analysis for temporal consistency}
    \label{tab:statAnalCons}
\end{table}
\begin{table}
    \centering
    \begin{tabular}{cccc}
        \toprule
        Sample for study                             & Mean      & 98\% c.i.        & t-value \\
        \hline
        Close mouth Fig.\ref{subfig:lion}          & 7.62      & (7.09, 8.15)     & 5.93 \\
        Enanrging window Fig.\ref{fig:probElab}      & 8.58      & (8.12, 9.05)     & 10.83 \\
        Shorten sleeves Fig.\ref{subfig:sleeves}      & 7.79      & (7.26, 8.33)     & 6.48 \\
        Connect island Fig.\ref{subfig:island}       & 8.15      & (7.67, 8.63)     & 8.64 \\
        Flatten hairline Fig.\ref{subfig:starwar}    & 7.49      & (6.84, 8.13)     & 4.49 \\
        \bottomrule
    \end{tabular}
    \caption{User study analysis for effectiveness}
    \label{tab:statAnalEffec}
\end{table}

\section{Additional Qualitative Results}

\begin{figure*}[!]
    \centering

    \vspace{\rsHeight}
    \begin{subfigure}{\linewidth}
        \subfloat{\includegraphics[width = 0.114\linewidth]{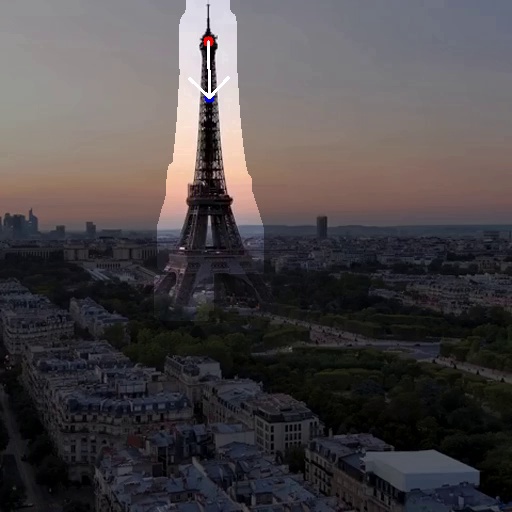}}\hspace{\rsWidth}
        \subfloat{\includegraphics[width = 0.114\linewidth]{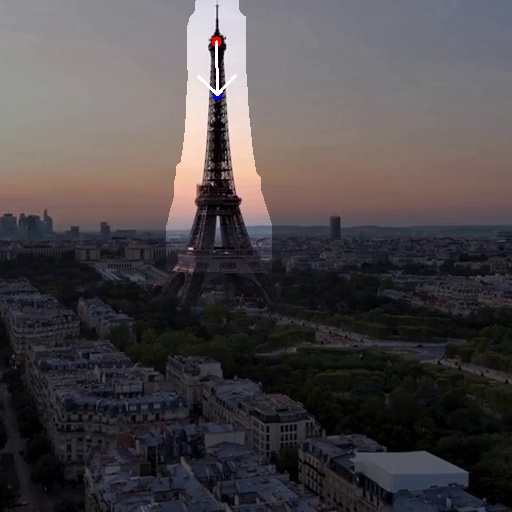}}\hspace{\rsWidth}
        \subfloat{\includegraphics[width = 0.114\linewidth]{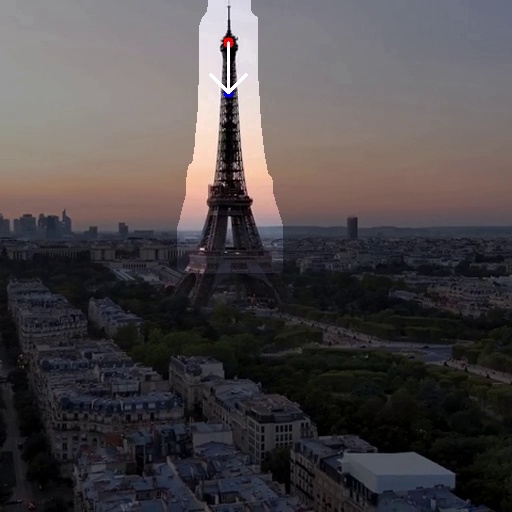}}\hspace{\rsWidth}
        \subfloat{\includegraphics[width = 0.114\linewidth]{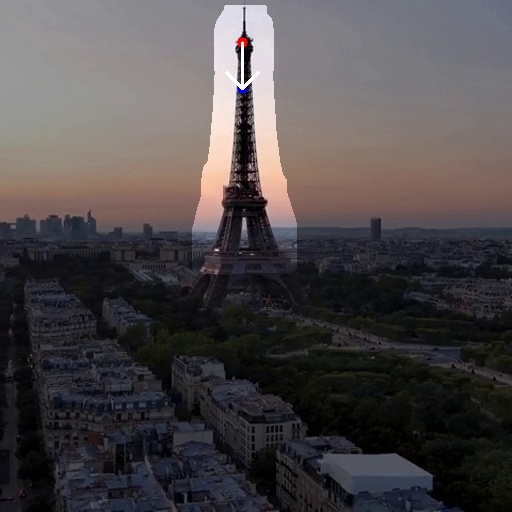}}\hspace{\rsmWidth}  
        \subfloat{\includegraphics[width = 0.114\linewidth]{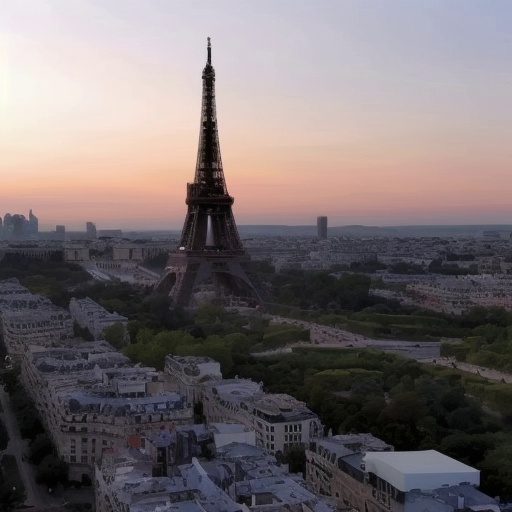}}\hspace{\rsWidth}
        \subfloat{\includegraphics[width = 0.114\linewidth]{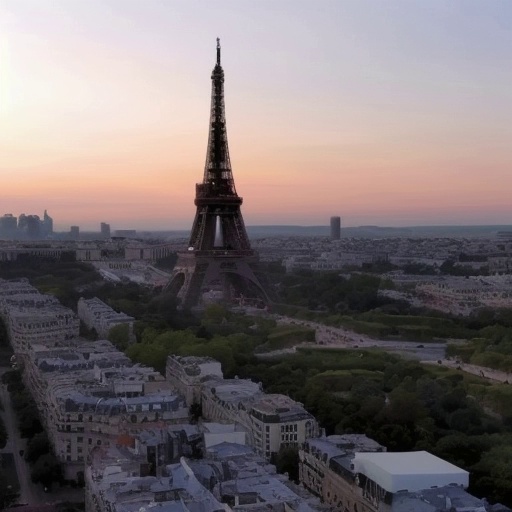}}\hspace{\rsWidth}
        \subfloat{\includegraphics[width = 0.114\linewidth]{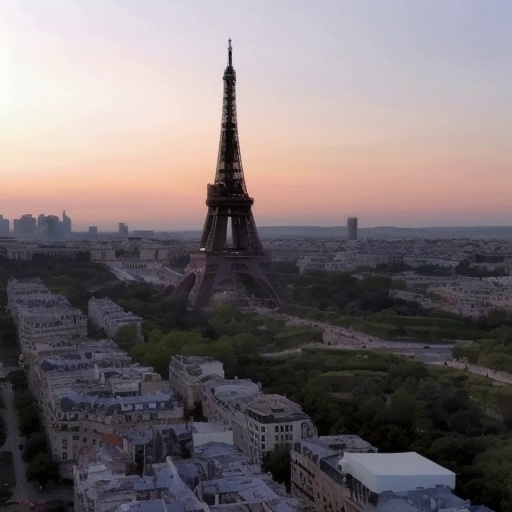}}\hspace{\rsWidth}  
        \subfloat{\includegraphics[width = 0.114\linewidth]{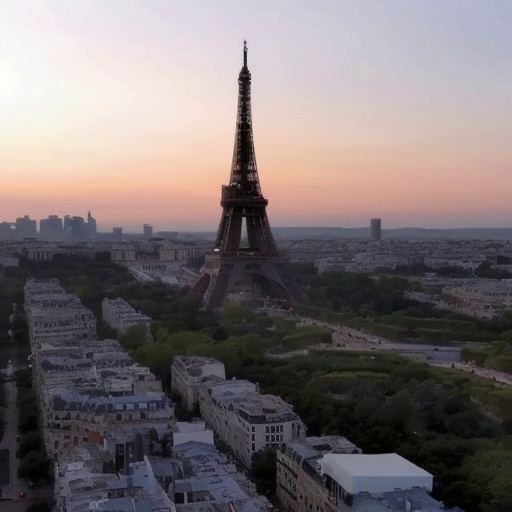}}\hspace{\rsWidth}
        \vspace{-0.1cm}
        \addtocounter{subfigure}{-8}
        \caption{Shorten Eiffel tower}
        \label{subfig:eiffel}
    \end{subfigure}

    \vspace{\rsHeight}
    \begin{subfigure}{\linewidth}
        \subfloat{\includegraphics[width = 0.114\linewidth]{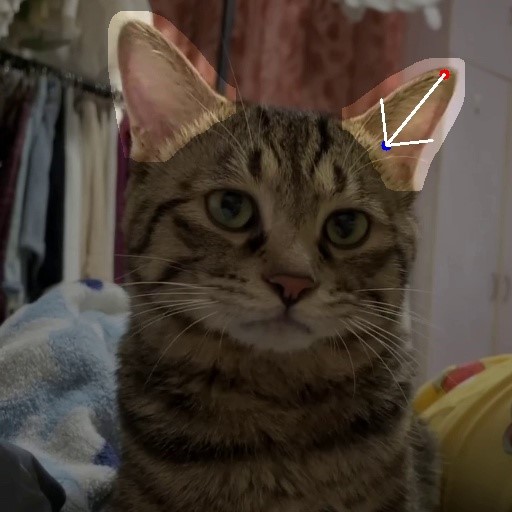}}\hspace{\rsWidth}
        \subfloat{\includegraphics[width = 0.114\linewidth]{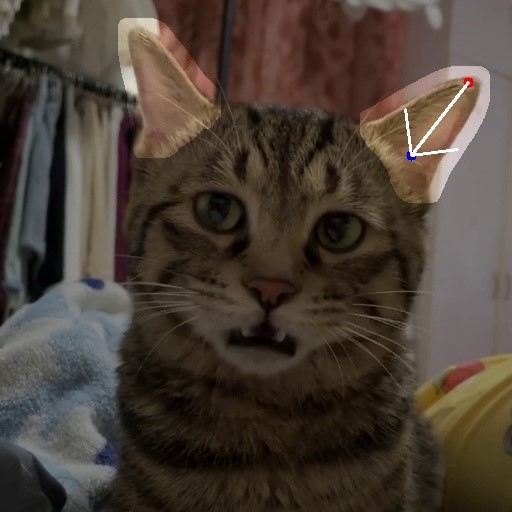}}\hspace{\rsWidth}
        \subfloat{\includegraphics[width = 0.114\linewidth]{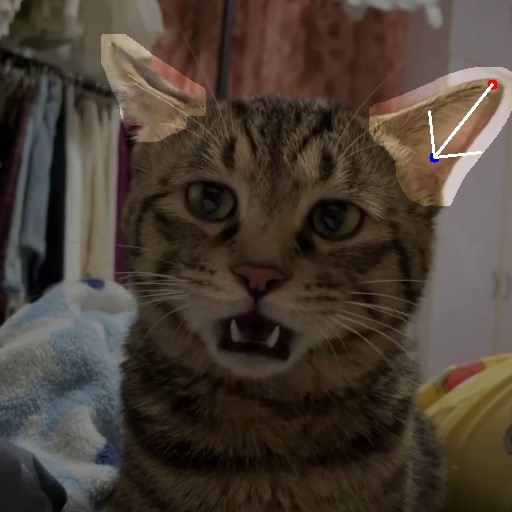}}\hspace{\rsWidth}
        \subfloat{\includegraphics[width = 0.114\linewidth]{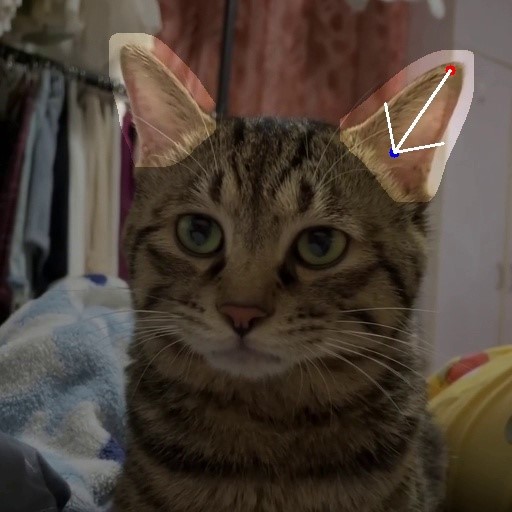}}\hspace{\rsmWidth}  
        \subfloat{\includegraphics[width = 0.114\linewidth]{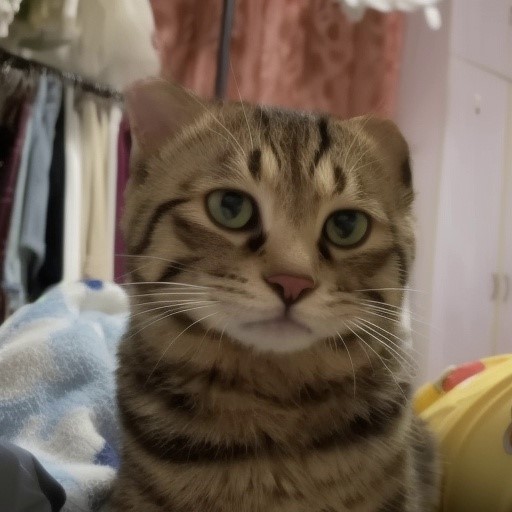}}\hspace{\rsWidth}
        \subfloat{\includegraphics[width = 0.114\linewidth]{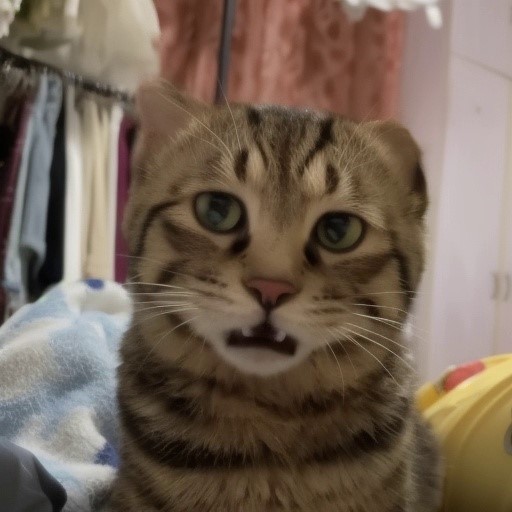}}\hspace{\rsWidth}
        \subfloat{\includegraphics[width = 0.114\linewidth]{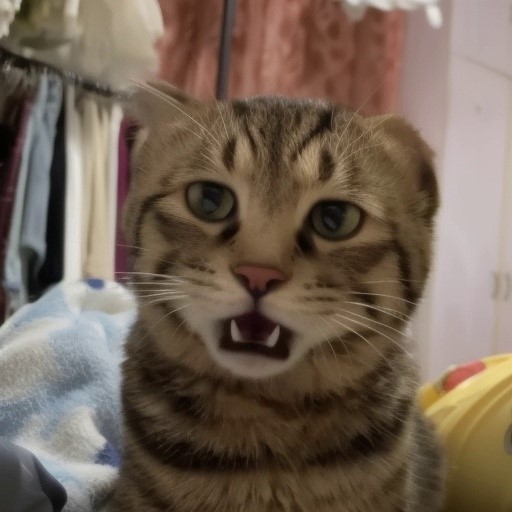}}\hspace{\rsWidth}  
        \subfloat{\includegraphics[width = 0.114\linewidth]{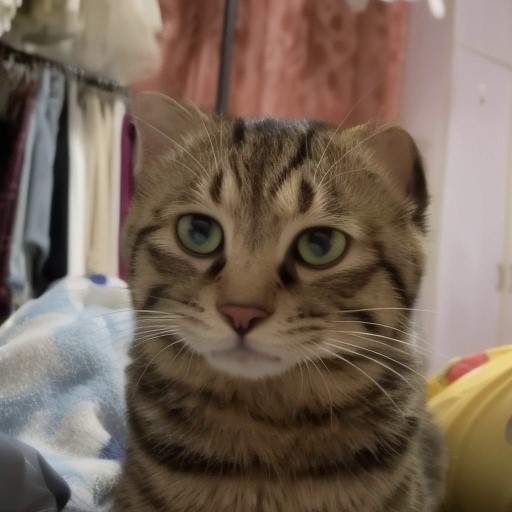}}\hspace{\rsWidth}
        \vspace{-0.1cm}
        \addtocounter{subfigure}{-8}
        \caption{Shorten the ears}
        \label{subfig:cat_ear}
    \end{subfigure}

    \vspace{\rsHeight}
    \begin{subfigure}{\linewidth}
        \subfloat{\includegraphics[width = 0.114\linewidth]{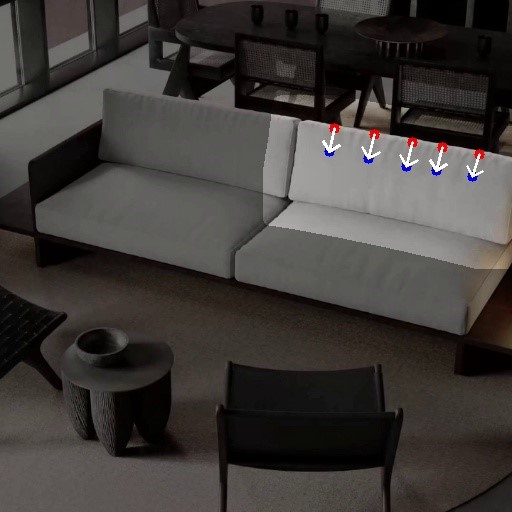}}\hspace{\rsWidth}
        \subfloat{\includegraphics[width = 0.114\linewidth]{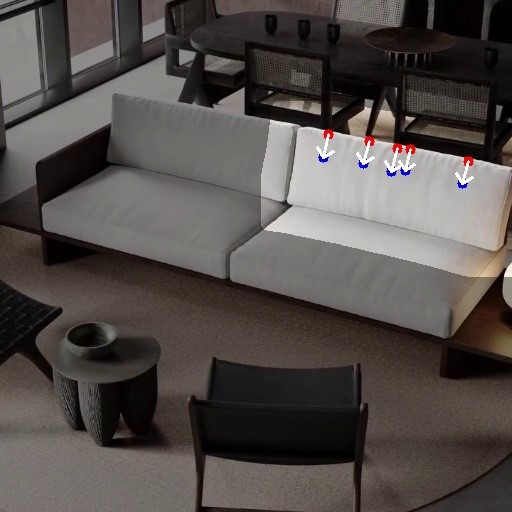}}\hspace{\rsWidth}
        \subfloat{\includegraphics[width = 0.114\linewidth]{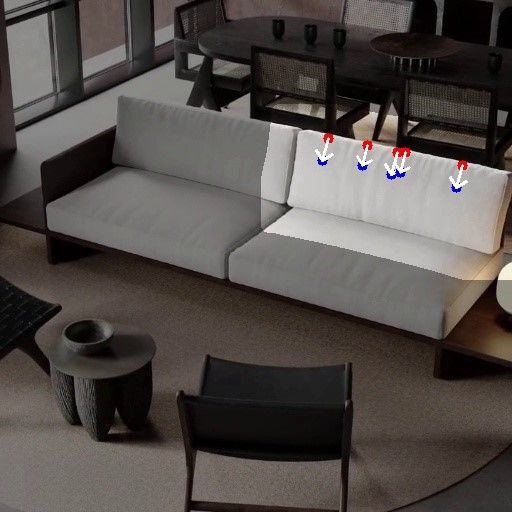}}\hspace{\rsWidth}
        \subfloat{\includegraphics[width = 0.114\linewidth]{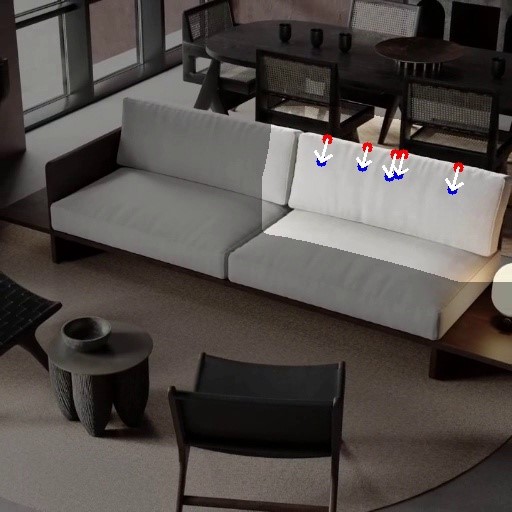}}\hspace{\rsmWidth}  
        \subfloat{\includegraphics[width = 0.114\linewidth]{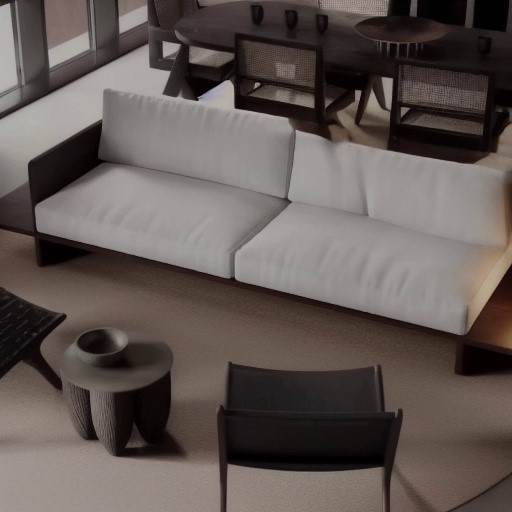}}\hspace{\rsWidth}
        \subfloat{\includegraphics[width = 0.114\linewidth]{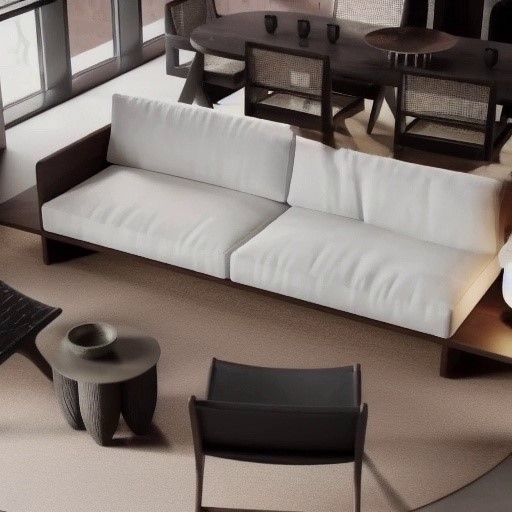}}\hspace{\rsWidth}
        \subfloat{\includegraphics[width = 0.114\linewidth]{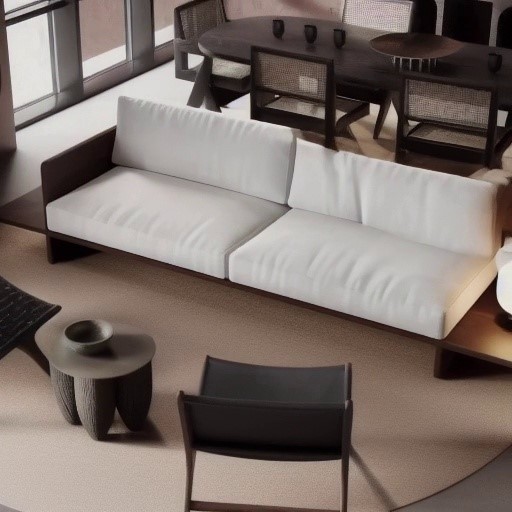}}\hspace{\rsWidth}  
        \subfloat{\includegraphics[width = 0.114\linewidth]{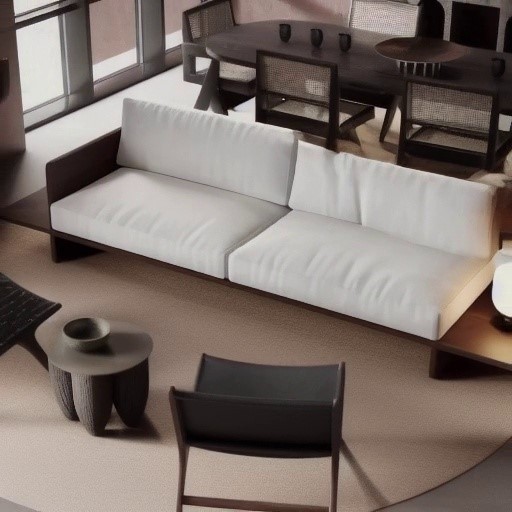}}\hspace{\rsWidth}
        \vspace{-0.1cm}
        \addtocounter{subfigure}{-8}
        \caption{Squeeze sofa}
        \label{subfig:sofa}
    \end{subfigure}

    \vspace{\rsHeight}
    \begin{subfigure}{\linewidth}
        \subfloat{\includegraphics[width = 0.114\linewidth]{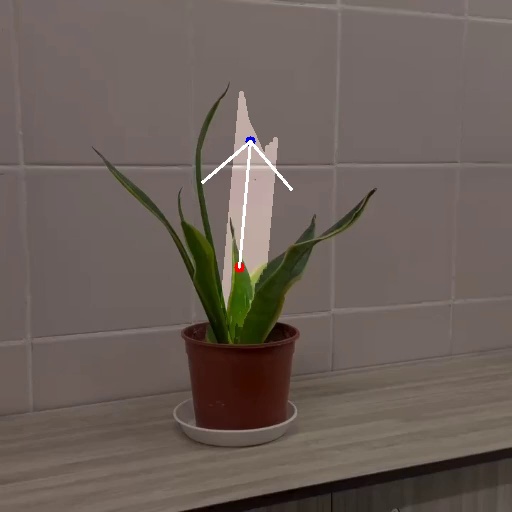}}\hspace{\rsWidth}
        \subfloat{\includegraphics[width = 0.114\linewidth]{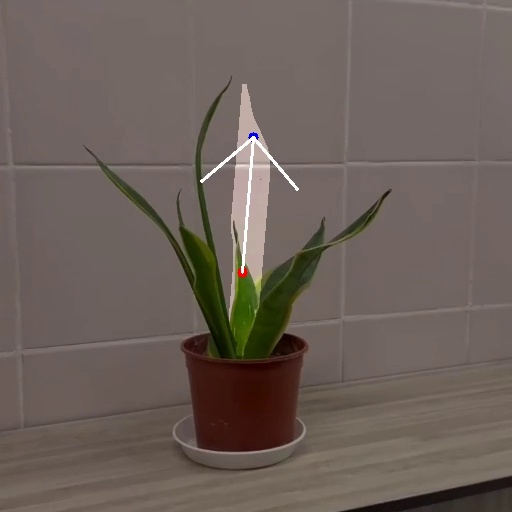}}\hspace{\rsWidth}
        \subfloat{\includegraphics[width = 0.114\linewidth]{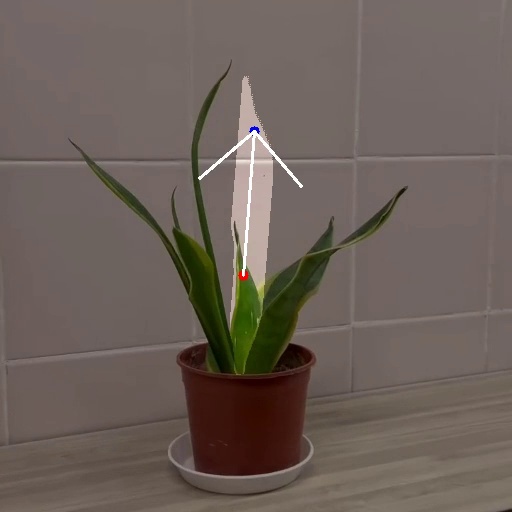}}\hspace{\rsWidth}
        \subfloat{\includegraphics[width = 0.114\linewidth]{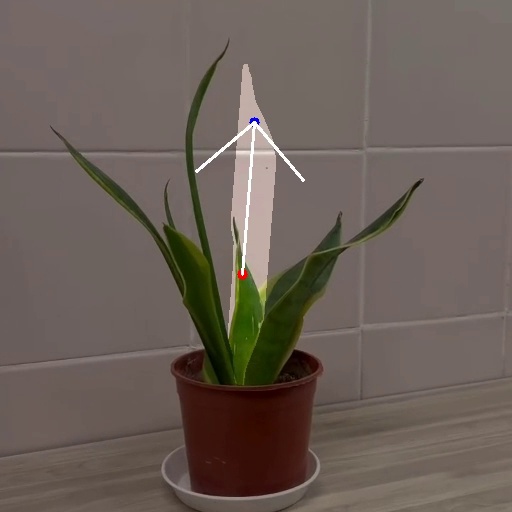}}\hspace{\rsmWidth}  
        \subfloat{\includegraphics[width = 0.114\linewidth]{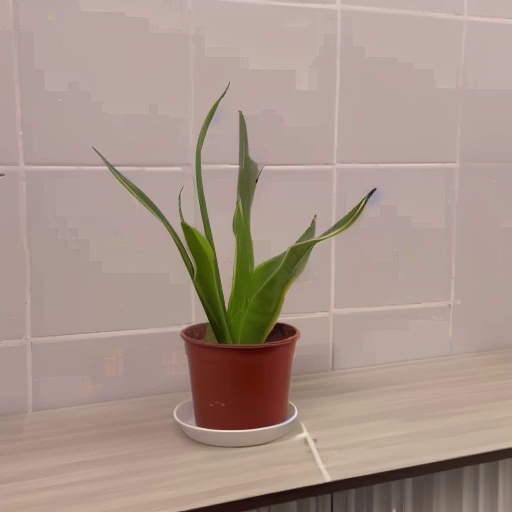}}\hspace{\rsWidth}
        \subfloat{\includegraphics[width = 0.114\linewidth]{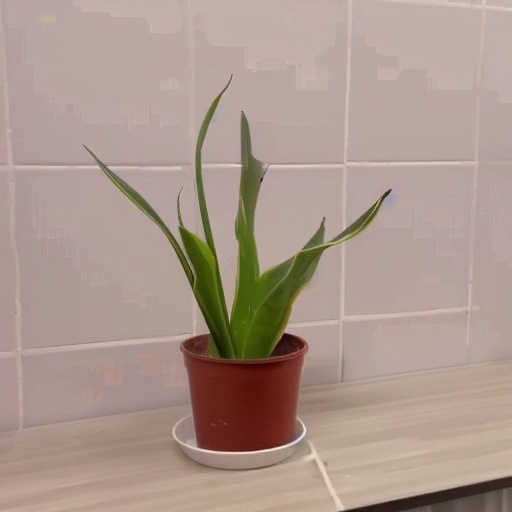}}\hspace{\rsWidth}
        \subfloat{\includegraphics[width = 0.114\linewidth]{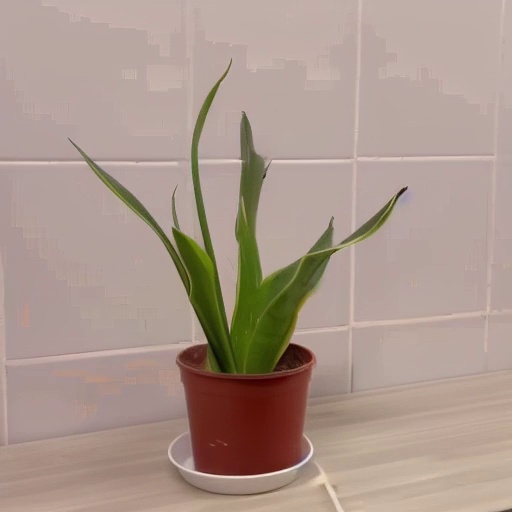}}\hspace{\rsWidth}  
        \subfloat{\includegraphics[width = 0.114\linewidth]{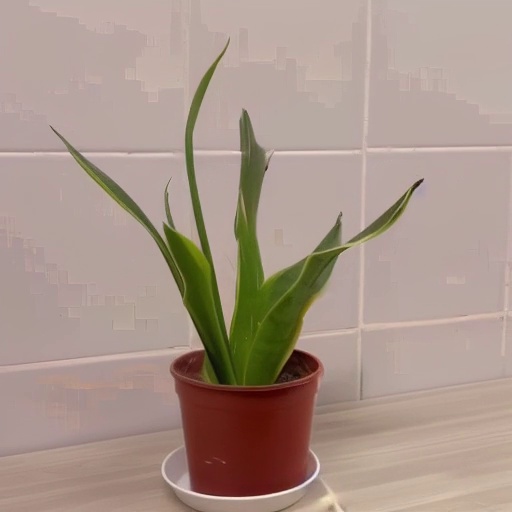}}\hspace{\rsWidth}
        \vspace{-0.1cm}
        \addtocounter{subfigure}{-8}
        \caption{Lengthen the plant}
        \label{subfig:plants}
    \end{subfigure}

    \vspace{\rsHeight}
    \begin{subfigure}{\linewidth}
        \subfloat{\includegraphics[width = 0.114\linewidth]{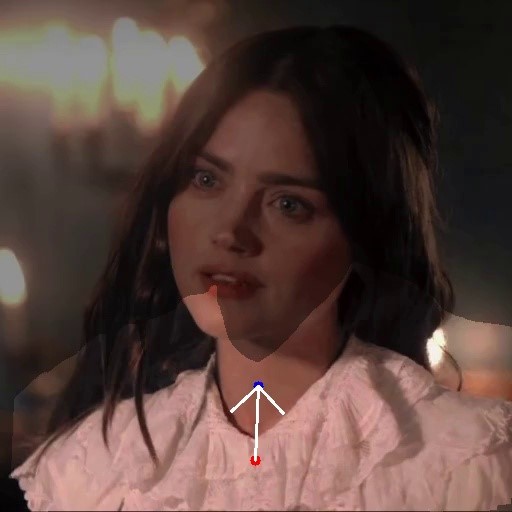}}\hspace{\rsWidth}
        \subfloat{\includegraphics[width = 0.114\linewidth]{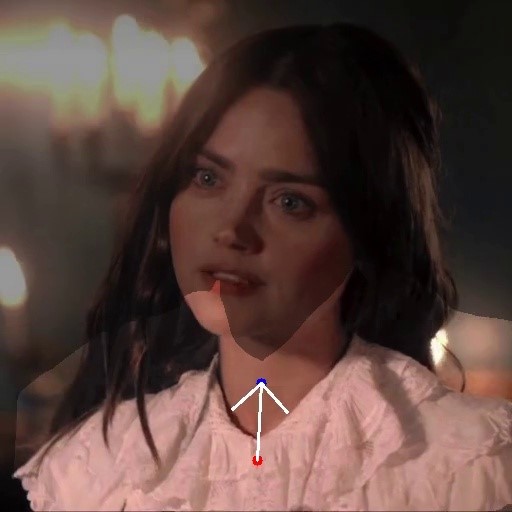}}\hspace{\rsWidth}
        \subfloat{\includegraphics[width = 0.114\linewidth]{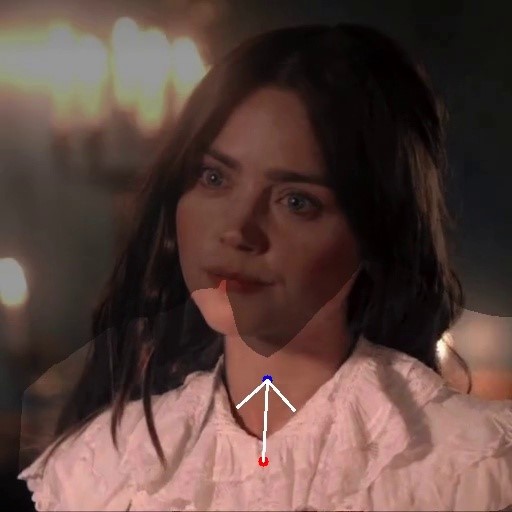}}\hspace{\rsWidth}
        \subfloat{\includegraphics[width = 0.114\linewidth]{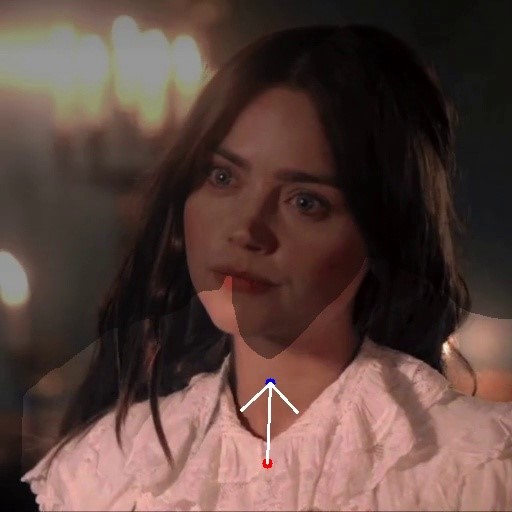}}\hspace{\rsmWidth}  
        \subfloat{\includegraphics[width = 0.114\linewidth]{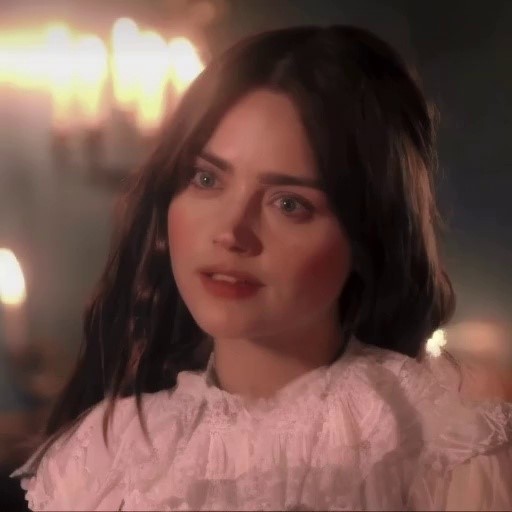}}\hspace{\rsWidth}
        \subfloat{\includegraphics[width = 0.114\linewidth]{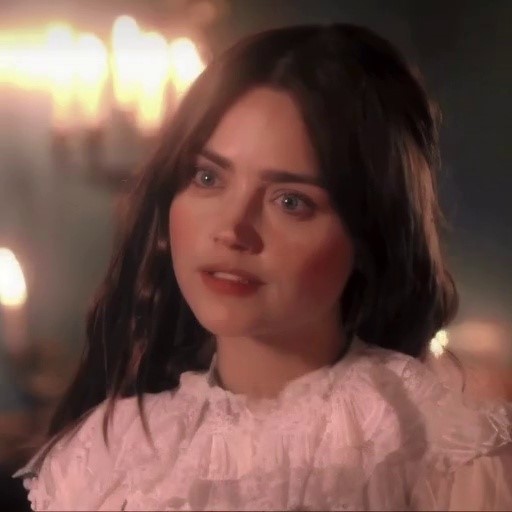}}\hspace{\rsWidth}
        \subfloat{\includegraphics[width = 0.114\linewidth]{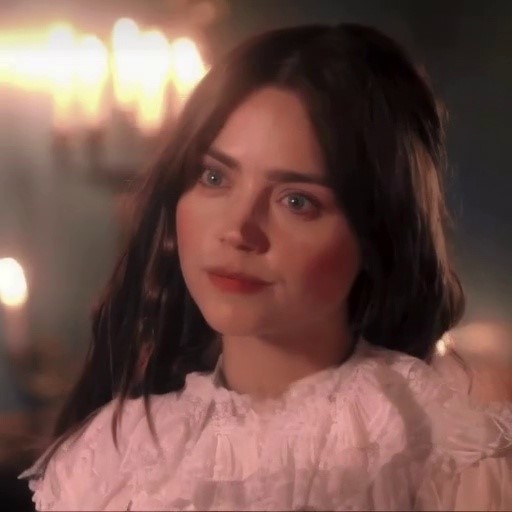}}\hspace{\rsWidth}  
        \subfloat{\includegraphics[width = 0.114\linewidth]{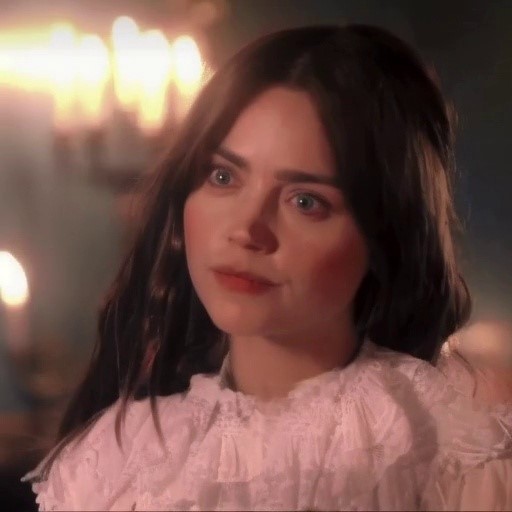}}\hspace{\rsWidth}
        \vspace{-0.1cm}
        \addtocounter{subfigure}{-8}
        \caption{Close neckline}
        \label{subfig:vic}
    \end{subfigure}

    \vspace{\rsHeight}
    \begin{subfigure}{\linewidth}
        \subfloat{\includegraphics[width = 0.114\linewidth]{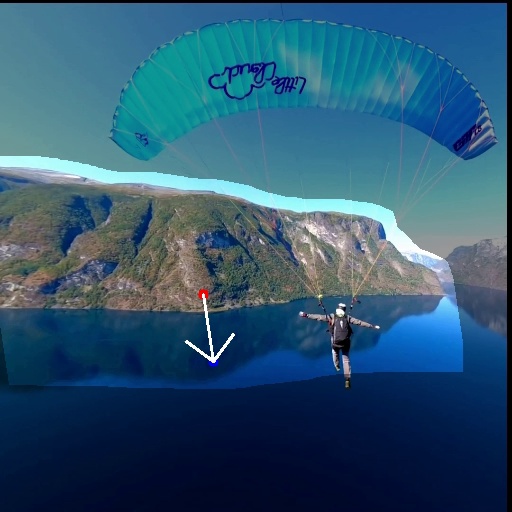}}\hspace{\rsWidth}
        \subfloat{\includegraphics[width = 0.114\linewidth]{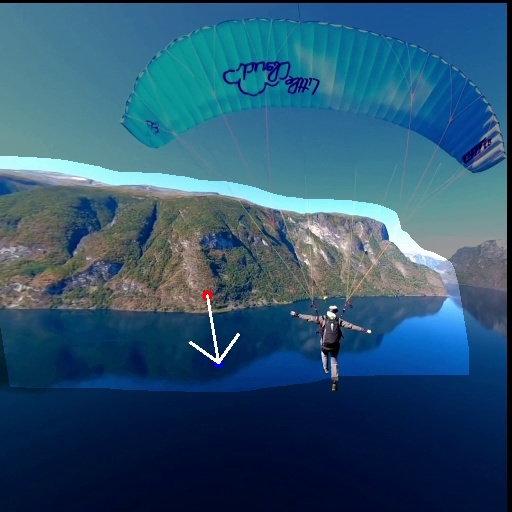}}\hspace{\rsWidth}
        \subfloat{\includegraphics[width = 0.114\linewidth]{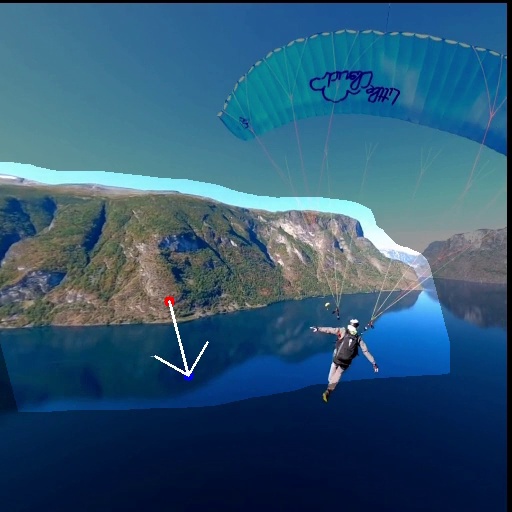}}\hspace{\rsWidth}
        \subfloat{\includegraphics[width = 0.114\linewidth]{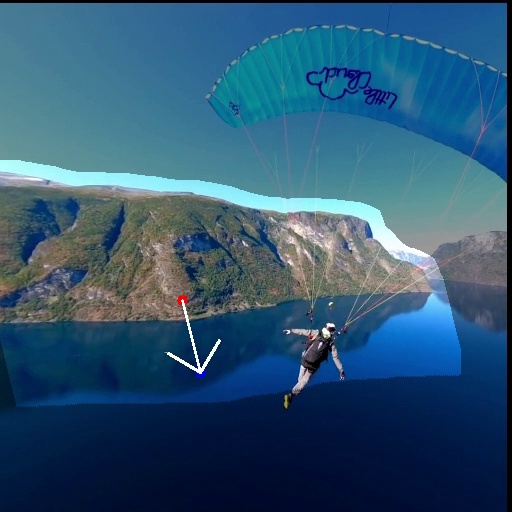}}\hspace{\rsmWidth}  
        \subfloat{\includegraphics[width = 0.114\linewidth]{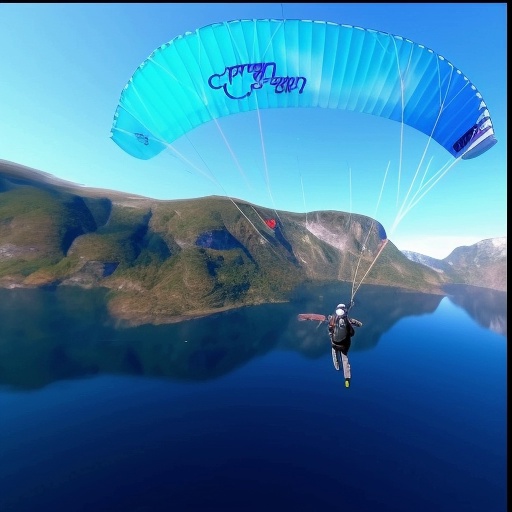}}\hspace{\rsWidth}
        \subfloat{\includegraphics[width = 0.114\linewidth]{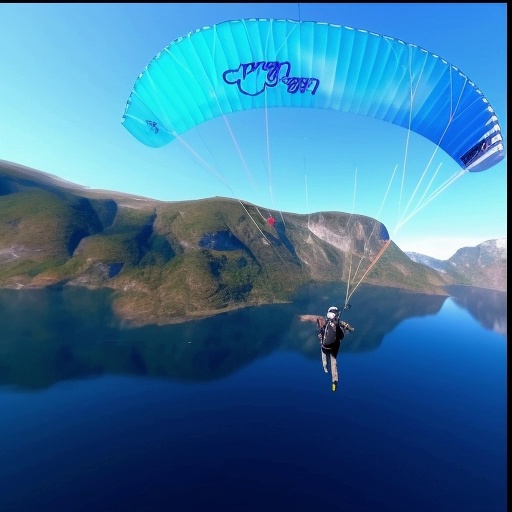}}\hspace{\rsWidth}
        \subfloat{\includegraphics[width = 0.114\linewidth]{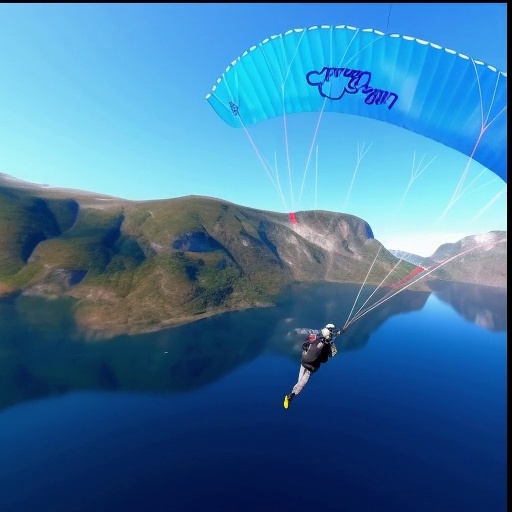}}\hspace{\rsWidth}  
        \subfloat{\includegraphics[width = 0.114\linewidth]{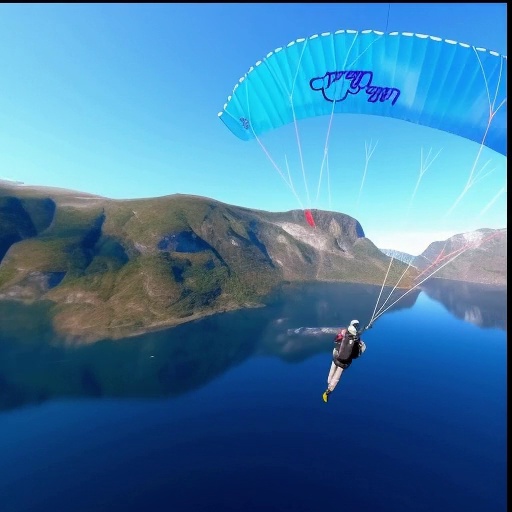}}\hspace{\rsWidth}
        \vspace{-0.1cm}
        \addtocounter{subfigure}{-8}
        \caption{Larger the island in front of skydiver}
        \label{subfig:scenery_0}
    \end{subfigure}

    \vspace{\rsHeight}
    \begin{subfigure}{\linewidth}
        \subfloat{\includegraphics[width = 0.114\linewidth]{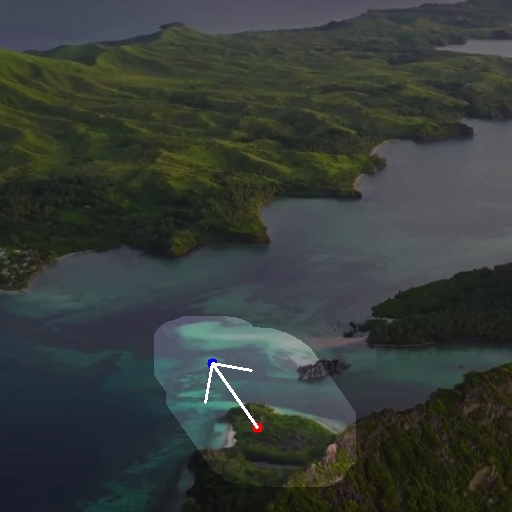}}\hspace{\rsWidth}
        \subfloat{\includegraphics[width = 0.114\linewidth]{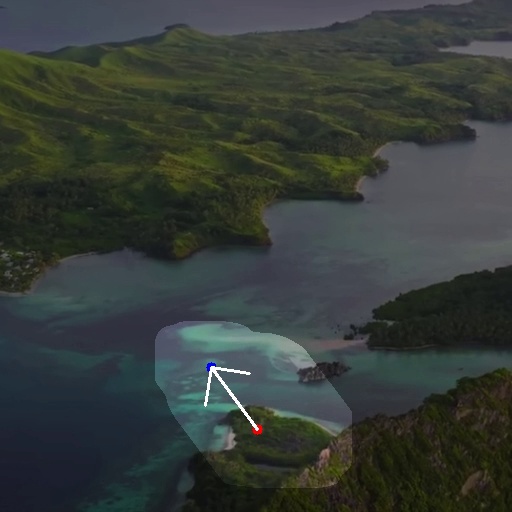}}\hspace{\rsWidth}
        \subfloat{\includegraphics[width = 0.114\linewidth]{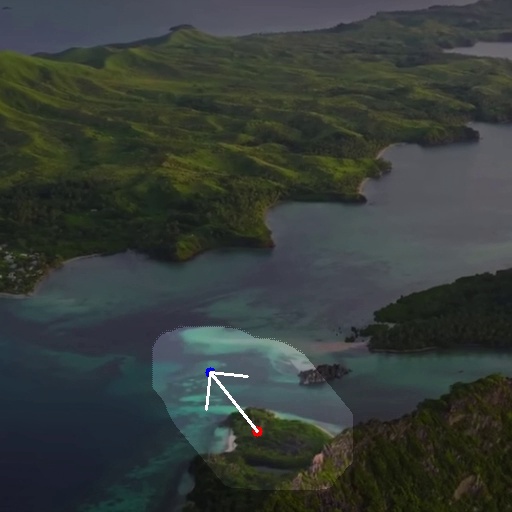}}\hspace{\rsWidth}
        \subfloat{\includegraphics[width = 0.114\linewidth]{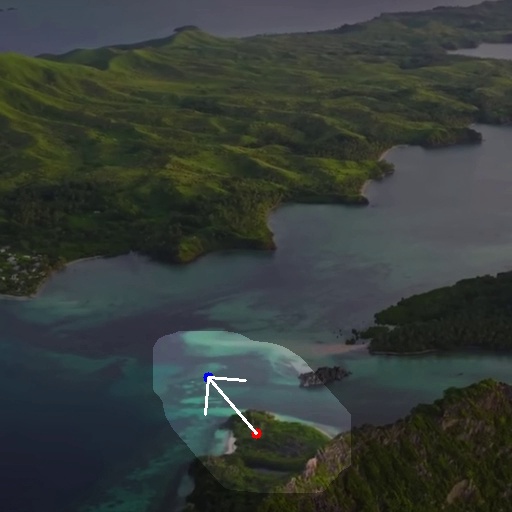}}\hspace{\rsmWidth}  
        \subfloat{\includegraphics[width = 0.114\linewidth]{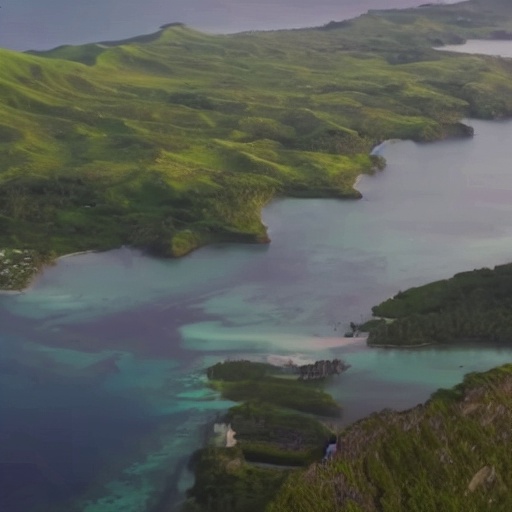}}\hspace{\rsWidth}
        \subfloat{\includegraphics[width = 0.114\linewidth]{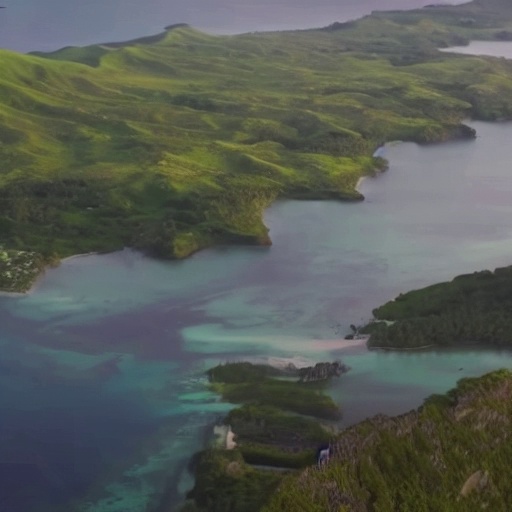}}\hspace{\rsWidth}
        \subfloat{\includegraphics[width = 0.114\linewidth]{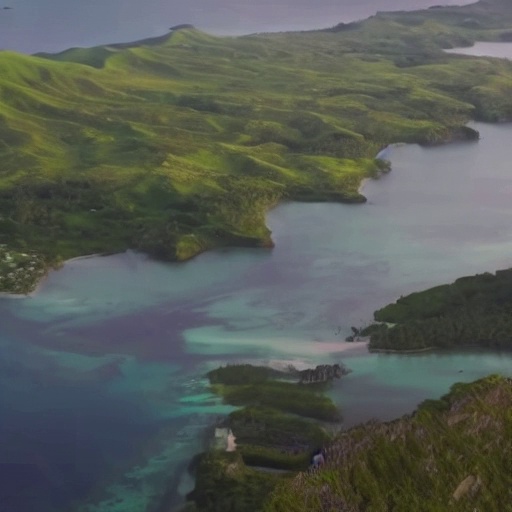}}\hspace{\rsWidth}  
        \subfloat{\includegraphics[width = 0.114\linewidth]{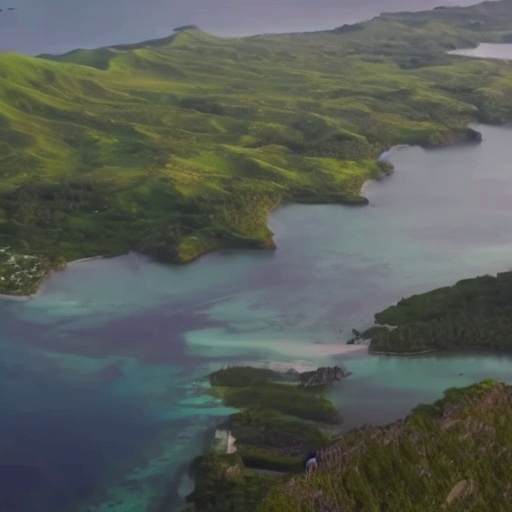}}\hspace{\rsWidth}
        \vspace{-0.1cm}
        \addtocounter{subfigure}{-8}
        \caption{Larger the continent}
        \label{subfig:scenery_1}
    \end{subfigure}

    \vspace{\rsHeight}
    \begin{subfigure}{\linewidth}
        \subfloat{\includegraphics[width = 0.114\linewidth]{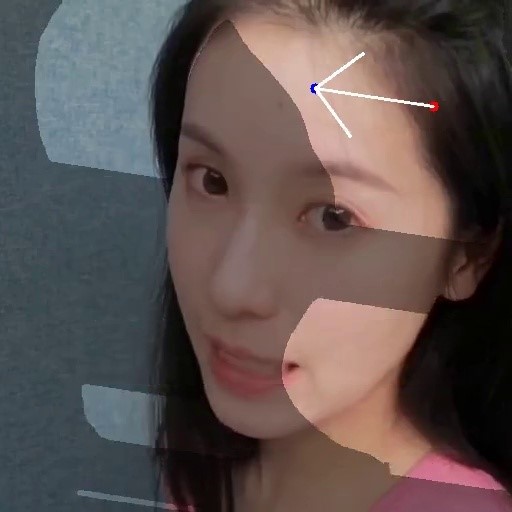}}\hspace{\rsWidth}
        \subfloat{\includegraphics[width = 0.114\linewidth]{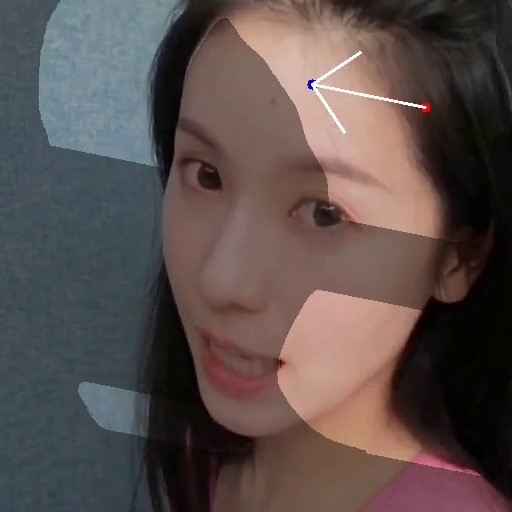}}\hspace{\rsWidth}
        \subfloat{\includegraphics[width = 0.114\linewidth]{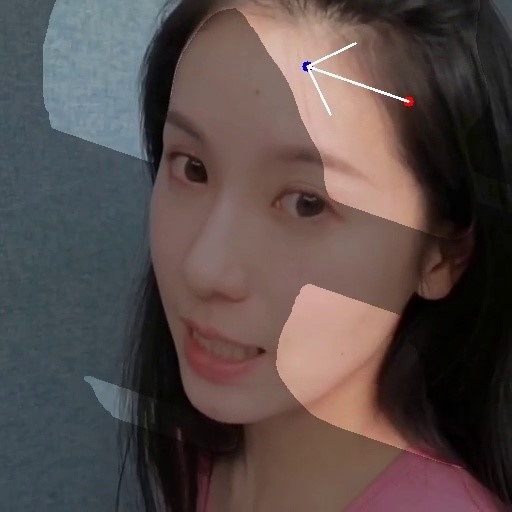}}\hspace{\rsWidth}
        \subfloat{\includegraphics[width = 0.114\linewidth]{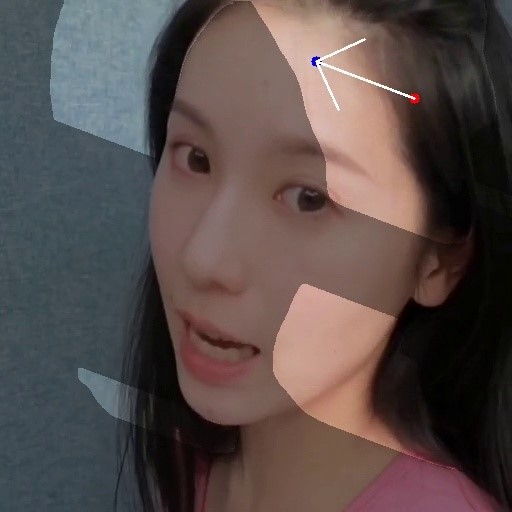}}\hspace{\rsmWidth}  
        \subfloat{\includegraphics[width = 0.114\linewidth]{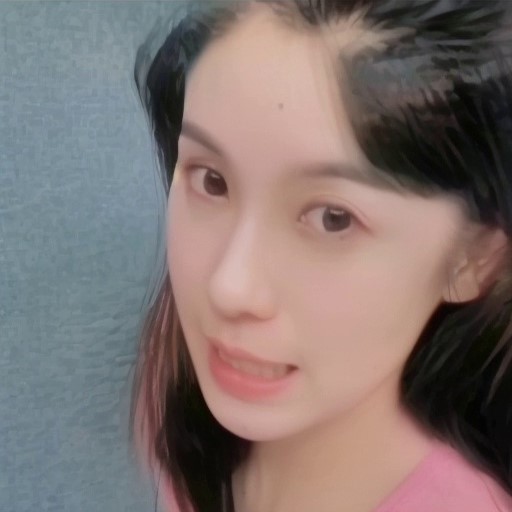}}\hspace{\rsWidth}
        \subfloat{\includegraphics[width = 0.114\linewidth]{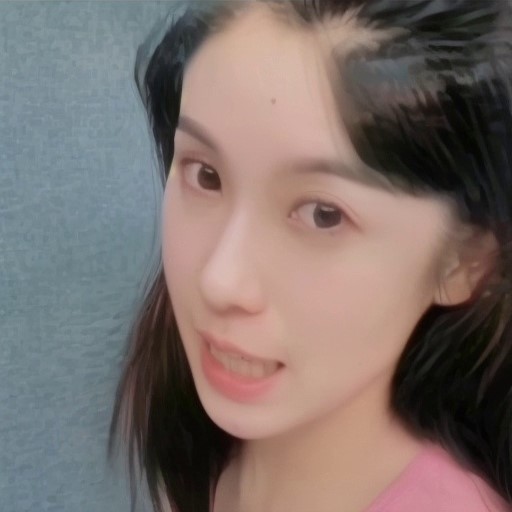}}\hspace{\rsWidth}
        \subfloat{\includegraphics[width = 0.114\linewidth]{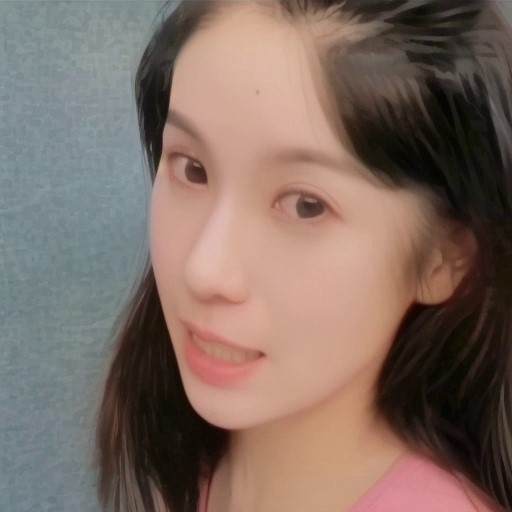}}\hspace{\rsWidth}  
        \subfloat{\includegraphics[width = 0.114\linewidth]{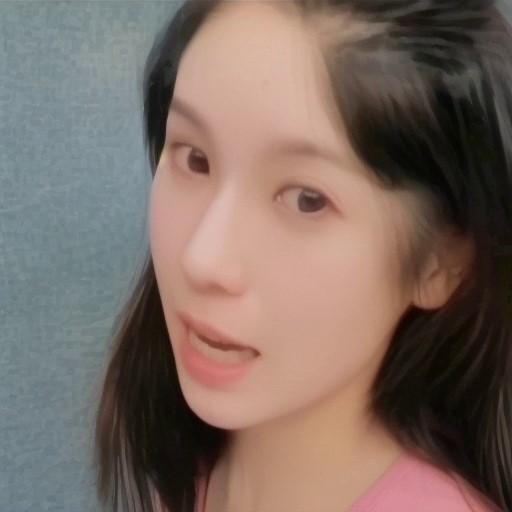}}\hspace{\rsWidth}
        \vspace{-0.1cm}
        \addtocounter{subfigure}{-8}
        \caption{Generate band}
        \label{subfig:band}
    \end{subfigure}

    \vspace{\rsHeight}
    \begin{subfigure}{\linewidth}
        \subfloat{\includegraphics[width = 0.114\linewidth]{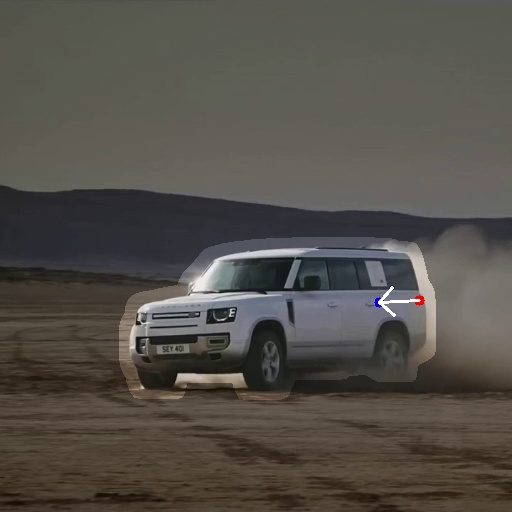}}\hspace{\rsWidth}
        \subfloat{\includegraphics[width = 0.114\linewidth]{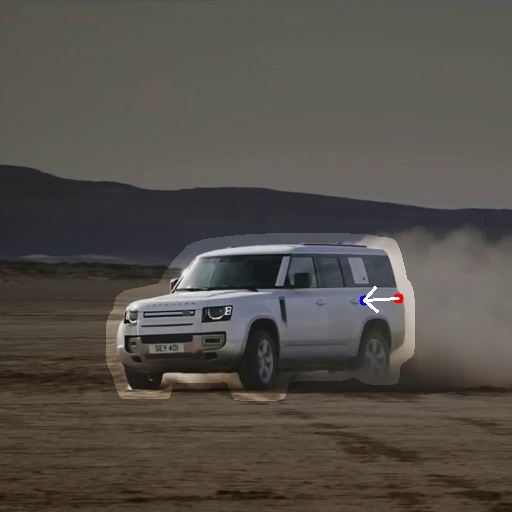}}\hspace{\rsWidth}
        \subfloat{\includegraphics[width = 0.114\linewidth]{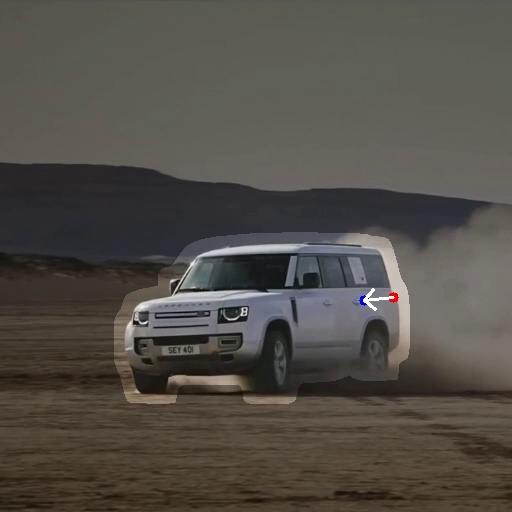}}\hspace{\rsWidth}
        \subfloat{\includegraphics[width = 0.114\linewidth]{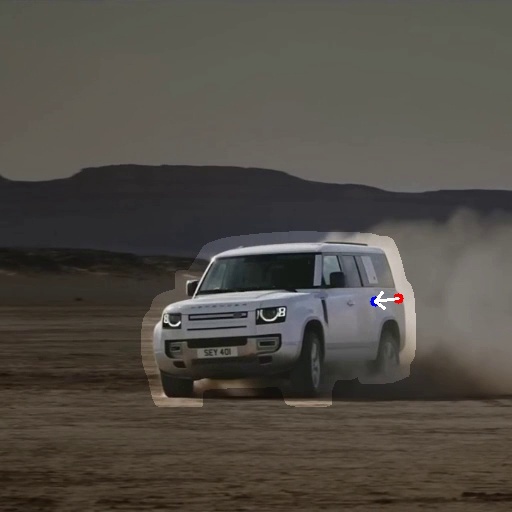}}\hspace{\rsmWidth}  
        \subfloat{\includegraphics[width = 0.114\linewidth]{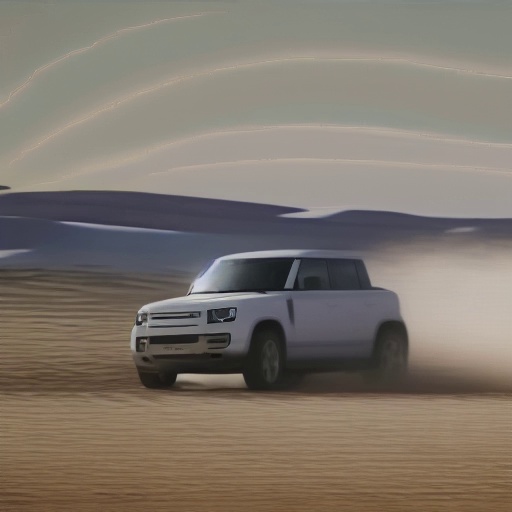}}\hspace{\rsWidth}
        \subfloat{\includegraphics[width = 0.114\linewidth]{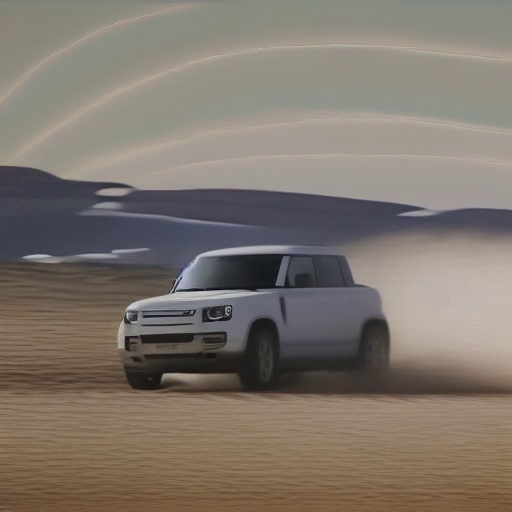}}\hspace{\rsWidth}
        \subfloat{\includegraphics[width = 0.114\linewidth]{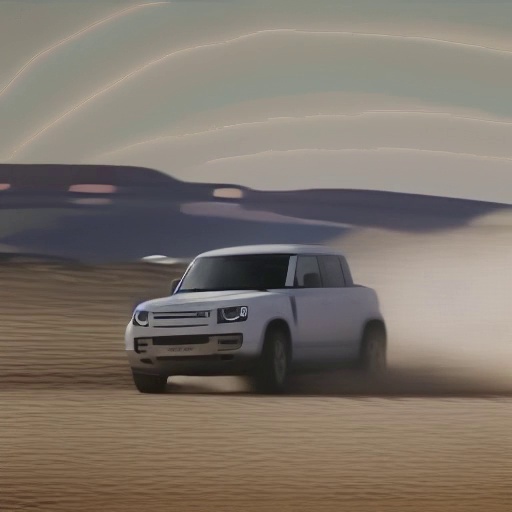}}\hspace{\rsWidth}  
        \subfloat{\includegraphics[width = 0.114\linewidth]{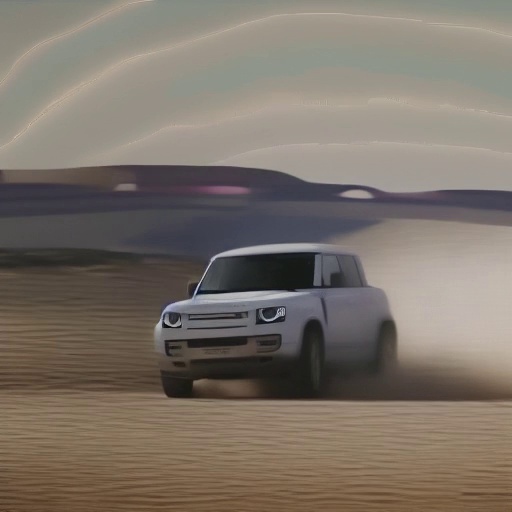}}\hspace{\rsWidth}
        \vspace{-0.1cm}
        \addtocounter{subfigure}{-8}
        \caption{Shorten the back of SUV}
        \label{subfig:suv}
    \end{subfigure}
    
    \vspace{\rsHeight}
    \begin{subfigure}{\linewidth}
        \subfloat{\includegraphics[width = 0.114\linewidth]{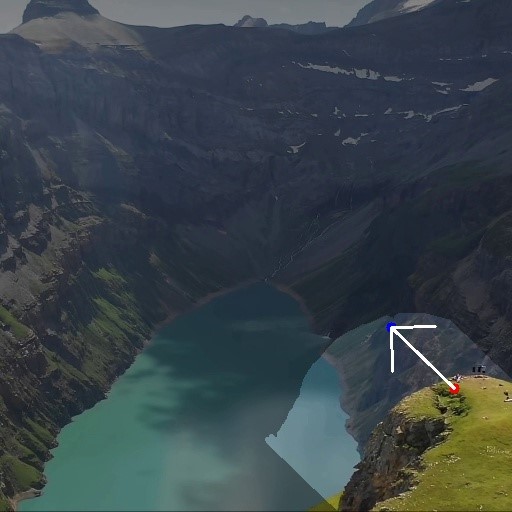}}\hspace{\rsWidth}
        \subfloat{\includegraphics[width = 0.114\linewidth]{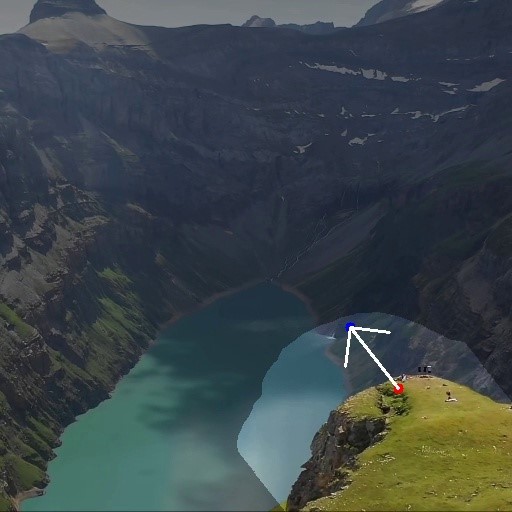}}\hspace{\rsWidth}
        \subfloat{\includegraphics[width = 0.114\linewidth]{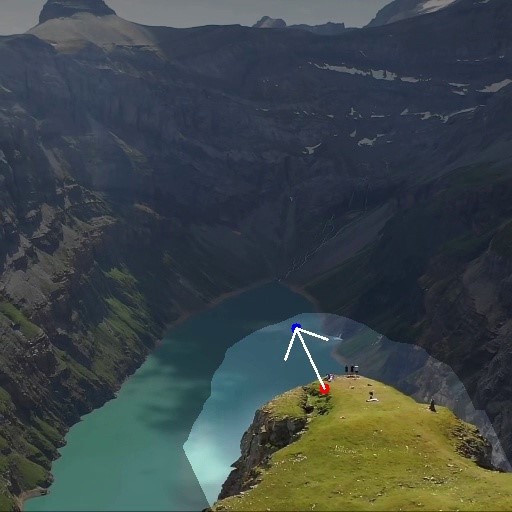}}\hspace{\rsWidth}
        \subfloat{\includegraphics[width = 0.114\linewidth]{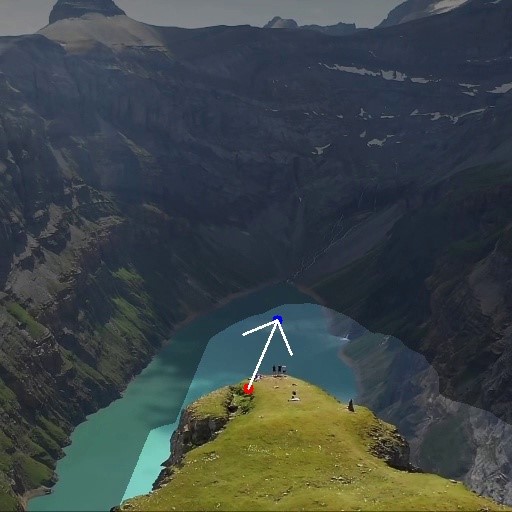}}\hspace{\rsmWidth}  
        \subfloat{\includegraphics[width = 0.114\linewidth]{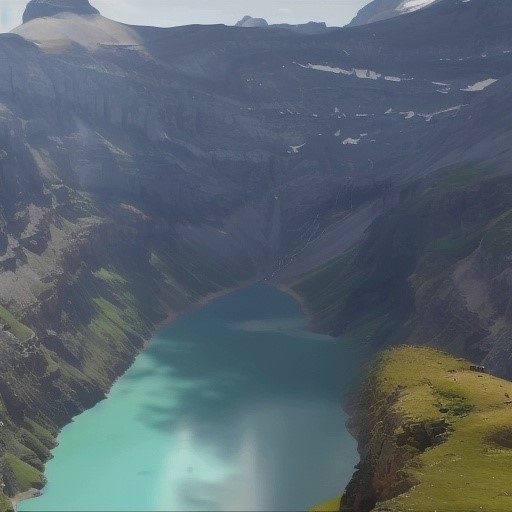}}\hspace{\rsWidth}
        \subfloat{\includegraphics[width = 0.114\linewidth]{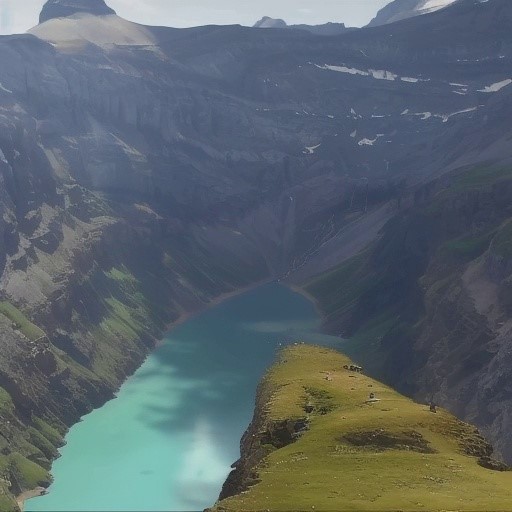}}\hspace{\rsWidth}
        \subfloat{\includegraphics[width = 0.114\linewidth]{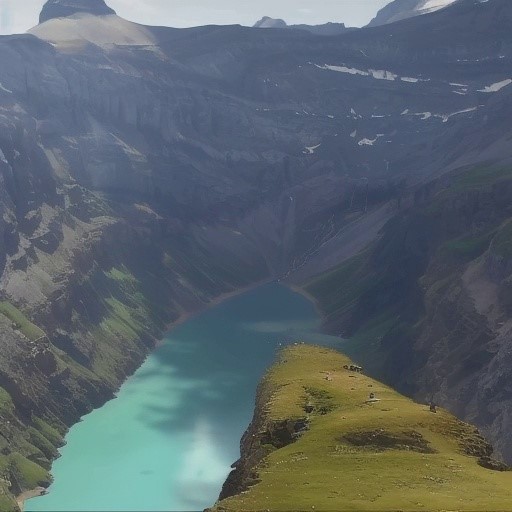}}\hspace{\rsWidth}  
        \subfloat{\includegraphics[width = 0.114\linewidth]{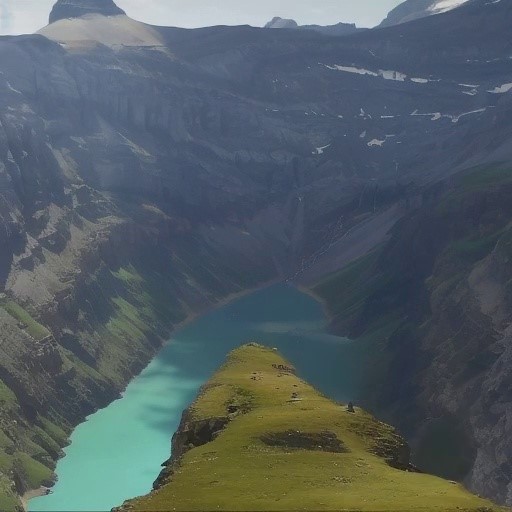}}\hspace{\rsWidth}
        \vspace{-0.1cm}
        \addtocounter{subfigure}{-8}
        \caption{Cliff extension}
        \label{subfig:cliff}
    \end{subfigure}
    \caption{More results of DragVideo. Left four frames are propagated editing instructions. Right four frames are edited output.}
    \label{fig:suppResult}
\end{figure*}

\begin{figure*}[t]
    \centering
    \begin{subfigure}{0.47\linewidth}
        \centering
        \includegraphics[width=\linewidth]{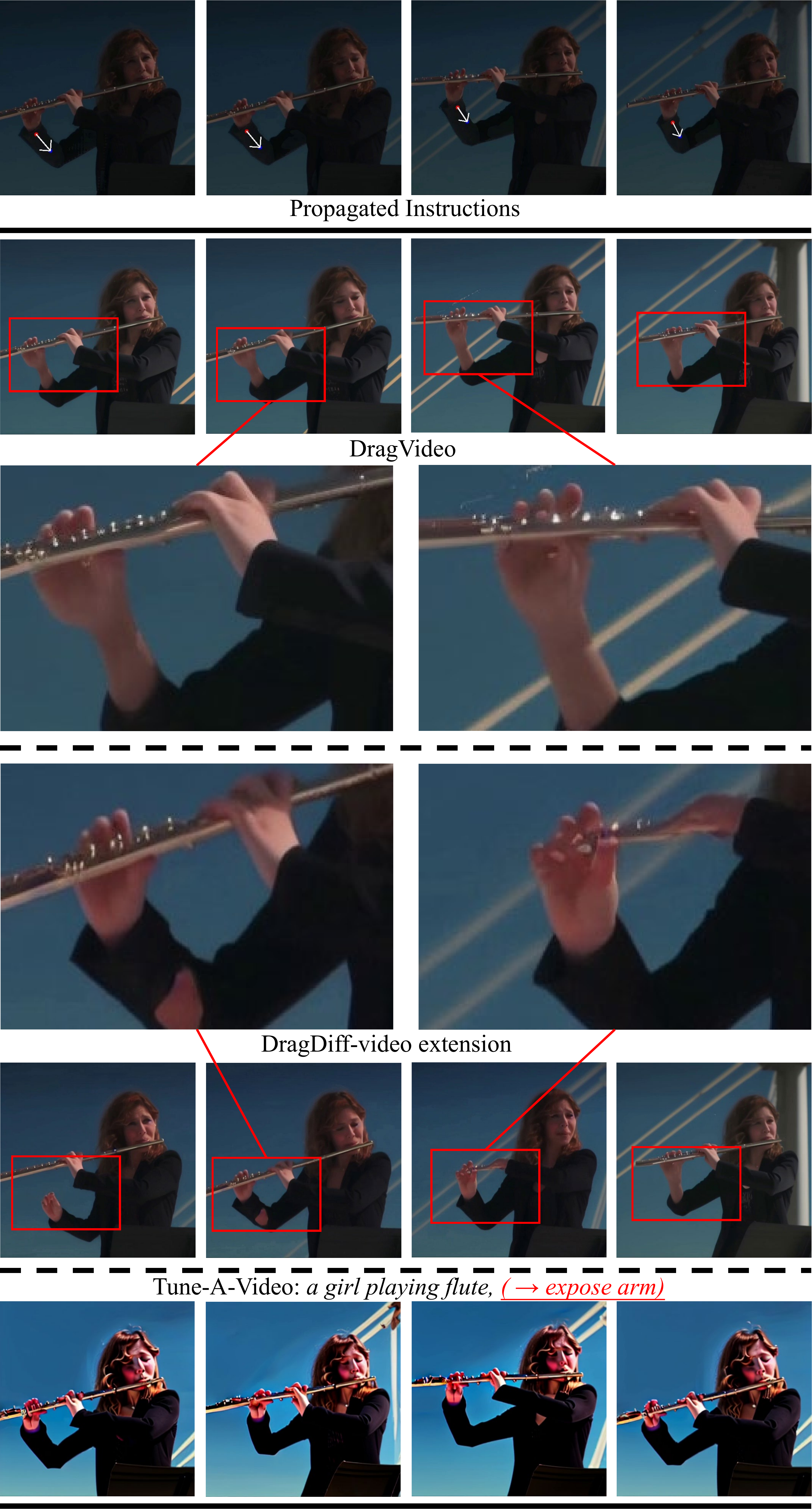}
        \caption{Remove sleeve of the suit}
        \label{suppBase:Paris}
    \end{subfigure}
    \begin{subfigure}{0.47\linewidth}
        \centering
        \includegraphics[width=\linewidth]{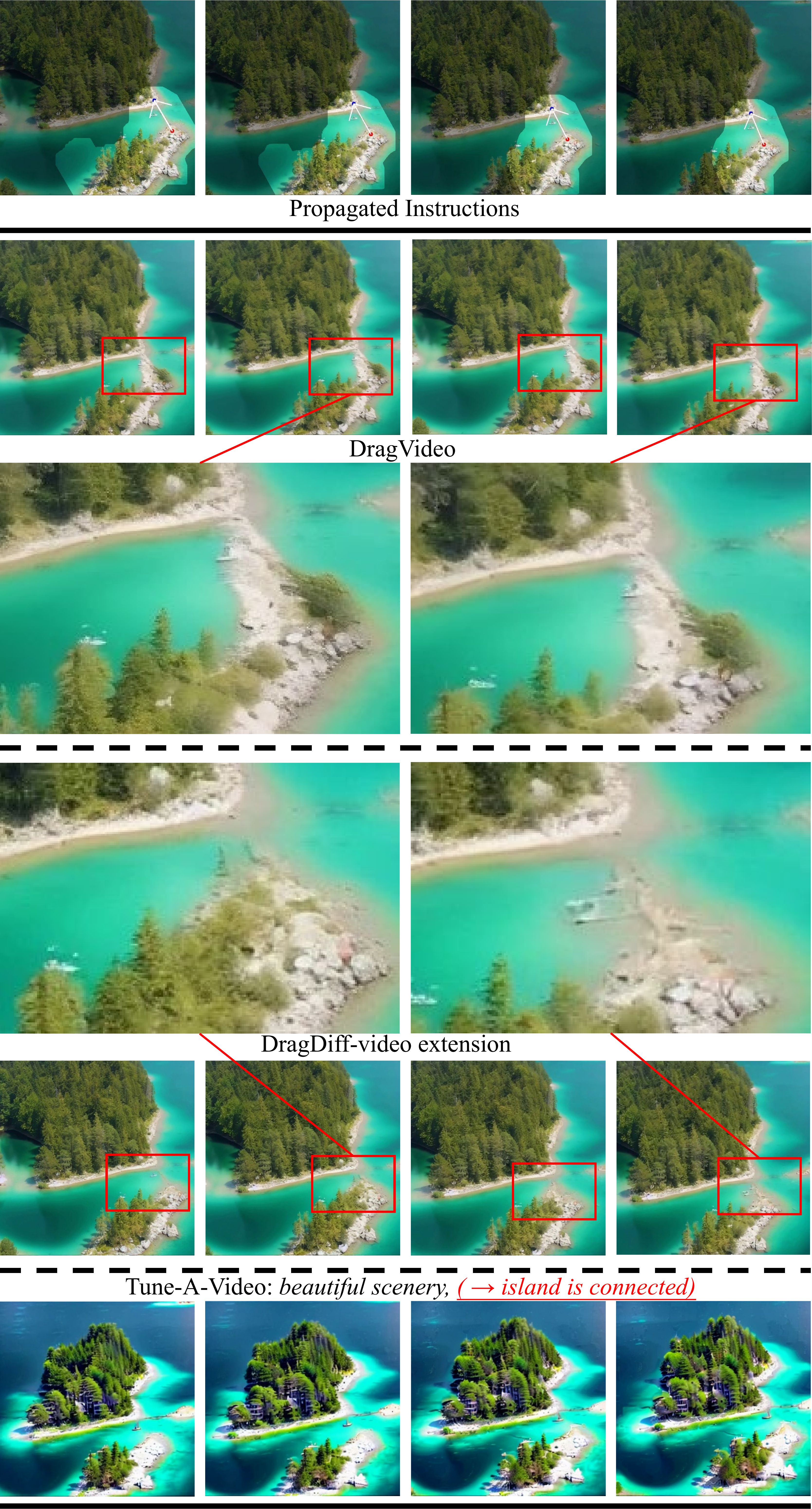}
        \caption{Connect island}
        \label{suppBase:Swiss_1}
    \end{subfigure}
    \caption{More comparison between two baselines and DragVideo. First row is original frames with propagated editing instructions. DragVideo editing (second row with zoom) achieves decent results. Directly extending DragDiff to video (third row with zoom) faces the temporal inconsistent problem. Using prompt-based editing Tune-A-Video (last row) can introduce unintended style alterations without achieving the goal.}
    \label{fig:suppBase}
\end{figure*}

In this section, we provide additional qualitative results that underscore the capability of DragVideo. The qualitative results can be referenced in Figure \ref{fig:suppResult}. Moreover, we offer more comparison examples between DragVideo and the frame-by-frame drag baseline; the comparisons are in Figure \ref{fig:suppBase}.

From Figure \ref{fig:suppResult}, it becomes evident that DragVideo consistently delivers accurate editing across a wide range of scenarios. Our framework exhibits a remarkable capability not merely to shift pixels from one location to another, but to exploit the wealth of information in the diffusion model. This results in the generation of fitting and credible content in alignment with the drag instructions, a characteristic evident in almost all qualitative results. When comparing video extension of DragDiffusion ~\cite{dragdiff} with DragVideo in Figure \ref{fig:suppBase}, the latter displays superior spatial-temporal consistency throughout the video. The DragDiffusion extension approach often results in varied characteristics in each frame, while DragVideo maintains high levels of consistency.

To further showcase DragVideo's efficacy and to provide a broader comparison with the baseline, we have compiled a video juxtaposing all DragVideo and baseline outcomes. Please refer to the $supp\_videos.mp4$ file in the supplementary materials folder.


\section{Additional Ablation Tests}
\begin{figure*}[t]
    \centering
    \captionsetup[subfloat]{labelformat=empty}
    \vspace{\abHeight}
    \subfloat[Propagated Instructions]{
        \subfloat{\includegraphics[width = 0.113\linewidth]{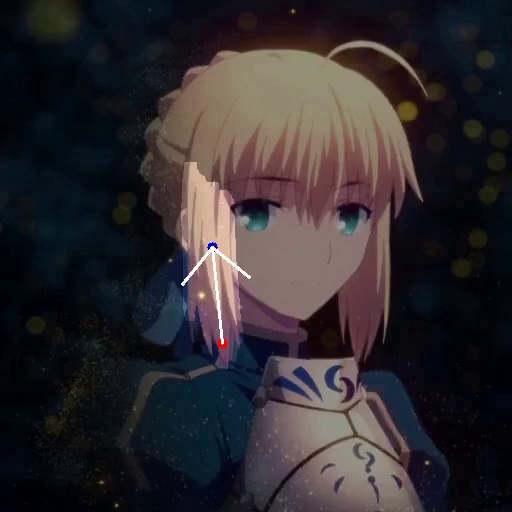}}\hspace{\abWidth}
        \subfloat{\includegraphics[width = 0.113\linewidth]{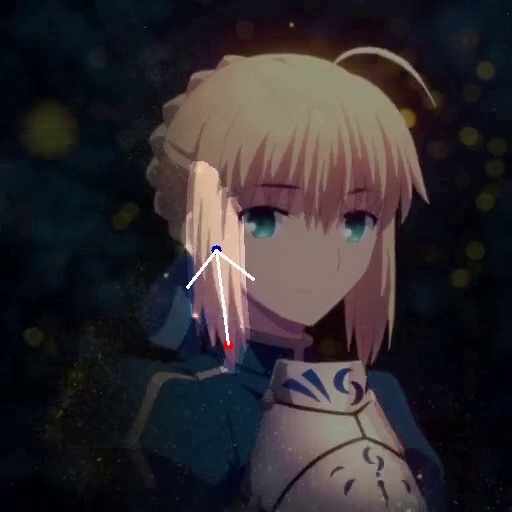}}\hspace{\abWidth}
        \subfloat{\includegraphics[width = 0.113\linewidth]{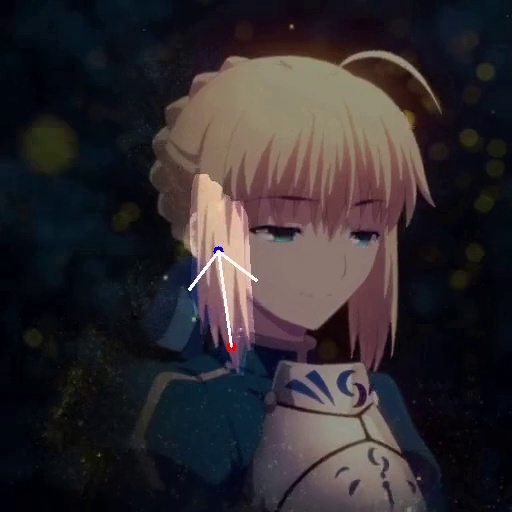}}\hspace{\abWidth}
        \subfloat{\includegraphics[width = 0.113\linewidth]{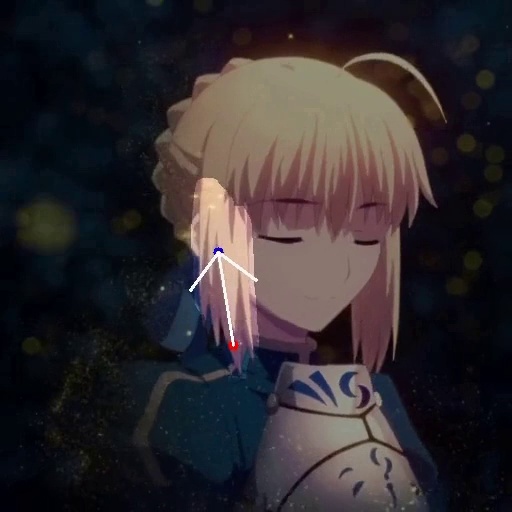}}\hspace{\abmWidth}
    }
    \subfloat[Edited Output]{
        \subfloat{\includegraphics[width = 0.113\linewidth]{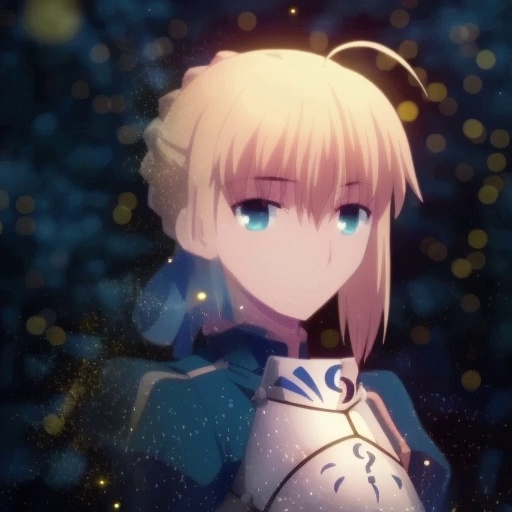}}\hspace{\abWidth}
        \subfloat{\includegraphics[width = 0.113\linewidth]{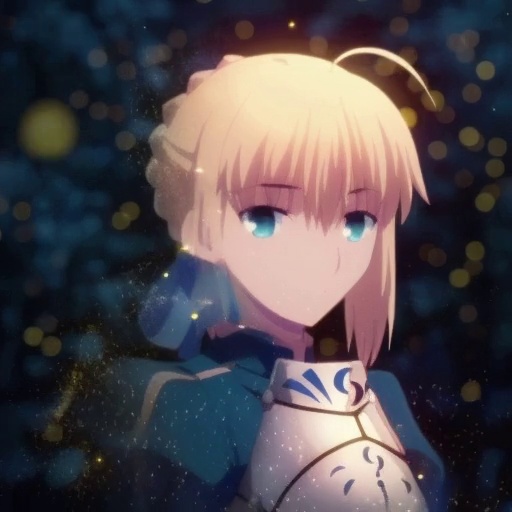}}\hspace{\abWidth}
        \subfloat{\includegraphics[width = 0.113\linewidth]{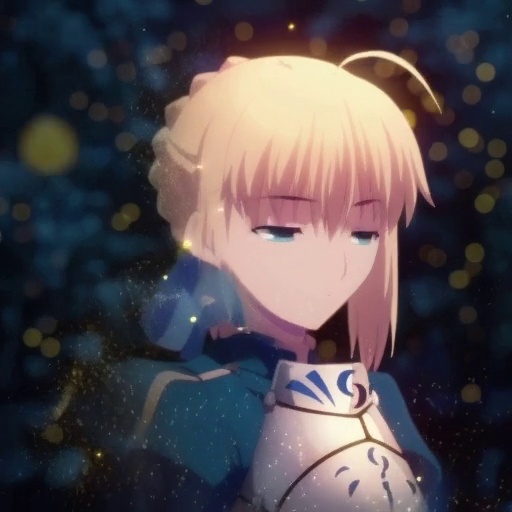}}\hspace{\abWidth}
        \subfloat{\includegraphics[width = 0.113\linewidth]{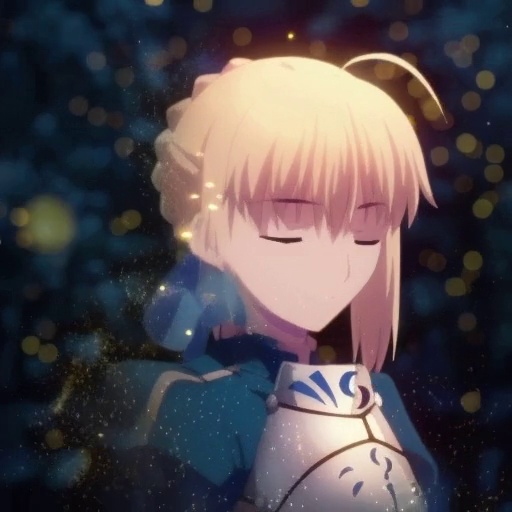}}\hspace{\abWidth}
    }
    
    \vspace{\abHeight}
    \subfloat[without LoRA]{
        \subfloat{\includegraphics[width = 0.113\linewidth]{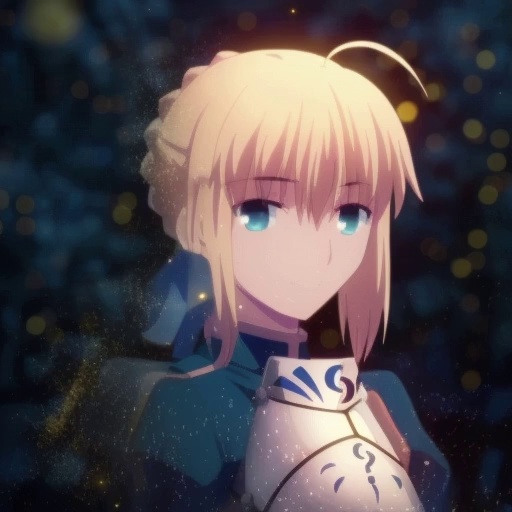}}\hspace{\abWidth}
        \subfloat{\includegraphics[width = 0.113\linewidth]{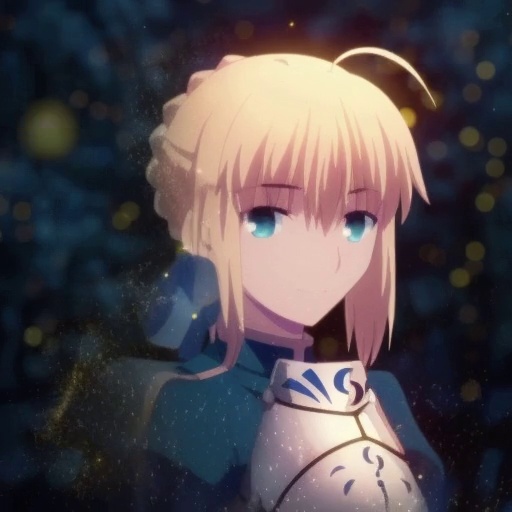}}\hspace{\abWidth}
        \subfloat{\includegraphics[width = 0.113\linewidth]{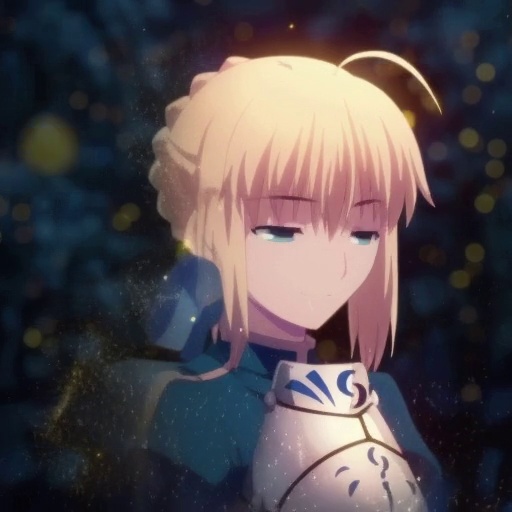}}\hspace{\abWidth}
        \subfloat{\includegraphics[width = 0.113\linewidth]{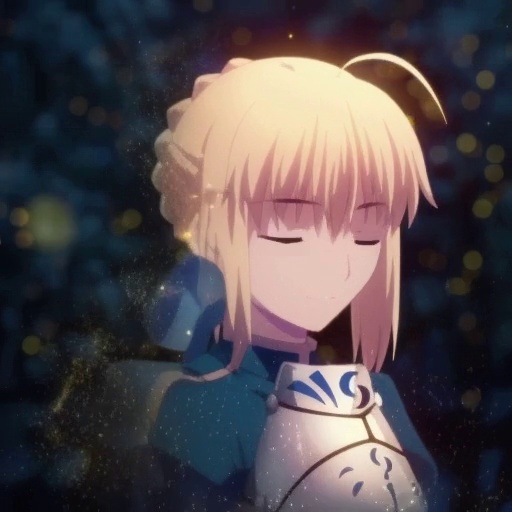}}\hspace{\abmWidth}
    }
    \subfloat[without MSA]{
        \subfloat{\includegraphics[width = 0.113\linewidth]{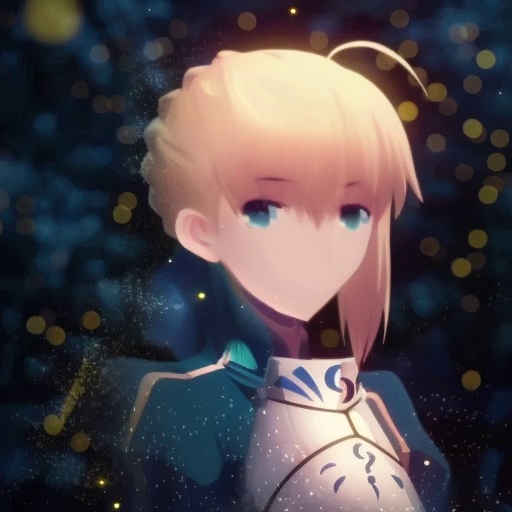}}\hspace{\abWidth}
        \subfloat{\includegraphics[width = 0.113\linewidth]{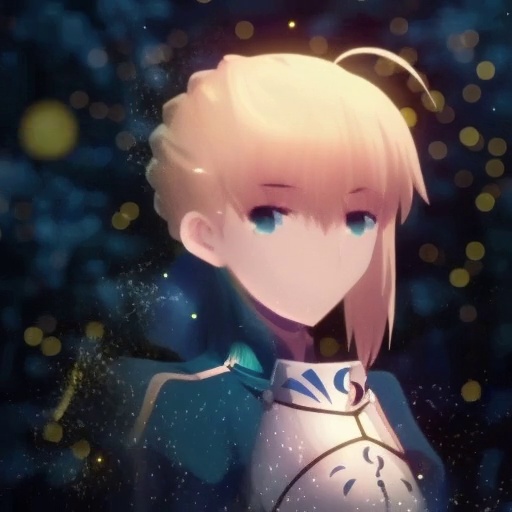}}\hspace{\abWidth}
        \subfloat{\includegraphics[width = 0.113\linewidth]{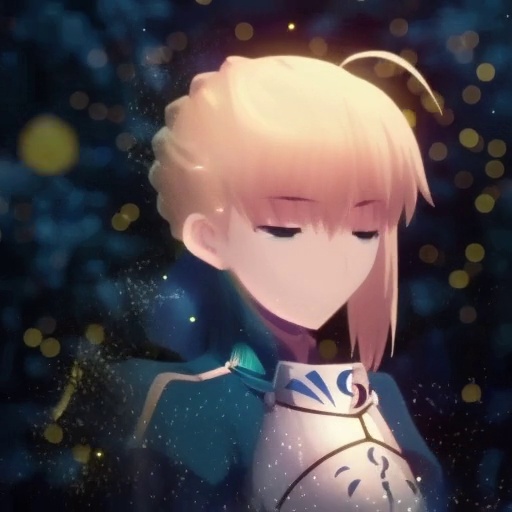}}\hspace{\abWidth}  
        \subfloat{\includegraphics[width = 0.113\linewidth]{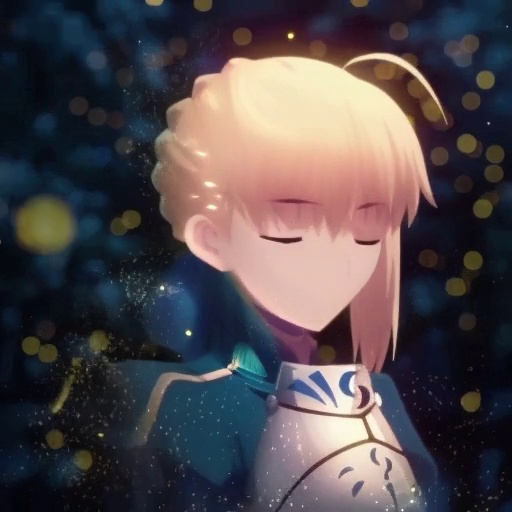}}\hspace{\abWidth}
    }
    
    \vspace{\abHeight}
    \subfloat[Propagated Instructions]{
        \subfloat{\includegraphics[width = 0.113\linewidth]{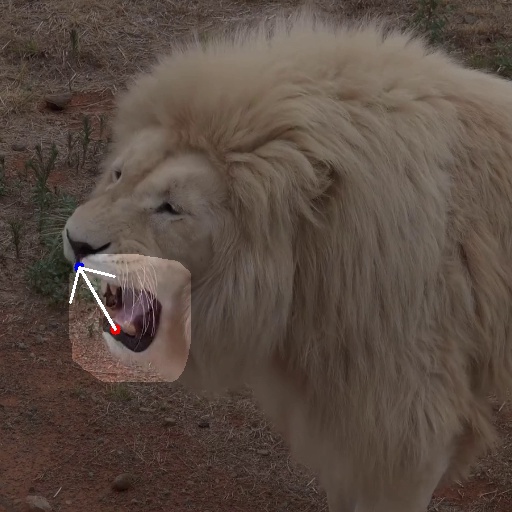}}\hspace{\abWidth}
        \subfloat{\includegraphics[width = 0.113\linewidth]{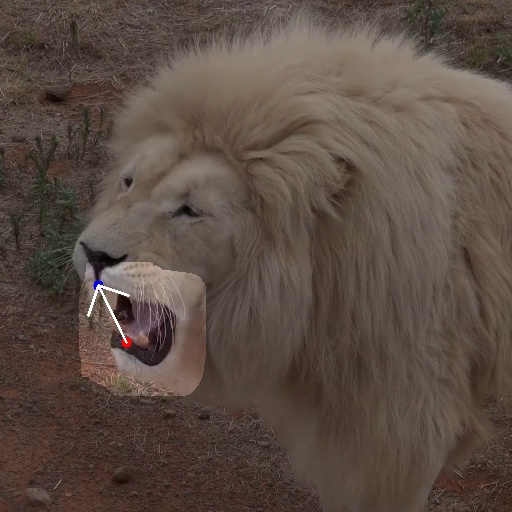}}\hspace{\abWidth}
        \subfloat{\includegraphics[width = 0.113\linewidth]{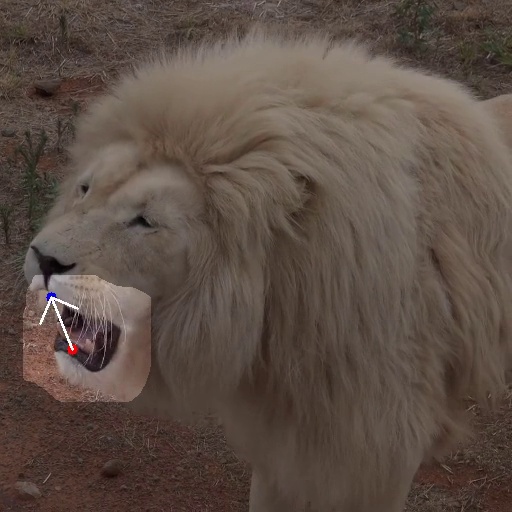}}\hspace{\abWidth}
        \subfloat{\includegraphics[width = 0.113\linewidth]{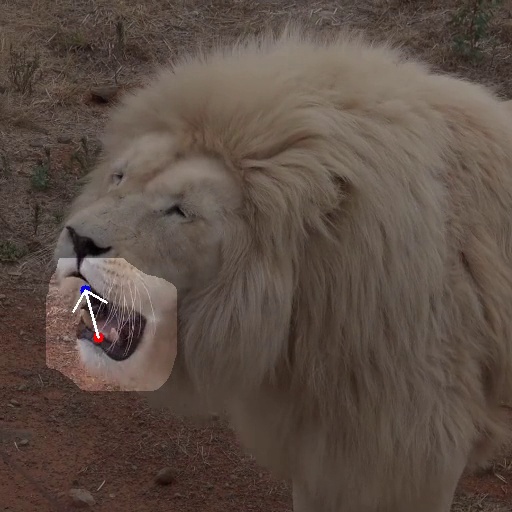}}\hspace{\abmWidth}
    }
    \subfloat[Edited Output]{
        \subfloat{\includegraphics[width = 0.113\linewidth]{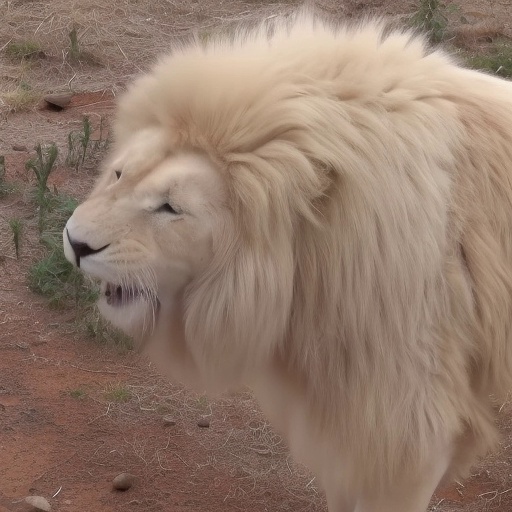}}\hspace{\abWidth}
        \subfloat{\includegraphics[width = 0.113\linewidth]{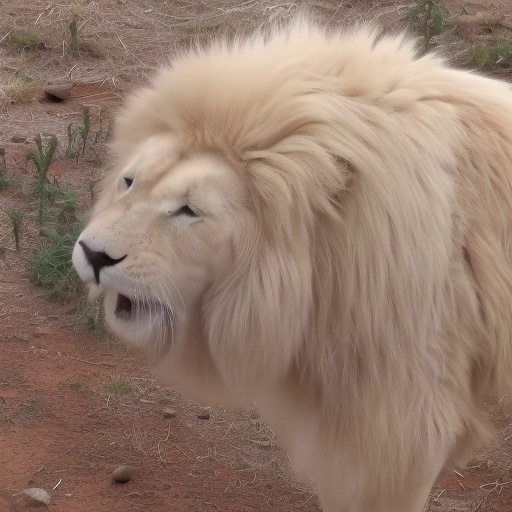}}\hspace{\abWidth}
        \subfloat{\includegraphics[width = 0.113\linewidth]{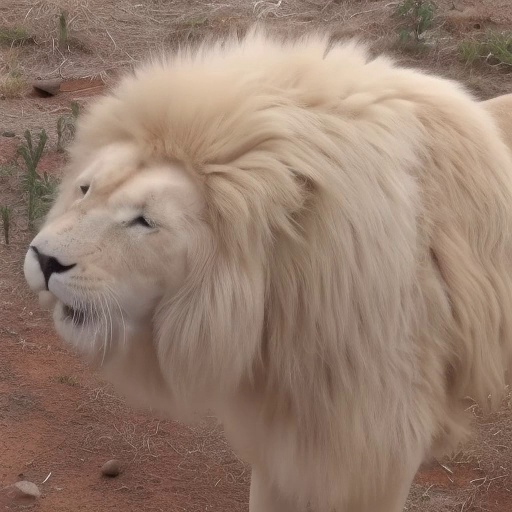}}\hspace{\abWidth}
        \subfloat{\includegraphics[width = 0.113\linewidth]{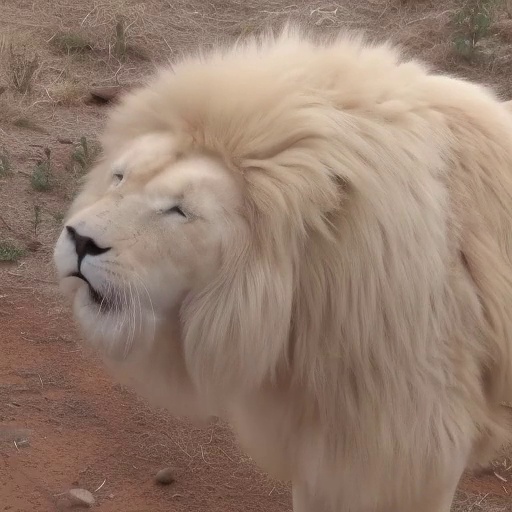}}\hspace{\abWidth}
    }
    
    \vspace{\abHeight}
    \subfloat[without LoRA]{
        \subfloat{\includegraphics[width = 0.113\linewidth]{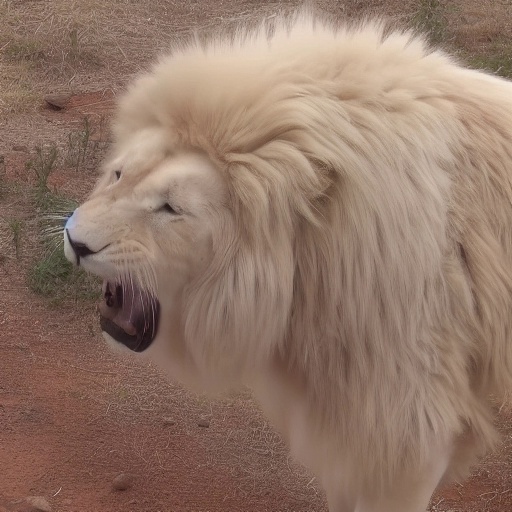}}\hspace{\abWidth}
        \subfloat{\includegraphics[width = 0.113\linewidth]{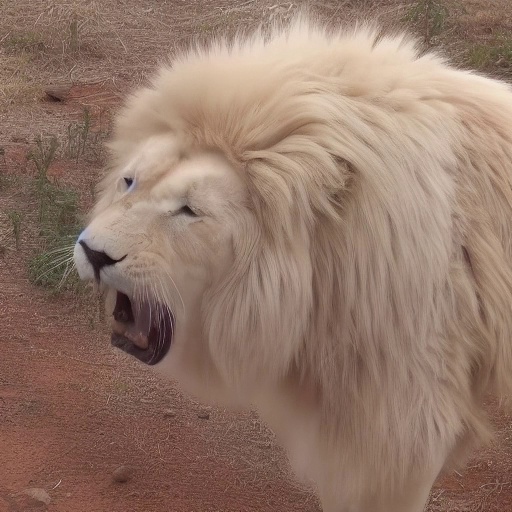}}\hspace{\abWidth}
        \subfloat{\includegraphics[width = 0.113\linewidth]{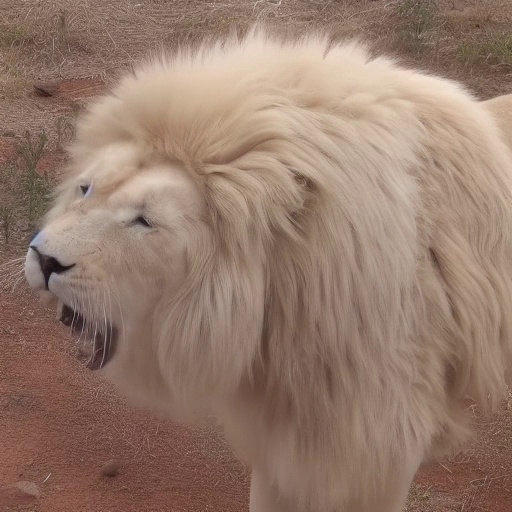}}\hspace{\abWidth}
        \subfloat{\includegraphics[width = 0.113\linewidth]{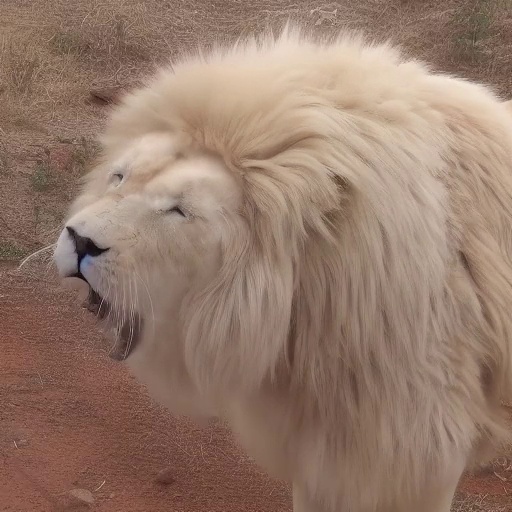}}\hspace{\abmWidth}
    }
    \subfloat[without MSA]{
        \subfloat{\includegraphics[width = 0.113\linewidth]{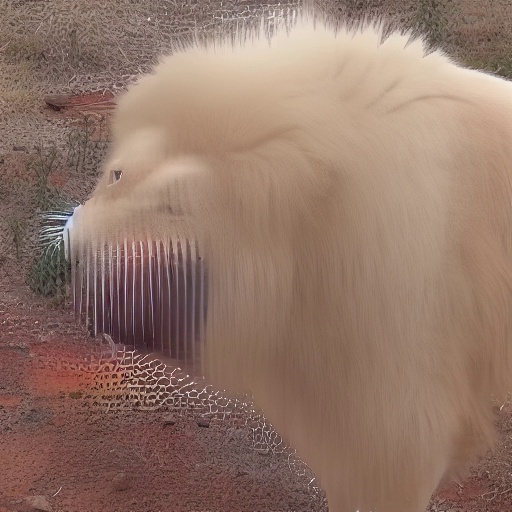}}\hspace{\abWidth}
        \subfloat{\includegraphics[width = 0.113\linewidth]{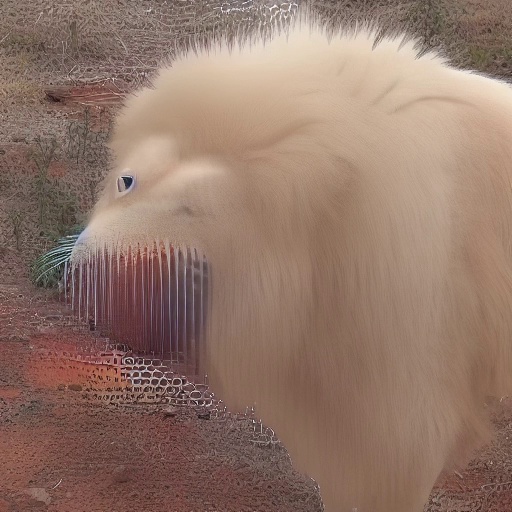}}\hspace{\abWidth}
        \subfloat{\includegraphics[width = 0.113\linewidth]{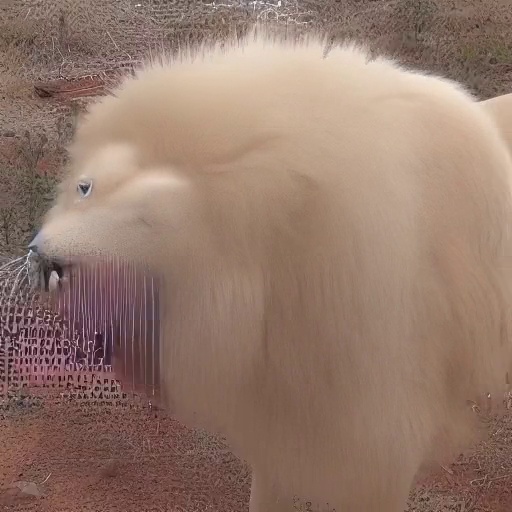}}\hspace{\abWidth}  
        \subfloat{\includegraphics[width = 0.113\linewidth]{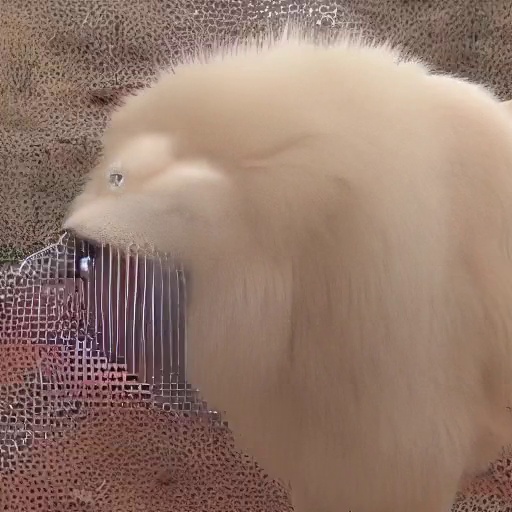}}\hspace{\abWidth}
    }
    \caption{More ablation study of DragVideo. Missing LoRA (bottom left) and missing MSA (bottom right) reduce the quality of DragVideo's outputs. DragVideo with all components (top right) achieves better results.}
    \label{fig:supp_ablation}
\end{figure*}

This section presents two further ablation test results, as depicted in Figure \ref{fig:supp_ablation}. The ablation tests reveal that the removal of LoRA tends to result in the dragged content not reaching the target location. This phenomenon can be attributed to the task-specific LoRA's role in encoding information from the original video. In its absence, the video U-Net lacks the necessary information to move the content logically, thereby making the drag effect less noticeable. On the other hand, when the MSA is removed, a significant deterioration in reconstruction is observed, characterized by a loss of information post-dragging. This is because the video latent is highly sensitive to change, where even slight alterations may result in substantial differences in the Keys and Values within attention blocks in video U-Net. Consequently, the output without MSA exhibits severe issues with reconstruction consistency. These ablation tests underscore both the validity and the critical importance of the LoRA and MSA components within DragVideo.

\section{UI for easy editing}

\begin{figure*}[t]
    \captionsetup[subfigure]{labelformat=empty}
    \centering
    
    \begin{subfigure}[b]{0.85\linewidth}
        \centering
        \includegraphics[width=0.85\linewidth]{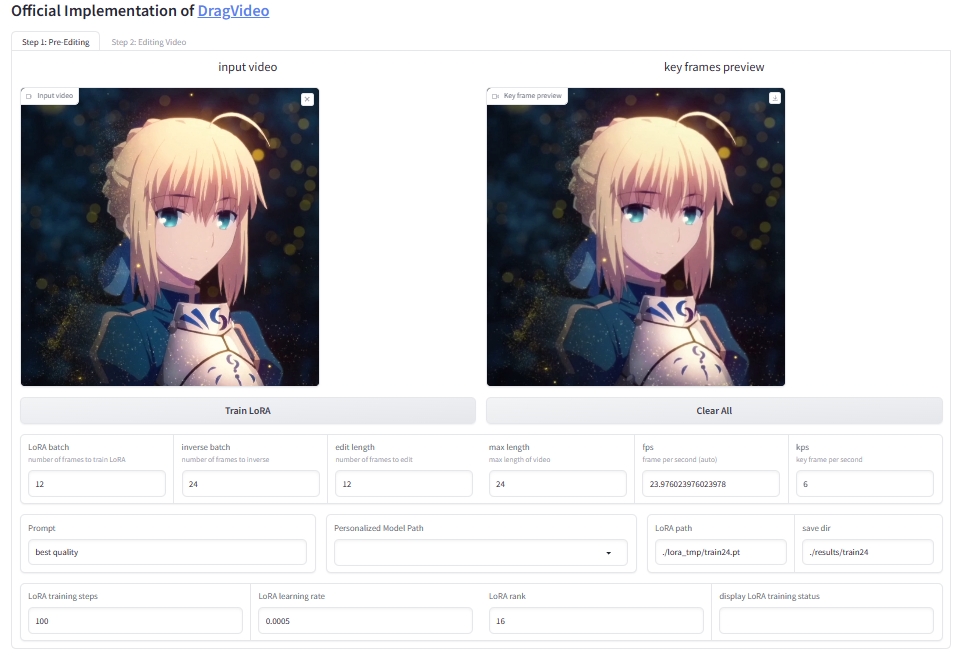}
        \caption{(a) Pre-editing process of video}
        \label{suppUI:preEdit}
    \end{subfigure}
    
    \begin{subfigure}[b]{0.85\linewidth}
        \centering
        \includegraphics[width=0.85\linewidth]{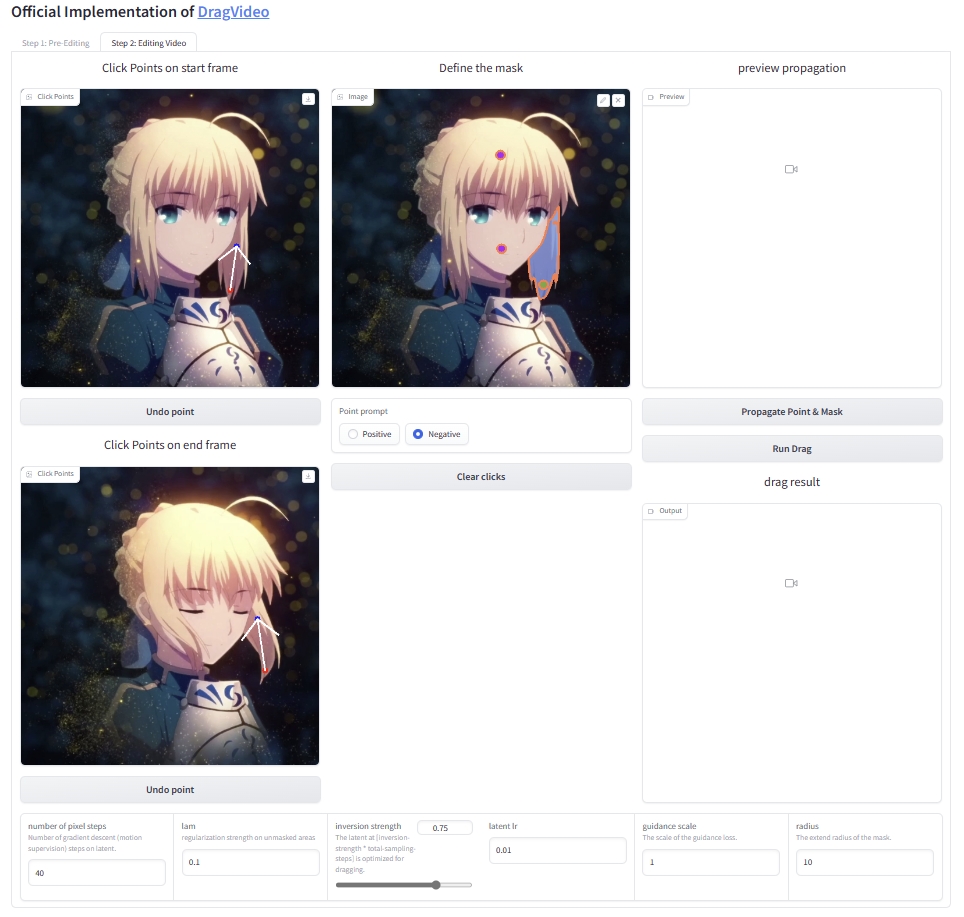}
        \caption{(b) Interactively edit the video}
        \label{suppUI:edit}
    \end{subfigure}
    \vspace{-0.1in}
    \caption{Overview of our implemented GUI, the first page is for pre-editing settings for our UI and the second page is for setting points, masks, and performing edit}
    \label{fig:suppUI}
    \vspace{-0.15in}
\end{figure*}

This section outlines the usability of our implemented User Interface (UI), the view of our UI can be found at Figure \ref{fig:suppUI}. Designed to facilitate effortless interactive video editing, our UI is composed of two primary stages. Initially, users have the convenience of setting keyframe process parameters and training LoRA parameters (Figure \ref{suppUI:preEdit}). For example, they can determine the desired frames per second (fps) post-video processing by assigning a specific value to the "kps" parameter. After uploading and processing the input video using the specified parameters, a preview of the processed video can be viewed. If the results are satisfactory, users can initiate the task-specific LoRA training by clicking the "Train LoRA" button.

Following the successful training of LoRA, users can navigate to the second stage of our UI. Here, they can set the drag instructions by clicking points on the first and last frames (Figure \ref{suppUI:edit}). Our UI also offers users the ability to design the mask in the first frame for precise mask tracking, utilizing positive and negative points. Additionally, users can manipulate the "radius" parameter to increase the mask size. Once these modifications are complete, the "Propagate point \& mask" button can be clicked to generate a preview video with the propagated drag instructions and masks. After scrutinizing the propagation outcomes, users have the choice to refine the latent optimization parameters, for example, the latent learning rate. Finally, the "run" button initiates the video editing process.

In conclusion, our UI offers a streamlined workflow for interactive video editing. Users can easily configure parameters, preview processed videos, train task-specific LoRA models, set drag instructions and masks, propagate them, and ultimately execute video editing tasks with minimal effort. To further illustrate the use of our UI for interactively dragging the video, please refer to the $ui\_demo.mp4$ file in the supplementary materials folder.

\end{appendix}

\end{document}